\documentclass[letterpaper,twocolumn,10pt]{article}
\usepackage{usenix}

\usepackage{tikz}
\usepackage{amsmath}

\usepackage{amsfonts} 
\usepackage{amsmath}
\usepackage{algorithm}
\usepackage{xcolor}
\usepackage{algpseudocode}
\algrenewcommand\algorithmiccomment[1]{\textit{\textcolor{gray}{// #1}}} %
\usepackage{subfigure}
\usepackage{gensymb}
\usepackage{multirow}
\usepackage{multicol}
\usepackage{booktabs}
\usepackage{soul}
\usepackage[colorinlistoftodos,prependcaption,textsize=tiny,textwidth=1.2cm]{todonotes}
\usepackage{setspace}
\usepackage{authblk}

\renewcommand{\vec}[1]{\mathbf{#1}} %
\newcommand{\vecg}[1]{\boldsymbol{#1}} %

\let\oldthebibliography\thebibliography
\renewcommand{\thebibliography}[1]{
  \oldthebibliography{#1}
  \fontsize{8pt}{4pt}\selectfont
}

\begin{document}


\date{}

\title{\Large \bf A First Physical-World Trajectory Prediction Attack via LiDAR-induced Deceptions in Autonomous Driving}

\author{\normalsize Yang Lou{$^{*1}$}, Yi Zhu{$^{*2}$}, Qun Song{$^{*3}$}, Rui Tan{$^{4}$}, Chunming Qiao{$^2$}, Wei-Bin Lee{$^{5,6}$}, Jianping Wang{$^{1}$}}
\affil{
 	\textit{\hspace{0.0in}$^{1}$\,City University of Hong Kong \hspace{0.2in}$^{2}$\,State University of New York at Buffalo} \\
        \textit{\hspace{0.0in}$^{3}$\,Delft University of Technology \hspace{0.2in}$^{4}$\,Nanyang Technological University} \\
	  \textit{\hspace{0.0in}$^{5}$\,Information Security Center, Hon Hai Research Institute} \\
        \textit{\hspace{0.0in}$^{6}$\,Department of Information Engineering and Computer Science, Feng Chia University}
}

\maketitle


\let\oldthefootnote\thefootnote
\let\thefootnote\relax\footnotetext{*Equal contribution}
\let\thefootnote\oldthefootnote

\vspace{-1em}
\begin{abstract}
Trajectory prediction forecasts nearby agents' moves based on their historical trajectories. Accurate trajectory prediction (or prediction in short) is crucial for autonomous vehicles (AVs). Existing attacks compromise the prediction model of a victim AV by \emph{directly} manipulating the historical trajectory of an attacker AV, which has limited real-world applicability. This paper, for the first time, explores an \emph{indirect} attack approach that induces prediction errors via attacks against the perception module of a victim AV. Although it has been shown that physically realizable attacks against LiDAR-based perception are possible by placing a few objects at strategic locations,  it is still an open challenge to find an object location from the vast search space in order to launch effective attacks against prediction under varying victim AV velocities. 

Through analysis, we observe that a prediction model is prone to an attack focusing on a single point in the scene. Consequently, we propose a novel two-stage attack framework to realize the single-point attack. The first stage of prediction-side attack efficiently identifies, guided by the distribution of detection results under object-based attacks against perception, the state perturbations for the prediction model that are effective and velocity-insensitive. In the second stage of location matching, we match the feasible object locations with the found state perturbations. 
Our evaluation using a public autonomous driving dataset shows that our attack causes a collision rate of up to $63\%$ and various hazardous responses of the victim AV.
The effectiveness of our attack is also demonstrated on a real testbed car~\footnote{A demo video of our attack on a real testbed car is available at \url{https://1drv.ms/v/s!Aoc_mWfaEyaGbrdCPNS9oKMjm9Q?e=f1GA6u}.}. To the best of our knowledge, this study is the first security analysis spanning from LiDAR-based perception to prediction in autonomous driving, leading to a realistic attack on  prediction. To counteract the proposed attack, potential defenses are discussed.

\end{abstract}

\vspace{-1.1em}
\section{Introduction}
\vspace{-1.2em}

\begin{figure}[t]
    \centering
    \hfill
    \subfigure[A scene with adversarial cardboards placed around a parked car.]{
        \label{fig:intro_attack_scene}
        \includegraphics[width=0.45\columnwidth]{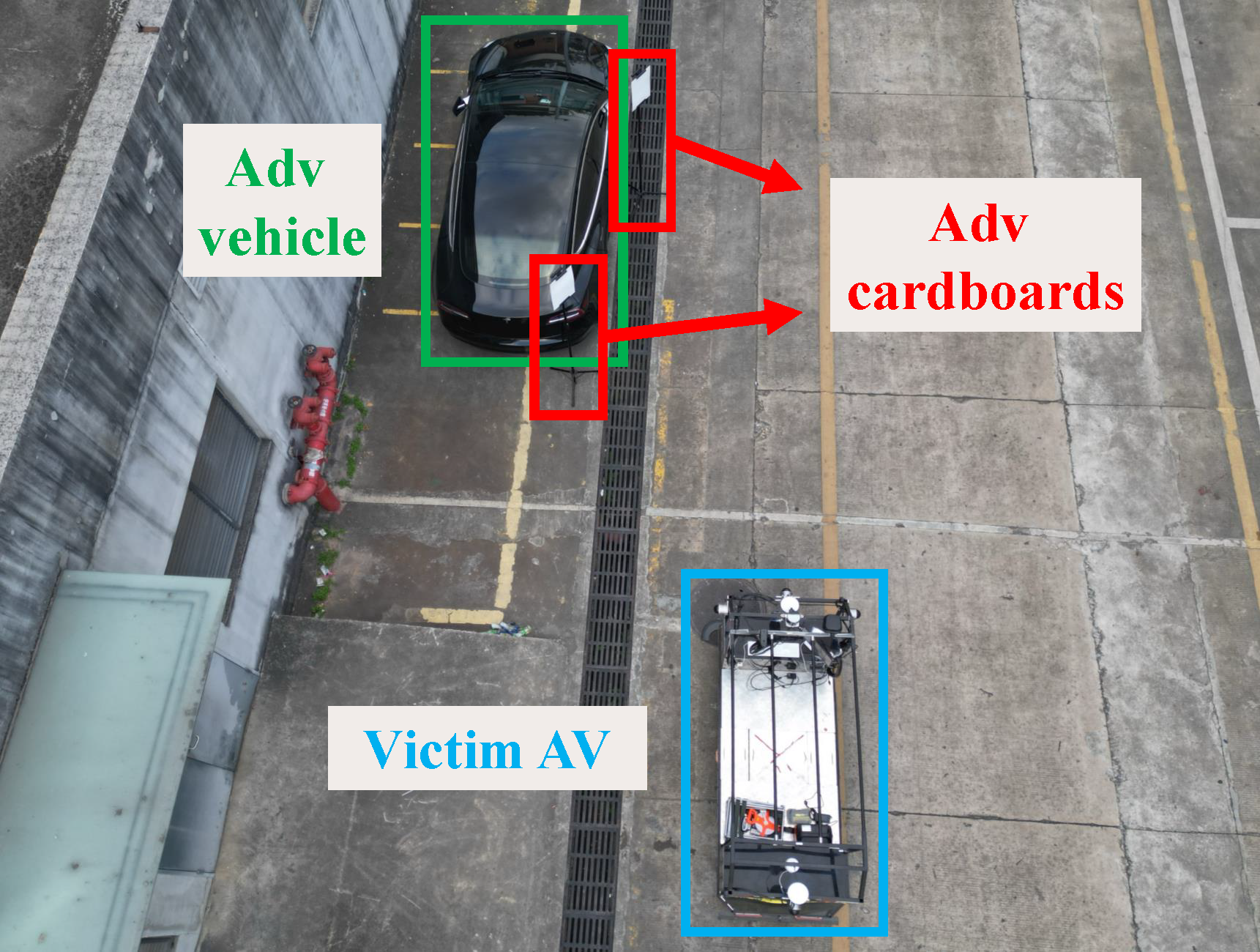}
    }
    \hfill
    \subfigure[Random object locations cause a prediction error posing no threat.]{
        \label{fig:intro_random_loc}
        \includegraphics[width=0.45\columnwidth]{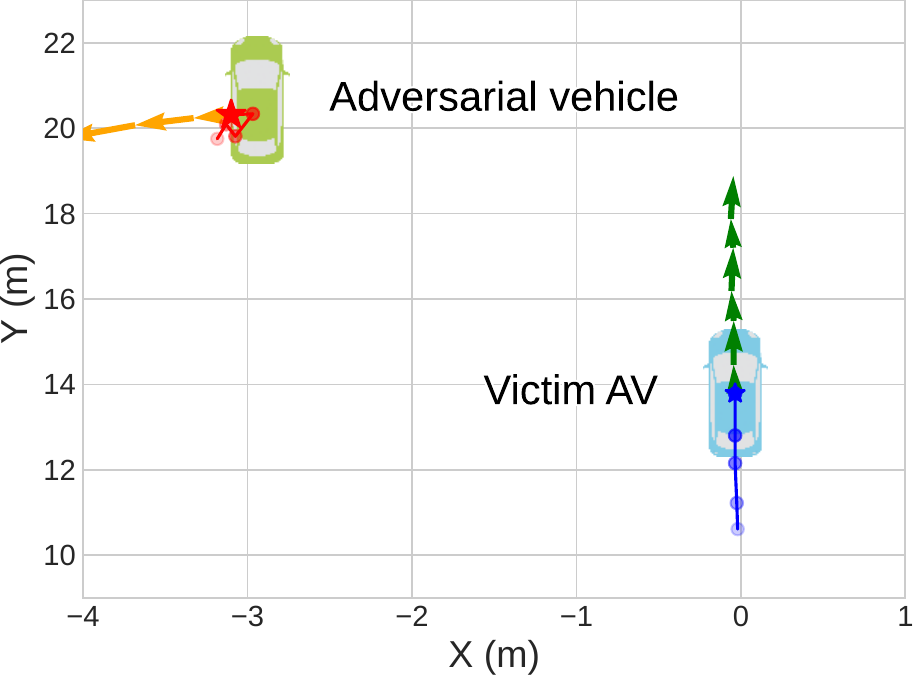}
    }\hfill\\
    \vspace{-0.2em}
    \hfill
    \subfigure[Brute-force search finds object locations threatening victim vehicle.]{
        \label{fig:intro_forward_mod}
        \includegraphics[width=0.45\columnwidth]{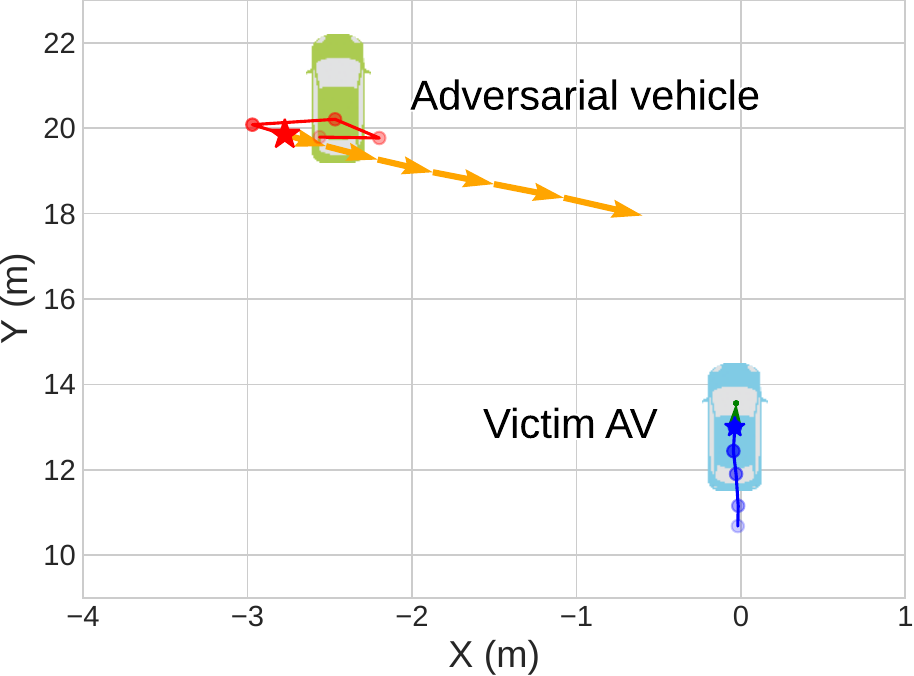}
    }
    \hfill
    \subfigure[No longer threatening for victim vehicle approaching at higher speed.]{
        \label{fig:intro_forward_high}
        \includegraphics[width=0.45\columnwidth]{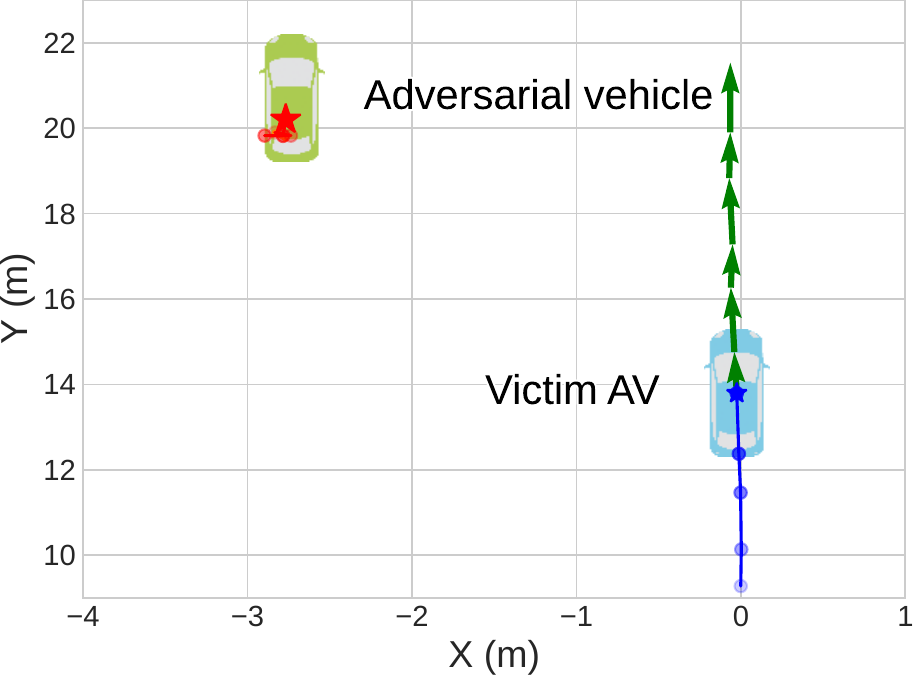}
    }
    \hfill
    \vspace{-1em}
    \caption{A motivation experiment.}
    \vspace{-1em}
    
\end{figure}

Autonomous vehicles (AVs) are transforming transportation systems worldwide. Autonomous driving (AD) systems, e.g., Autoware.AI~\cite{autoware_ai}, typically consist of perception, prediction, and planning modules. In this AD system pipeline, the perception module utilizes sensors, such as LiDAR, to detect and track on-road objects; the prediction module forecasts the future trajectories of nearby agents; the planning module determines the AV's future driving behavior. The predicted trajectories are particularly vital for the subsequent planning module as they significantly influence an AV's driving behavior. 
The existing studies~\cite{cao2022advdo,zhang2022adversarial,tan2023targeted} directly compromise the prediction module of a victim AV by manipulating the historical trajectory of an attacker agent nearby, which is usually a vehicle. In these attacks, the attacker's vehicle must drive according to a specific {\em adversarial driving trajectory} to mislead the victim AV's prediction module into generating wrong future trajectory prediction for the attacker's vehicle. However, it is hard to precisely drive the attacker's vehicle along a pre-computed trajectory due to kinematic constraints. Moreover, such adversarial trajectories are designed based on a restrictive assumption that the victim AV would drive at a certain velocity.



To address these limitations, in this paper, we explore the possibility of indirectly compromising the prediction module via perturbing the perception module of the victim AV such that the perception errors can induce the prediction module to predict a wrong trajectory of a nearby agent. The existing physically-realizable attacks against the LiDAR perception~\cite{petit2015remote,shin2017illusion,cao2019adversarial,sun2020towards} provide a basis for such indirect attack on the prediction module. Among them, object-based attacks, which use specific shapes or common objects to introduce additional adversarial LiDAR points captured by the victim AV, have lower logistics overhead in attack implementation. 
However, the impact of the perturbed bounding box attributes (including coordinates, dimensions, and heading) as a result of the object-based attack on the subsequent prediction in the AD pipeline has never been systematically studied.

To study the feasibility of compromising the prediction model indirectly via object-based attacks to induce dangerous driving decisions, we conduct a motivation experiment on our custom-built real testbed AV. We employ an object-based attack method from~\cite{zhu2021can}, due primarily to its practicality, to plan the locations for placing two {\em adversarial cardboards} around a car parked roadside, called {\em adversarial vehicle}, as shown in Fig.~\ref{fig:intro_attack_scene}. Then, the victim AV running the LiDAR-based AD pipeline drives by.
Fig.~\ref{fig:intro_random_loc} shows that random object placement may result in a wrongly predicted trajectory of the adversarial vehicle, which stays stationary in reality. However, as the predicted trajectory points to the left, it imposes no threat on the victim AV. Then, we employ brute-force search to find the locations of the two adversarial cardboards that lead to a collision between the victim AV's original trajectory and the adversarial vehicle's trajectory predicted by the victim AV. The search is based on the victim AV's original trajectory. As shown in Fig.~\ref{fig:intro_forward_mod}, the attack-induced illusive collision leads to a sudden brake decided by the victim AV's planning module. This result shows the existence of object-based attacks that generate threatening effects penetrating the AV pipeline. However, when the victim AV runs at a varied velocity causing a deviation from the spatio-temporal trajectory assumed by the attacker's brute-force search, the attack becomes no threatening as shown in Fig.~\ref{fig:intro_forward_high}.
This paper aims to design an efficient attack approach to identify locations for placing adversarial objects that lead to hazardous responses of the victim AV regardless of its driving velocity.

Our vulnerability analysis via experiments shows that the AV's trajectory prediction is prone to substantial perception errors when the moving victim AV is at a point in the scene. Based on this observation, we choose to reduce the attack design space to focus on a fixed {\em attack point} and perturb the victim AV's perception to induce prediction errors, when it arrives at the attack point. This strategy greatly reduces the search space yet largely preserves the potential of the attack in leading to the victim AV's hazardous responses.
However, as the historical LiDAR frames captured before the victim AV arrives at the attack point form a part of the prediction model's input, the victim AV's velocity still affects the effectiveness of the single-point attack.
In addition, the exhaustive search for the locations of multiple adversarial objects in the 3D proximity space of the adversarial vehicle, while considering all possible victim AV's velocity values for robustness to velocity variations, still incur undesirable overhead.


To address the above two issues
related to the single-point attack strategy, we design a novel two-stage attack framework. In the stage of \textit{prediction-side attack}, we maximize attack effectiveness by searching for input state perturbations for the prediction model that 
maximize the interference between the predicted trajectory of the adversarial vehicle and the victim AV's planned trajectory. To maintain attack effectiveness across various velocity conditions, we employ the Expectation over Transformation (EoT) technique. 
Our experiment shows that, with object-based attacks, the perturbed detection results (i.e., bounding boxes) of the adversarial vehicle exhibit distributional patterns. Thus, we utilize the distribution to guide the optimization of the state perturbations for the prediction model, which improves the efficiency of attack design.
We use the Projected Gradient Descent (PGD) method to iteratively search for the state perturbations and select the perturbations that lead to collisions. However, the optimized state perturbations may not translate to feasible perturbed detection results caused by the placement of adversarial objects. To achieve attack feasibility, in the stage of \textit{location matching}, we employ the Hungarian algorithm to match the adversarial locations with the optimized state perturbations and further refine the candidate adversarial locations by probing in their close vicinity for reducing the matching cost.

We conduct experiments on both the nuScenes~\cite{caesar2020nuscenes} autonomous driving dataset and our real testbed AV.
In the dataset-based experiments, our attack achieves a collision rate of up to 63\%, effectively inducing various hazardous responses of the victim, including sudden brakes, accelerations, and unexpected lane changes, in the presence of object displacement errors and object size variations.
In the physical-world experiments with our custom-built testbed AV, our attack is effective in various velocity conditions and achieves the highest attack success rate compared with the random location and the brute-force sampling attacks. The experiments also show that our attack is robust against object displacement errors and variations in the victim AV's direction. 

Our main contributions can be summarized as follows:
\vspace{-0.5em}
\begin{itemize}
    \setlength\itemsep{-0.5em}
    \item We design and launch the first physical-world attack on the AD system's trajectory prediction by strategically placing objects to indirectly induce prediction errors, leading to hazardous responses of the victim AV.
    \item We find that the prediction model is susceptible to the single-point attack. Built upon this, we design a novel two-stage attack framework that efficiently identifies adversarial locations for object placement, leading to effective, velocity-insensitive, and feasible attacks.
    \item Our attack is evaluated on both a public autonomous driving dataset and a custom-built real-world testbed AV, achieving consistently higher threats compared with the baselines and attack robustness. Additionally, we propose potential defenses to mitigate the threat.
\end{itemize}

\vspace{-0.5em}
\section{Background and Related Work}

\subsection{Autonomous Driving System}
A typical AD system consists of perception, trajectory prediction, and motion planning modules, as detailed below.


\textbf{LiDAR-based Perception}\label{sec:lidar_based_perception},
known for its precision in acquiring environmental data, is widely adopted in the default pipeline of open-source AD systems~\cite{autoware_universe} and in commercial solutions~\cite{vueron_vueone, velodyne_idriverplus_2019}.
This process starts with 3D object detection, which analyzes a LiDAR point cloud to identify objects around the ego vehicle.
Modern AD perception utilizes deep learning models, denoted by $M_{det}(\cdot)$, to process LiDAR point cloud $D$ and 
produce a set of 3D bounding boxes $\vec{B}=M_{det}(D)$. A bounding box $\vec{b} \in \vec{B}$, characterized by coordinates, dimensions, heading, and a confidence score, represents a detected object.
Existing 3D object detection methods can be divided into 
Bird's Eye View (BEV)-based~\cite{yang2018pixor}, voxel-based~\cite{yan2018second,yin2021center}, and point-based~\cite{shi2019pointrcnn}. They respectively map point clouds into 2D representation, discretize 3D space into voxels, and operate on raw point cloud directly. Detected objects are tracked over time by an object tracking model $M_{track}$.

\textbf{Trajectory Prediction} forecasts the future trajectories of road agents (vehicles and pedestrians) based on their current and historical states perceived over $H$ time steps denoted by
$\vec{X} = (\vec{X}^{-H+1}, \ldots, \vec{X}^{0})$, where $\vec{X}^t = (\vec{x}^t_{1}, \ldots, \vec{x}^t_{N_a})$ represents the states of $N_a$ agents at time step $t$ and each $\vec{x}^t$ includes an agent's coordinates, velocity, acceleration, and heading.
The trajectory prediction for a horizon of $T$ time steps by a predictor $M_{pred}(\cdot)$ is $\vec{Y} = (\vec{Y}^{1}, \ldots, \vec{Y}^{T}) = M_{pred}(\vec{X})$, where $\vec{Y}^{t} = (\vec{y}^t_{1}, \ldots, \vec{y}^t_{N_a})$ is the predicted coordinates of the $N_a$ agents at time step $t$.
Existing trajectory predictors employ recurrent neural networks~\cite{lee2017desire,luo2020probabilistic}, graph neural networks~\cite{liang2020learning,zeng2021lanercnn}, and transformers~\cite{liu2021multimodal,zhou2022hivt}.
Recent generative predictors such as Trajectron++~\cite{salzmann2020trajectron++} and AgentFormer~\cite{yuan2021agentformer}, which perform conditional sampling and selection, have shown superior performance and are therefore employed in this work.

\textbf{Motion Planning} takes into account road agents' current states $\vec{X}^0$ and trajectory prediction results $\vec{Y}$ to plan a safe, efficient, and feasible future trajectory $\vec{P} = M_{plan}(\vec{X}^0, \vec{Y})$ for the ego vehicle. Existing solutions can be divided into sampling-based~\cite{kuwata2009real}, control-based~\cite{chen2022interactive}, graph search-based~\cite{dolgov2010path}, and learning-based methods~\cite{scheel2022urban,zeng2019end}.



\vspace{-0.5em}
\subsection{Adversarial Attacks}\label{sec:related_work_adv_attack}
Deep learning models are shown to be vulnerable to adversarial examples~\cite{goodfellow2014explaining,papernot2017practical,tsai2020robust,sharif2016accessorize,lee2019physical,ma2023slowtrack,xie2017adversarial,carlini2018audio,zhang2021argot,suya2020hybrid}, which are perturbed input samples that mislead the model to generate erroneous outputs.
Specifically, given a model $M(\cdot)$, for a benign input $x$ with ground truth label $y$, the attack searches for a minimal perturbation $\delta$ such that $M(x+\delta) \neq y$ (non-targeted attack) or $M(x+\delta) = y'$ (targeted attack), where $y'$ is the target label. Adversarial examples may pose great threat on the safety-critical AD systems that employ deep learning models for perception and prediction.
In what follows, we review the existing adversarial attacks on LiDAR-based perception, trajectory prediction, and motion planning, respectively.

The attacks on LiDAR-based perception are studied in both simulations~\cite{hau2021object, hau2022using} and physical environments.
Physical-space attacks fall into two groups: \textit{laser-based}~\cite{petit2015remote,shin2017illusion,cao2019adversarial,sun2020towards,sato2023wip,jin2023pla,cao2023you} and \textit{object-based}~\cite{tu2020physically,zhu2021can,cao2021invisible,zhu2021adversarial}. The former, also known as spoofing, injects fake points by intercepting LiDAR's emitted laser pulses using a receiver and then transmitting counterfeit pulses back to the LiDAR sensor with a manipulated delay.
However, such attacks require precise timing and specialized equipment, making them costly and conspicuous.
Object-based attacks use physical objects to mislead LiDAR detection models. For example, in~\cite{zhu2021can}, a vehicle can be hidden from LiDAR detection by placing common objects at computed adversarial locations near the vehicle.
Both laser-based and object-based attacks usually hide or create objects by manipulating bounding box confidence scores. Yet, their effects on bounding box parameters (coordinates, dimensions, and heading) and the impact of induced errors in estimating these parameters on subsequent AD modules remain largely unexplored.
This work considers object-based attacks that are more realistic for real-world deployments.

In the tracking module, tracker hijacking attacks~\cite{jia2020fooling,jha2020ml,jia2020robust,yan2020hijacking,chen2021unified,muller2022physical} perturb camera image inputs to manipulate bounding box parameters. 
A primary goal of these attacks is object move-in, which manipulates the tracker of a roadside object towards the road center, potentially inducing falsified trajectories similar to ours due to pipeline effects.
However, these studies focus exclusively on the tracking module, while our attack optimizes such effects in the downstream trajectory prediction module. Unlike these studies that focus on camera-only systems, our attack applies to LiDAR-based perception, introducing unique challenges and attack vectors that necessitate new methodologies.

The attacks on trajectory prediction perturb the input states of the adversarial vehicle and mislead the victim AV's prediction model. 
However, existing methods either neglect the kinematic laws governing the adversarial vehicle~\cite{zhang2022adversarial,tan2023targeted} or disregard the driving uncertainties of the victim AV, including speed variations~\cite{cao2022advdo,cao2023robust}, which could alter the adversarial vehicle's input states during the attack.
Differently, this paper plans the object-based attacks that generate pipeline effect on LiDAR-based perception and then trajectory prediction.  
There are a few studies focusing on physical attacks against motion planning. The work~\cite{wan2022too} uses common road objects to trigger a Semantic Denial-of-Service in the motion planning module, causing emergency stops or critical driving decision failures. In comparison, this paper targets the trajectory prediction module in AD systems. Regarding methods, the work in~\cite{wan2022too} exploits flaws in programming code logic, while our attack focuses on vulnerabilities in learning-based perception and prediction models, highlighting a distinct approach in exploiting AD systems' weaknesses.

\vspace{-0.5em}
\section{Threat Model}\label{sec:problem_definition}
\begin{figure}[t]
    \centering
    \includegraphics[width=0.8\columnwidth]{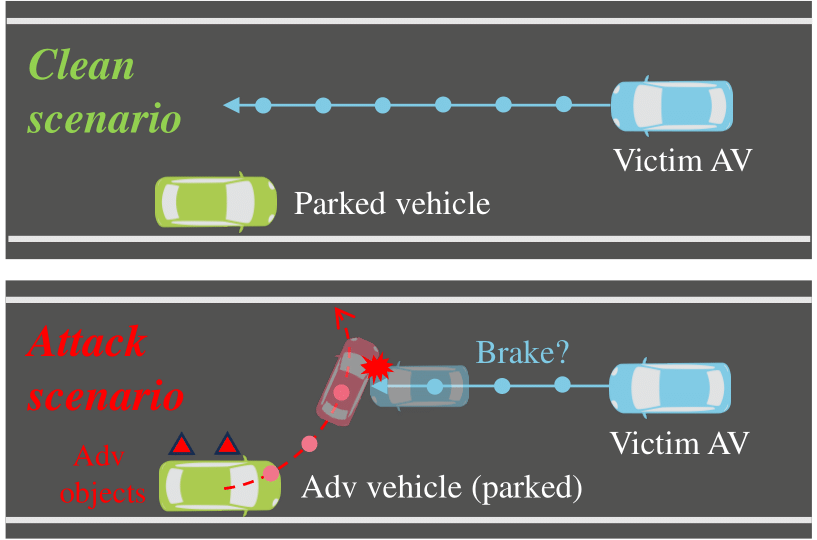}
    \caption{Our attack scenario. Solid blue arrows are victim AV's planned trajectories; dash red arrow is adversarial vehicle's future trajectory predicted by the victim AV.}
    \label{fig:attack_scenario}
    \vspace{-0.5em}
\end{figure}

\textbf{Attack Goal.}
We consider an attack scenario depicted in Fig.~\ref{fig:attack_scenario}, where a static adversarial vehicle is parked on the roadside, while a victim AV is driving and approaching it on the adjacent lane.
Such a scenario is common in urban driving environments, such as street parking, industrial areas, and parking lots.
We assume that the victim AV runs the AD pipeline of LiDAR-based perception, trajectory prediction, and motion planning. 
The attack goal is to manipulate the victim AV's perception, such that the prediction forecasts an erroneous moving trajectory for the static adversarial vehicle and causes the victim AV's motion planning to decide a hazardous driving strategy. 
This can be motivated by intentional harm towards an individual, unhealthy competition among solution vendors aiming to undermine the safety reputation of rivals’ self-driving products, or the intent to disrupt traffic mobility by causing traffic jams~\cite{npr2023traffic}.
The manipulation of the victim AV's perception is achieved by placing some adversarial common objects, e.g., cardboards, around the adversarial vehicle. 
The key for the perception manipulation is to identify the \textit{adversarial locations} for placing the adversarial objects, such that the victim AV's LiDAR-based perception generates inaccurate bounding boxes over time for the adversarial vehicle.
The attack aims at misleading the victim AV to produce a falsified trajectory of the adversarial vehicle that intersects the victim AV's planned trajectory. As a result, the victim AV re-plans the motion to avoid the illusive collision with safety-undermining driving decisions, including emergent braking, sudden acceleration and/or lane change.

\textbf{Attacker Capabilities.}
We assume that the attacker is capable of placing objects around an adversarial vehicle, which can be the attacker's own car or a random car parked on the roadside. 
We consider a realistic but challenging setting in which the attacker has no access to the real-time data (e.g., LiDAR point cloud) collected by the victim AV. We consider the white-box attacker that has comprehensive knowledge of the victim AV's AD pipeline. This enables the attacker to query and analyze $M_{det}(\cdot)$, $M_{track}$, and $M_{pred}(\cdot)$ for crafting attacks.
To obtain these three models, the attacker can use social engineering on the system designers of the car manufacturers or reverse-engineer a vehicle identical to the victim AV. For the vehicles employing open-source AD systems~\cite{autoware_ai,baidu_apollo}, the overhead of obtaining the three models $M_{det}(\cdot)$, $M_{track}$, and $M_{pred}(\cdot)$ can be lower.
Before launching the attack, the attacker gathers data by driving a LiDAR-equipped vehicle to mimic the victim AV's behavior. The attacker uses the data to determine the locations for placing the objects.

\vspace{-0.5em}
\section{Challenges and Problem Definition}
\begin{figure*}[ht]
\centering
\includegraphics[width=0.85\textwidth]{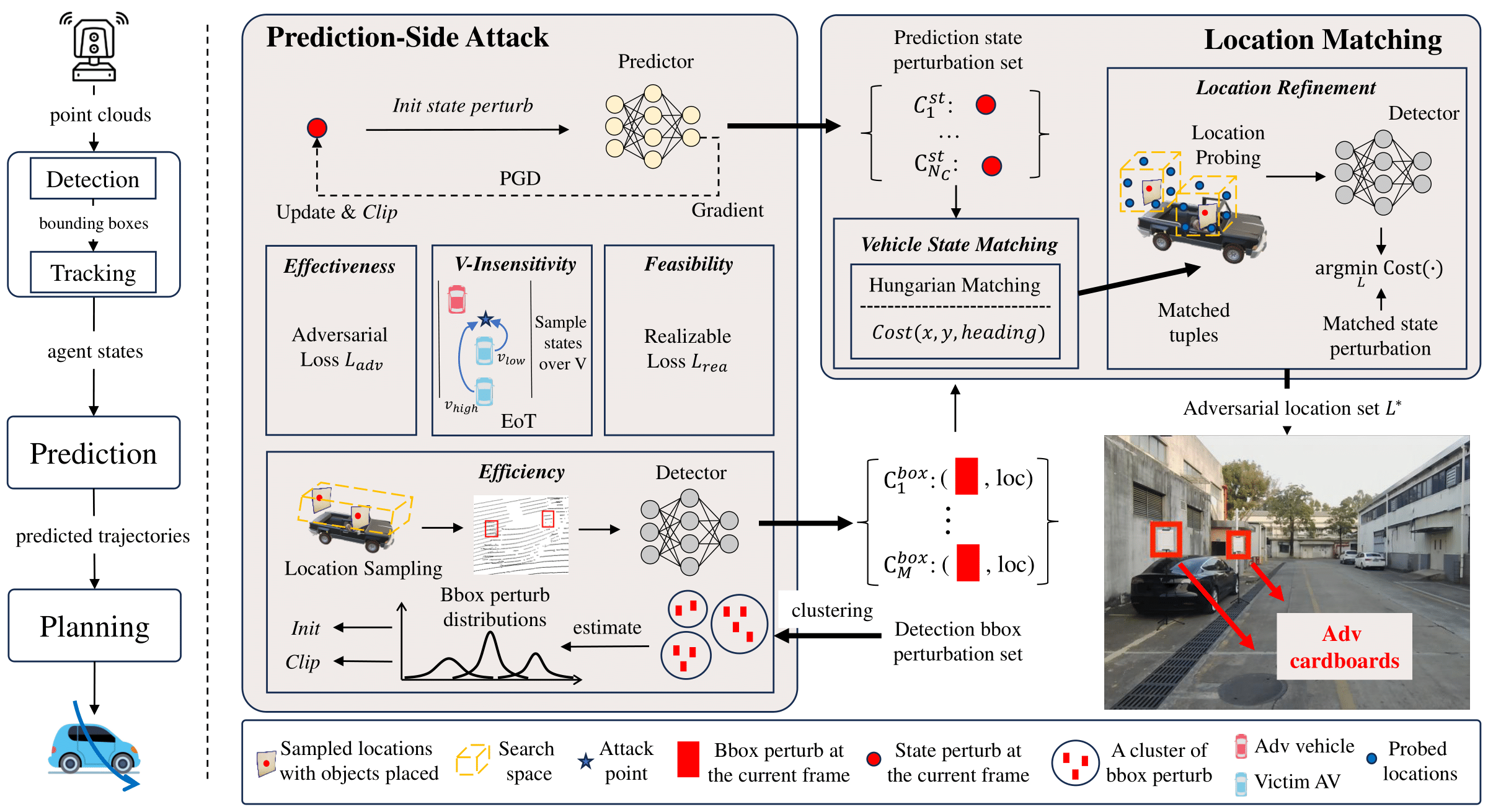}
\vspace{-0.5em}
\caption{Our proposed inverse attack framework for identifying the adversarial locations to place objects (cardboards in red boxes). The framework operates within the AD system pipeline, which is outlined on the left for reference.}
\label{fig:inverse_attack_framework}
\vspace{-0.5em}
\end{figure*}

\subsection{Attack Challenges}
\label{sec:challenges} 


Challenges to identify effective adversarial locations include:

\textbf{Challenge C1: Maximizing Attack Effectiveness.} 
Inducing errors in predicting the adversarial vehicle's trajectory, quantified by metrics such as average displacement error (ADE) and final displacement error (FDE), as in the literature, does not necessarily imply a threat to the victim AV. An erroneously predicted moving trajectory for an adversarial vehicle that is stationary in reality, if not intersecting the victim's planned trajectory, poses no actual danger. To maximize attack effectiveness, a meaningful objective is to minimize the distance between the adversarial vehicle's predicted trajectory and the victim AV's planned trajectory.

\textbf{Challenge C2: Velocity-Induced Uncertainty in Attack Effectiveness.} 
The victim AV may drive by the adversarial vehicle at varying velocities. When the victim AV's velocity is different from that used for attack construction, the victim AV's perception results and the consequent predicted trajectory of the adversarial vehicle may differ from those expected by the attacker.
Thus, should the victim AV deviate from its expected trajectory, the effectiveness of the pre-determined adversarial locations may be reduced.

\textbf{Challenge C3: Vast Search Space for Adversarial Locations.} 
Given the non-differentiable processes within the AD system's independent modules, employing gradient-based optimization to achieve the attack goal presents significant difficulties. A straightforward approach would be exhaustively sampling a multitude of locations in the vicinity of the adversarial vehicle in order to identify the most effective set of adversarial locations.
However, such a brute-force sampling approach faces efficiency challenges, particularly in time-sensitive scenarios. For instance, if an attacker can only determine the victim AV's route after it departs, it has limited time to plan and deploy the attack.
First, even if we restrict the search area to only the top of the adversarial vehicle and its immediate surroundings, the search space is still extensive and grows exponentially with the number of adversarial objects. Second, if the search further accounts for the victim AV's velocity as an additional dimension to address \textbf{C2}, the search space is even larger.

\vspace{-1em}
\subsection{Problem Definition} 
\label{sec:problem-def}
We aim to identify a set of adversarial locations $L=\{l_{n} | n=1,...,N_L\}$, where $ l_{n} \in \mathbb{R}^3$ is the 3D coordinates of the $n$-th adversarial location and $N_L$ is the number of locations we consider, to achieve the attack goal.
Inspired by the finding \textbf{F1} in Section~\ref{sec:findings}, we propose to establish a fixed \textit{attack point} and manipulate the prediction of the adversarial vehicle when the victim AV reaches this point. In this way, the perception of the adversarial vehicle at the attack point is independent of the victim AV's velocity. We let $t=0$ denote the time step when the victim AV arrives at the attack point. The point cloud frame captured by the victim AV's LiDAR at $t=0$ is referred to as the {\em current frame}.
We formulate the problem of identifying $L$ as:
\begin{equation}
\label{eq:problem_definition}
\begin{split}
\arg\max_{L}\quad &\mathbb{E}_{v\in V}[\text{Interference}(\tilde{\vec{Y}}^{v}, \vec{P}^{v})], \\ 
\text{s.t.}\quad & \tilde{\vec{Y}}^{v} = M_{pred}(\tilde{\vec{X}}^v), \\
& \tilde{\vec{X}}^v = M_{track}(\{M_{det}(D^{v}_{t}(L)) | t = -H+1, \ldots, 0\}), \\
\end{split}
\end{equation}
where $v$ is the victim AV's velocity sampled from a range $V$;
$D_t^v(L)$ is the LiDAR point cloud captured by the victim AV with velocity $v$ at time step $t$ when the adversarial objects are placed at $L$;
the $\{M_{det}(D^{v}_{t}(L)) | t = -H+1,...,0\}$ is the sequence of the victim AV's detection results regarding the adversarial vehicle over $H$ time steps;
$\tilde{\vec{X}}^v$ represents the adversarial vehicle's states observed by the victim AV, determined by the tracking model $M_{track}(\cdot)$ based on the detection results; $\tilde{\vec{Y}}^{v}$ is the trajectory of the adversarial vehicle predicted by the victim AV;
$\vec{P}^{v}$ is the original planned trajectory of the victim AV;
$\text{Interference}(\tilde{\vec{Y}}^{v}, \vec{P}^{v})$ represents the interference between the trajectories $\tilde{\vec{Y}}^{v}$ and $\vec{P}^{v}$. One meaningful definition of the interference is the reciprocal of the average distance between the adversarial vehicle moving on $\tilde{\vec{Y}}^{v}$ and the victim AV moving on $\vec{P}^{v}$. In the above formulation, the quantities with the superscript $v$ are affected by the victim AV's velocity $v$.

\vspace{-0.5em}
\section{Attack Design}
\vspace{-0.5em}
To achieve the attack goal defined in Eq.~\ref{eq:problem_definition} while addressing the attack challenges summarized in Section~\ref{sec:challenges}, we design a novel two-stage attack framework as shown in Fig.~\ref{fig:inverse_attack_framework}. This framework aims to identify adversarial locations for placing objects that mislead the victim AV into forecasting a falsified future trajectory of the adversarial vehicle, inducing dangerous driving behaviors conducted by the victim AV.

\vspace{-1em}
\subsection{Approach Overview}\label{sec:attack_overview}
Our attack framework is built upon two key insights. First, the widespread use of learning-based models in AD systems, which are vulnerable to adversarial inputs, presents a potential attack surface to impact downstream modules such as trajectory prediction through pipeline effects. Second, in contrast to the brute-force forward sampling attack that is inefficient and often gets stuck in local optima, our inverse attack strategy effectively exploits vulnerabilities in the prediction module for enhanced effectiveness and efficiency.

At a high level, our framework compromises modules from prediction to perception. It initiates by conducting adversarial attacks on the prediction module to generate adversarial states, i.e., the adversarial inputs of the trajectory prediction module that can mislead the module into forecasting a falsified future trajectory of the adversarial vehicle. Then, we conduct object-based attacks on the perception module, which generates adversarial locations. Placing objects at these locations can mislead the perception module to generate the adversarial bounding box perturbations, resulting in the desired adversarial states in the previous step.


The first stage, called \textit{prediction-side attack} as detailed in Section~\ref{sec:prediction_side_attack}, is designed to find state perturbations that mislead the victim AV's prediction model.
Firstly, we address the attack effectiveness challenge \textbf{C1} by minimizing the distance between the predicted trajectory of the adversarial vehicle and the planned trajectory of the victim AV. Secondly, we propose a fix-point attack based on a key finding \textbf{F1} presented in Section~\ref{sec:findings} that the trajectory prediction models are especially susceptible to the adversarial attack at the current frame. Together with the Expectation over Transformation (EoT) technique applied to various velocities of the victim AV, we address the velocity-insensitivity challenge \textbf{C2}. Thirdly, we address the vast search space challenge \textbf{C3} by utilizing a feasible set of detection results to guide the searching for effective state perturbations against the prediction model, which is based on a key finding \textbf{F2} presented in Section~\ref{sec:findings}, i.e., the detection results under object-based attacks are limited and exhibit distributional patterns. Lastly, we employ the Projected Gradient Descent (PGD) method to iteratively update the state perturbations and select the perturbations that lead to collisions.

The second stage, called \textit{location matching} as detailed in Section~\ref{sec:loc-match}, finds the adversarial locations for placing common objects that can implement the state perturbations found by the prediction-side attack.

\begin{table}[t]
    \centering
    \caption{Average distance between predicted trajectory of adversarial vehicle and victim AV's planned trajectory. A smaller distance indicates a higher collision risk.}
    \label{tab:critical_frame}
    \vspace{0.5em}
    \resizebox{0.75\columnwidth}{!}{
        \begin{tabular}{@{}ccccccc@{}}
        \toprule
        \multirow{3}{*}{Model}            & \multicolumn{6}{c}{Average Distance (m)}             \\ \cmidrule(l){2-7} 
                                          & \multirow{2}{*}{Clean} & \multicolumn{5}{c}{Attacked Frame Index}  \\ \cmidrule(l){3-7} 
                                          &                        & -4 & -3 & -2 & -1 & 0   \\ \midrule
        \multicolumn{1}{c|}{Trajectron++} & 6.3                    & 6.1 & 6.1 & 5.9 & 3.0 & 2.7 \\
        \multicolumn{1}{c|}{AgentFormer}                       & 6.3                    & 5.9 & 5.8 & 5.8 & 5.0 & 3.9 \\ \bottomrule
        \end{tabular}
    }
    \vspace{-1em}
\end{table}

\vspace{-1em}
\subsection{Findings}\label{sec:findings}
\vspace{-0.5em}
We conduct experiments to gain key insights for the design of our attack, as detailed below.

\textbf{Finding F1: Trajectory prediction models are vulnerable to single-point attack at the current frame.} 
To address challenge \textbf{C2}, we conduct a vulnerability analysis of two representative trajectory prediction models:
Trajectron++~\cite{salzmann2020trajectron++} and AgentFormer~\cite{yuan2021agentformer}. Both models take five historical states as input to predict the future trajectory. We apply the PGD method to perturb each of the five historical states. We add random noises to the remaining states to simulate noisy conditions in practice.
The PGD attack minimizes the average distance between the predicted trajectory of the adversarial vehicle and the victim AV's planned trajectory. 
Then, we measure this average distance under both clean and single-point attack scenarios. The results presented in Table~\ref{tab:critical_frame} show that perturbing the current state (i.e., $t=0$) achieves the smallest average distance, indicating the highest threat for collision. 
The reasons are two-fold. First, models such as Trajectron++ sequentially process input states using LSTM, which relies heavily on the most recent input state. Second, models including AgentFormer transform input states into a local coordinate system centered on the current frame. Thus, a small perturbation on the current state has the largest impact on the prediction result.
The finding reveals that for a trajectory prediction model $M_{pred}(\cdot)$ and a sequence of $H$ input states $\mathbf{X} = (\mathbf{X}^{-H+1}, \cdots, \mathbf{X}^0)$, the model is particularly susceptible to adversarial attacks targeting the current state $\mathbf{X}^0$, despite variations in historical states due to say different velocities. 
Therefore, by focusing the design of the attack on a fixed {\em attack point} that the victim AV arrives at, we can simplify the problem and preserve the potential of the attack in inducing victim AV's hazardous responses.



\begin{figure}[!t]
    \centering

    \subfigure[nuScenes Dataset.]{
        \label{fig:perturb_dist_nuscs}
        \includegraphics[width=0.9\columnwidth]{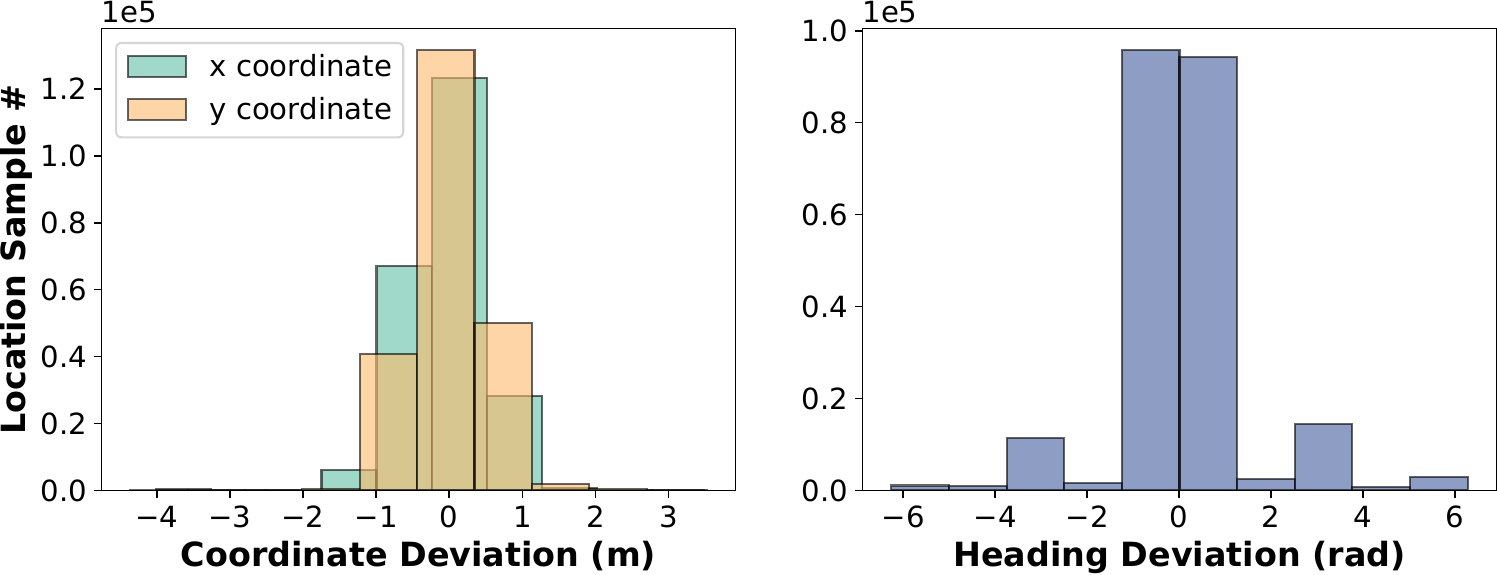}
    }\\
    \vspace{-0.5em}
    \subfigure[Self-Collected Scene.]{
        \label{fig:perturb_dist_realworld}
        \includegraphics[width=0.9\columnwidth]{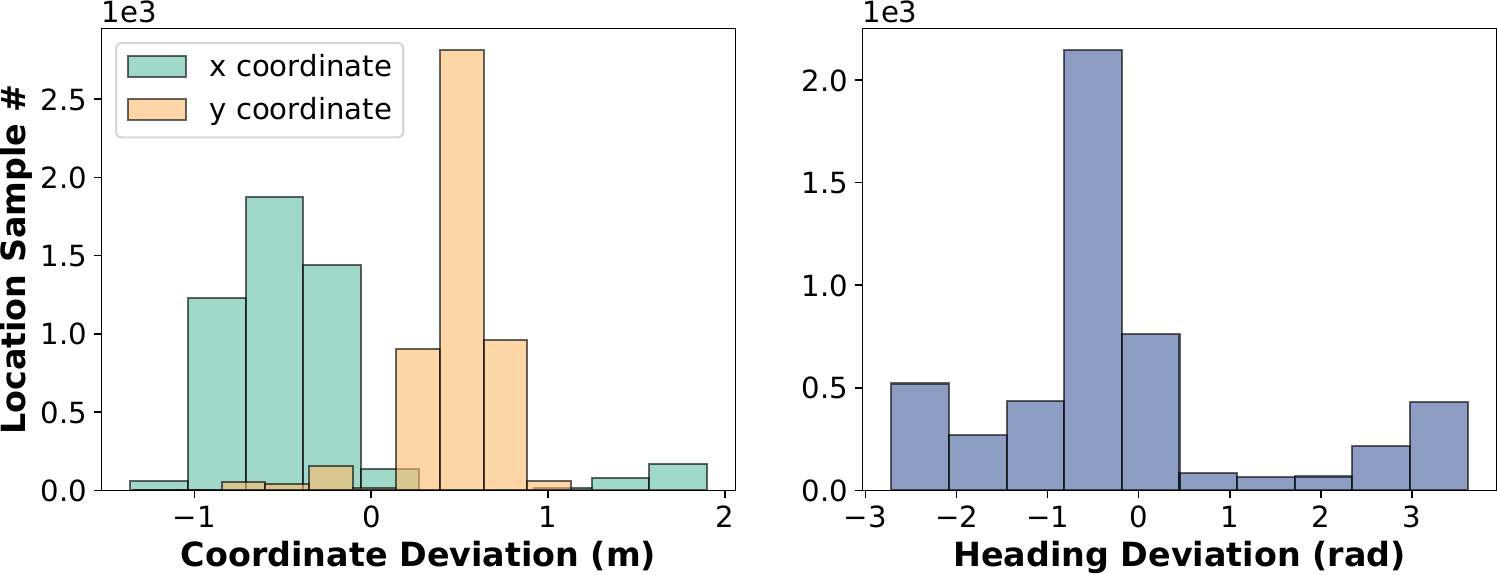}
    }
    \vspace{-1em}
    \caption{Histogram of coordinates and heading perturbations in nuScenes dataset scenes and a self-collected real-world scene under object-based attacks.}
    \label{fig:perturb_distrib}
    \vspace{-1em}
\end{figure}


\textbf{Finding F2: 3D object detection results under attack exhibit distributional patterns that help improve attack efficiency.} 
We follow the approach in~\cite{zhu2021can} to compromise LiDAR-based 3D object detection systems. The principle behind this attack is that strategic placement of adversarial objects distorts the vehicle’s perceived shape, perturbing the object detector’s geometric feature recognition.
To analyze the resulting bounding box perturbations, we conduct experiments to launch attacks on 100 driving scenes from the nuScenes dataset~\cite{caesar2020nuscenes} and a self-collected real-world driving scene. For each driving scene, we uniformly sample 500 candidate locations around the adversarial vehicle to place objects and record the resulted \textit{bounding box perturbation} in five consecutive frames, which is defined as the deviation of the bounding box for the adversarial vehicle with and without attack. To be more specific, we measure the deviations of the $x$ and $y$ coordinates and the heading $h$ of the bounding box, which serve as the inputs to the downstream trajectory prediction. Based on the experiment results in Fig.~\ref{fig:perturb_distrib}, we can observe that the bounding box perturbations caused by placing adversarial objects exhibit distributional patterns.
The deviations of $x$ and $y$ coordinates are primarily distributed within the interval [-1, 1].
The distribution of heading deviations exhibits distinct clusters. Note that heading deviations of around $\pm 1 \text{rad}$ mean that the adversarial vehicle is perceived to have a heading close to its actual heading, while deviations of $\pm 3 \text{rad}$ indicate a heading in the opposite direction. 
This finding is key to address the vast search space challenge \textbf{C3}. This is because, instead of exhaustive location sampling, we only need a limited number of location probes to recognize the bounding box perturbation distribution, thus improving efficiency in finding adversarial locations.

\vspace{-1em}
\subsection{Prediction-Side Attack}
\label{sec:prediction_side_attack}

To mislead victim AV's trajectory prediction model, we need to identify the \textit{state perturbation} at the current time step, which is added to the input states of the prediction model. It is denoted by $\vecg{\delta}^{\text{st}}=\{\delta^{\text{st}}_{x},\delta^{\text{st}}_{y},\delta^{\text{st}}_{h}\}$, where $x$, $y$, and $h$ indicate x coordinate, y coordinate, and heading, respectively. Now, we discuss the considerations when designing $\vecg{\delta}^{\text{st}}$ to address the three challenges presented in Section~\ref{sec:challenges}.

\textbf{Attack Effectiveness.} To find the state perturbation at the current time step that can address challenge \textbf{C1} and achieve the attack goal defined in Eq.~\ref{eq:problem_definition}, we propose \textit{Adversarial Loss} $\mathcal{L}_{adv} = \| \tilde{\vecg{Y}}^{v} - \vecg{P}^{v} \|_{2}$, which maximizes attack effectiveness by minimizing the $\ell_2$ distance between the predicted trajectory of the adversarial vehicle $\tilde{\vecg{Y}}^{v}$ and the original planned trajectory $\vecg{P}^{v}$ of the victim AV. 

\begin{algorithm}[t]
\caption{Prediction-side Attack}
\label{alg:prediction_side_attack}
\begin{algorithmic}[1]
\Require Prediction model $M_{pred}(\cdot)$, victim AV's original planned trajectory $\vecg{P}$, bbox perturbation set $C^{\text{box}}$, bbox perturbation clusters $R$, victim AV's velocity range $V$
\For{$r$ in $R$}
    \For{$e \gets 1$ to $E$}
        \State Init $\delta^{\text{st}}_{i} \sim \mathcal{N}(\mu_{i}^{r}, \sigma_{i}^{r}), i \in \{x, y, h\}$;
        \For{$iter \gets 1$ to $I$}
            \For{$v \sim V$} 
                \State Simulate historical states to estimate $\tilde{\vecg{X}}^{v}$;
                \State Trajectory prediction $\tilde{\vecg{Y}}^{v}=M_{pred}(\tilde{\vecg{X}}^{v})$;
                \State Compute loss $\mathcal{L} = \mathcal{L}_{adv}+\mathcal{L}_{rea}$; 
            \EndFor
            \State Update $\vecg{\delta}^{\text{st}}$ with PGD using avg. $\mathcal{L}$ over $V$;
            \State Clip $\delta^{\text{st}}_i$ into $[\mu_{i}^{r}-2\sigma_{i}^{r}, \mu_{i}^{r}+2\sigma_{i}^{r}], i \in \{x, y, h\}$;
        \EndFor
        \State Inference $\tilde{\vecg{Y}}=M_{pred}(\vecg{X}+\vecg{\delta}^{\text{st}})$;
        \If{collision between $\tilde{\vecg{Y}}$ and $\vecg{P}$}
            \State $C^{\text{st}} \gets C^{\text{st}} \cup \{\vecg{\delta}^{\text{st}}\}$;
        \EndIf
    \EndFor
\EndFor
\State \Return State perturbation set $C^{\text{st}}$
\end{algorithmic}
\end{algorithm}

\textbf{Attack Efficiency.}
Given the vast search space as mentioned in challenge \textbf{C3} in Section~\ref{sec:challenges}, finding a physically feasible state perturbation achieved by placing objects around the adversarial vehicle is costly. To address this, we rely on the finding \textbf{F2} and leverage a set of physically feasible detection results under object-based attack to efficiently guide the searching for the effective state perturbation.
Specifically, we evenly sample a small number of locations around the adversarial vehicle to derive a \textit{bounding box perturbation} set $C^{\text{box}}$. Each element in this set consists of a bounding box perturbation $\vecg{\delta}^{\text{box},m}=\{\delta^{\text{box}}_{x},\delta^{\text{box}}_{y},\delta^{\text{box}}_{h}\}$ and its corresponding location set $L_{m}$. 
Based on the distribution of the bounding box heading perturbation, we divide $C^{\text{box}}$ into a set of clusters $R$.
For each cluster $r$ in $R$, the bounding box perturbation $\delta^{\text{box}}_i$ follows the distribution $\mathcal{N}(\mu_i^{r}, \sigma_i^{r})$, where $i \in \{x, y, h\}$.
Then, for each $r$ in $R$, we leverage the estimated bounding box perturbation distribution to constrain the searching for the state perturbation by the following two steps.
(1) At the start of each attack epoch, we initialize the state perturbation $\delta_i^{\text{st}}$ from the estimated distribution $\mathcal{N}(\mu_i^{r}, \sigma_i^{r})$, $i \in \{x, y, h\}$, and then conduct iterative updates for attack optimization. At the end of each update, we clip $\delta^{\text{st}}_i$ within the range $[\mu_{i}^{r}-2\sigma_{i}^{r}, \mu_{i}^{r}+2\sigma_{i}^{r}]$ for $i \in \{x, y, h\}$. 
(2) We define a \textit{Realizable Loss}  
as $\mathcal{L}_{rea} = \| \vecg{\delta}^{\text{st}} - \underset{\vecg{\delta}^{\text{box}} \in C^{\text{box}}}{\text{argmin}} \| \vecg{\delta}^{\text{st}} - \vecg{\delta}^{\text{box}} \|_1 \|_1$, aiming to minimizes the $\ell_1$ distance between $\vecg{\delta}^{\text{st}}$ and its closest $\vecg{\delta}^{\text{box}}$ in $C^{\text{box}}$. An advantage of constraining the state perturbation within multiple clusters is that the generated perturbations are diversified, which enhances the probability of finding adversarial locations capable of achieving the desired perturbation.

\textbf{Velocity-Insensitivity.} 
To address challenge \textbf{C2}, we rely on the finding \textbf{F1} and aim to identify a state perturbation $\vecg{\delta}^{\text{st}}$ effective for the victim AV passing the attack point at various velocities. 
To achieve this, we employ the Expectation over Transformation (EoT) technique~\cite{athalye2018synthesizing}.
Specifically, we compute the expected sum of adversarial and realizable losses over a range of possible adversarial and victim AV input states associated with various velocities. 
To simulate the historical states of the victim AV, we fix the current frame at the attack point and backtrack the vehicle's motion using a kinematic model to obtain historical frames, ensuring compliance with vehicle dynamics.
For the adversarial vehicle, we randomly sample states from the bounding box perturbation distribution.
In each iteration, we update the state perturbation using the average losses across sampled velocities. 

\textbf{Optimization Procedure.} 
To summarize, we propose the optimization problem in Eq.~\ref{eq:prediction_side_opt_problem}. 
\begin{equation}
\label{eq:prediction_side_opt_problem}
\begin{split}
\arg\min_{\vecg{\delta}^{\text{st}}}\quad &\mathbb{E}_{v\in V} [\mathcal{L}_{adv}(\tilde{\vecg{Y}}^{v},\vecg{P}^{v}) + \mathcal{L}_{rea}(\tilde{\vecg{X}}^{v},C^{\text{box}})], \\
\text{s.t.}\quad &\tilde{\vecg{Y}}^{v} = M_{pred}(\tilde{\vecg{X}}^{v}), \quad \tilde{\vecg{X}}^{v} = \vecg{X}^{v} + \vecg{\delta}^{\text{st}}, \\
&\delta^{\text{st}}_i \in [\mu^{i}_{r}-2\sigma^{i}_{r}, \mu^{i}_{r}+2\sigma^{i}_{r}], \forall i \in \{x, y, h\}.
\end{split}
\end{equation}
The attack procedure is summarized in Alg.~\ref{alg:prediction_side_attack}. The goal is to identify a set of candidate state perturbations, denoted as $C^{\text{st}}=\{\vecg{\delta}^{\text{st}, n}\}^{N_C}_{n=1}$, that are effective and velocity-insensitive.
Given the bounding box perturbation set $C^{\text{box}}$ and corresponding distribution clusters $R$, for each cluster $r$ in $R$, we conduct multiple attack epochs starting from initializing $\vecg{\delta}^{\text{st}}$ within its respective distribution (line 3). We compute the average loss on different victim AV's velocities and use it to iteratively update $\vecg{\delta}^{\text{st}}$ with the PGD method (lines 5-10). At the end of each iteration, we clip  $\vecg{\delta}^{\text{st}}$ to its distribution range (line 11).
Due to the randomness in initialization and optimization, the candidate $\vecg{\delta}^{\text{st}}$ may not lead to a successful attack. 
To select effective state perturbations for forming the set $C^{\text{st}}$, we utilize the criteria in~\cite{chen2022scept} to evaluate collision likelihood between the adversarial vehicle's predicted trajectory $\tilde{\vecg{Y}}$ and the victim AV's planned trajectory $\vecg{P}$. In particular, a collision is defined when the outer circumferences of both vehicles intersect. 
We choose only perturbations that cause collisions to form the final set $C^{\text{st}}$ (lines 13-15).
\begin{algorithm}[t]
\caption{Location Matching}
\label{alg:location matching}
\begin{algorithmic}[1]
\Require State perturb set $C^{\text{st}}$, bbox perturb set $C^{\text{box}}$

\State \Comment{Vehicle State Matching}
\State $N_C \gets \text{size of } C^{\text{st}}$, $M \gets \text{size of } C^{\text{box}}$;
\State Init. cost matrix of size $N_C \times M$;
\For{$n \gets 1$ to $N_{C}$}
    \For{$m \gets 1$ to $M$}
        \State Calculate $cost(m,n) \gets \sum_{i \in \{x, y, h\}} w_i \cdot |\delta^{\text{box},m}_{i} - \delta^{\text{st},n}_{i}|$;
    \EndFor
\EndFor
\State Apply Hungarian Algorithm to $cost$;
\State Generate matched tuples $(\hat{\vecg{\delta}}^{\text{box},n},\hat{L}_{n}, \vecg{\delta}^{\text{st},n})_{n=1}^{N_C}$;


\State

\State \Comment{Location Refinement}
\State $L^{*} \gets \text{null}$;
\For{each tuple $(\hat{\vecg{\delta}}^{\text{box},n},\hat{L}_{n}, \vecg{\delta}^{\text{st},n})$}
    \State Init. $f_{\text{min}} \gets \text{current cost}$, $L_{\text{min}} \gets \hat{L}_{n}$;
    \For{each location set $L$ near $\hat{L}_n$}
        \State Calculate cost $f = \sum_{i \in \{x, y, h\}} w_i \cdot | \delta^{\text{box},L}_{i} - \delta^{\text{st},n}_{i} |$;
        \If{$f < f_{\text{min}}$}
            \State $f_{\text{min}} \gets f$, $L_{\text{min}} \gets L$;
        \EndIf
    \EndFor
    \State $L^* \gets L^* \cup \{L_{\text{min}}\}$;
\EndFor
\State \Return Adversarial location set $L^{*}$
\end{algorithmic}
\end{algorithm}

\vspace{-0.2em}
\subsection{Location Matching}
\label{sec:loc-match}
In prediction-side attack, we efficiently identify a set of state perturbations of the adversarial vehicle that are effective and velocity-insensitive. 
To launch a successful attack in the real world, we must also identify an adversarial location set $L^{*}$ that can achieve the desired state perturbations.
To achieve this feasibility, we adopt a two-step solution, including the \textit{Vehicle State Matching} and \textit{Location Refinement}. The procedure is illustrated in Alg.~\ref{alg:location matching}.

\textbf{Vehicle State Matching.} 
In this step, we utilize the sampled bounding box perturbation set $C^{\text{box}}$ in Section~\ref{sec:prediction_side_attack} to estimate the object locations that can achieve the desired state perturbations.
Specifically, we employ the Hungarian algorithm to find the matched bounding box perturbation $\hat{\vecg{\delta}}^{\text{box},m}$ and its corresponding adversarial location set $\hat{L}_m$ in $C^{\text{box}}$ with each desired state perturbation $\vecg{\delta}^{\text{st},n}$ in $C^{\text{st}}$. The cost matrix for the Hungarian algorithm is defined as:
\begin{equation}
\label{eq:matching_cost}
\text{cost}(m, n) = \sum_{i \in \{x, y, h\}} w_i \cdot |\delta^{\text{box},m}_{i} - \delta^{\text{st},n}_{i}|.
\end{equation}
Here, $\text{cost}(m, n)$ is the matching cost between the $m$-th bounding box perturbation $\delta^{\text{box},m}_i$ and the $n$-th state perturbation $\delta^{\text{st},n}_i$, for $i \in \{x, y, h\}$. The $w_i$ represents the pre-set weights corresponding to $\{x, y, h\}$, where elements with larger normalized perturbation in $\vecg{\delta}^{\text{st}}$ are assigned greater weights. For instance, if a desired state perturbation has a larger normalized $\delta^{\text{st}}_x$ value compared with the normalized $\delta^{\text{st}}_y$ and $\delta^{\text{st}}_h$, we will assign a larger weight to $x$. This is inspired by our empirical observation that a larger perturbation
is more related to the change of a prediction result. 
For each desired state perturbation, this step finds all the matched bounding box perturbation and its corresponding location set, generating matched tuples $(\hat{\vecg{\delta}}^{\text{box},n},\hat{L}_{n}, \vecg{\delta}^{\text{st},n})$, for $n = 1, \ldots, N_c$. 


\begin{figure*}[ht]
    \centering
    \hspace{0.5em}
    \includegraphics[width=0.3\textwidth]{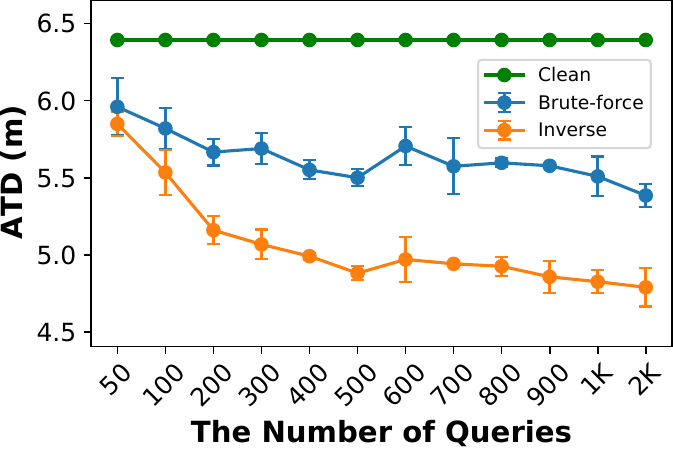}
    \hfill
    \includegraphics[width=0.3\textwidth]{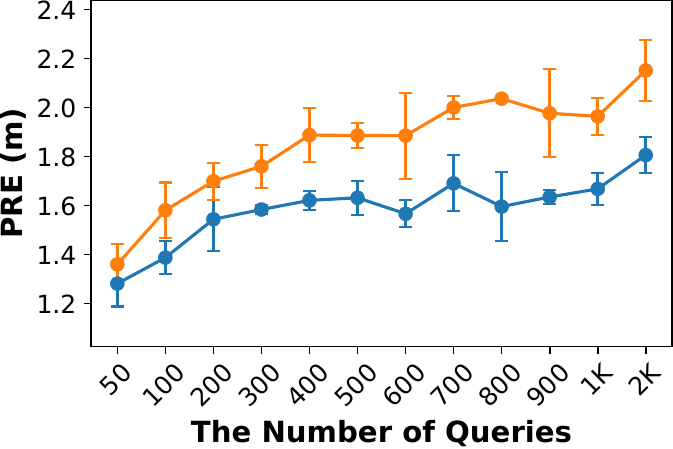}
    \hfill
    \includegraphics[width=0.3\textwidth]{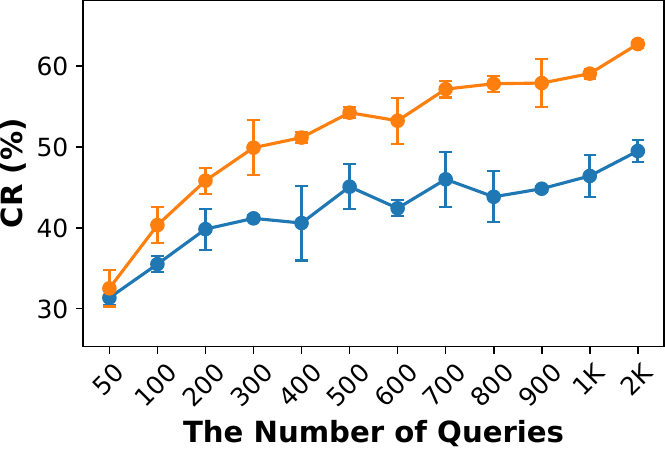}
    \hspace{1em}
    \vspace{-0.5em}
    \caption{Average Trajectory Distance (ATD), Planning-Response Error (PRE), and Collision Rate (CR) under clean, brute-force sampling attack, and our inverse attack scenarios.}
    \label{fig:eval_errorbar}
    \vspace{-1em}
\end{figure*}

\textbf{Location Refinement.} 
Due to the limited samples in $C^{\text{box}}$, the bounding box perturbation $\hat{\vecg{\delta}}^{\text{box},n}$ matched in the above Vehicle State Matching step might not precisely align with the desired state perturbations $\vecg{\delta}^{\text{st},n}$. This step aims to refine candidate adversarial locations, thereby further aligning the corresponding bounding boxes towards the desired state perturbations.
Specifically, we search for refined adversarial location set in close proximity to $\hat{L}_{n}$, aiming to reduce the cost $f = \sum_{i \in \{x, y, h\}} w_i \cdot | \delta^{\text{box},L}_{i} - \delta^{\text{st},n}_{i} |$, where $\delta^{\text{box},L}_{i}$ is a bounding box perturbation after placing objects at $L$ near $\hat{L}_{n}$.
We randomly probe multiple locations in the vicinity of $\hat{L}_{n}$. The probing space is defined as small cubes centered around each $\hat{l}_{n}$ in $\hat{L}_{n}$, as shown in Fig.~\ref{fig:inverse_attack_framework}. We set the cube size to be nearly equal to the voxel size of the LiDAR detector, allowing for moderate detection shifts. 
Finally, the refined location set leading to the smallest cost value is selected to form the final adversarial location set $L^{*}$.

\vspace{-0.5em}
\section{Experiments on Dataset}\label{sec:experiment_dataset}

\begin{figure*}[!t]
    \centering

    \subfigure[Sudden brake scene.]{
        \includegraphics[width=0.2\textwidth]{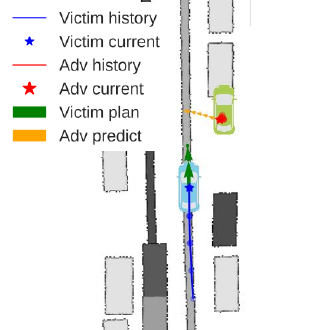}
        \label{fig:vis_brake}
    }\hspace{0pt}
    \subfigure[Sudden acceleration scene.]{
        \includegraphics[width=0.2\textwidth]{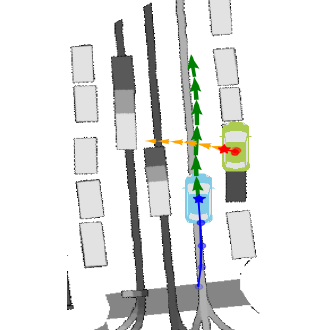}
        \label{fig:vis_acc}
    }\hspace{0pt}
    \subfigure[Lane change scene.]{
        \includegraphics[width=0.2\textwidth]{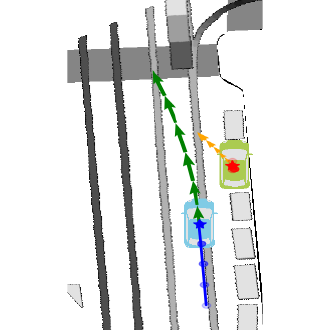}
        \label{fig:vis_lane_change}
    }\hspace{0pt}
    \subfigure[Attack impacts categorization.]{
    \label{fig:attack_impact_categorisation}
    \includegraphics[width=0.34\textwidth]{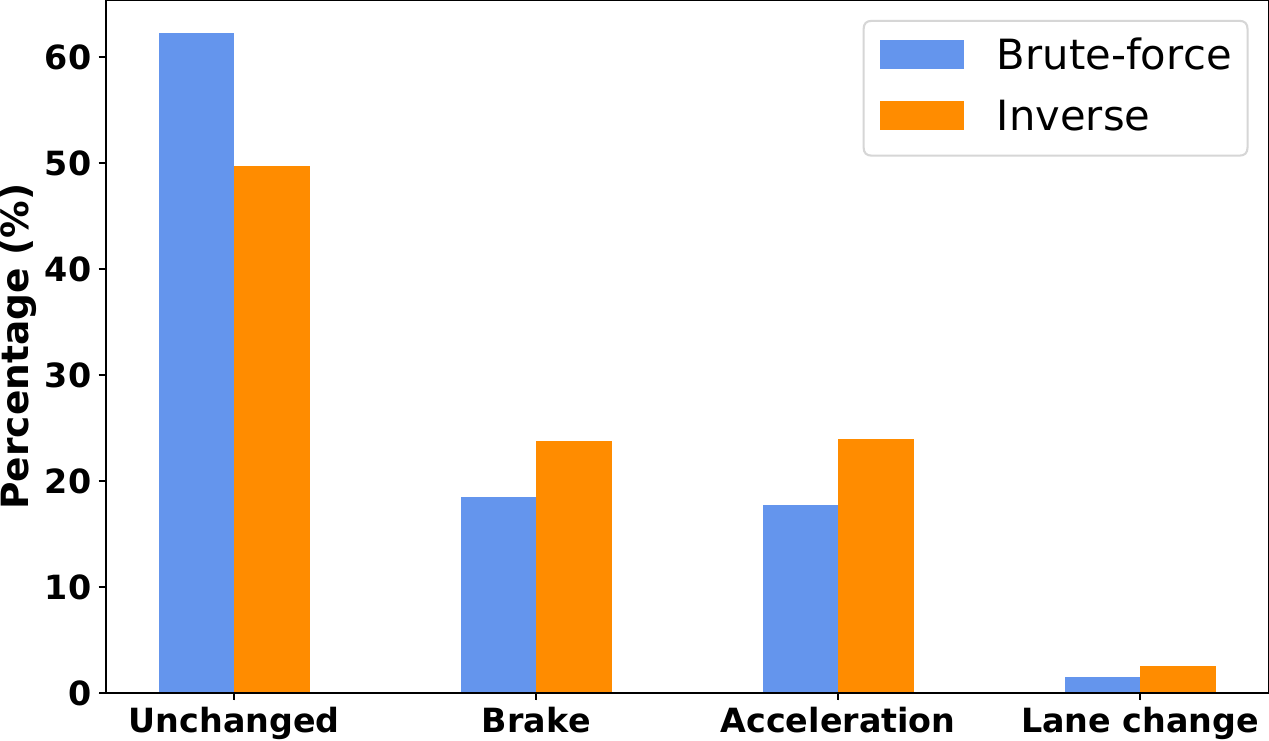}
    }\hspace{0pt}
    \caption{The attack impacts are categorized into {\em unchange}, {\em sudden brake}, {\em acceleration}, and {\em lane change}, with the latter three indicate hazardous driving behaviors.}
    \label{fig:digital_attack_impact_cate_vis}
    \vspace{-1em}
    
\end{figure*}

\begin{figure}[t]
    \centering
    \hfill
    \begin{minipage}{0.48\columnwidth}
        \includegraphics[width=\linewidth]{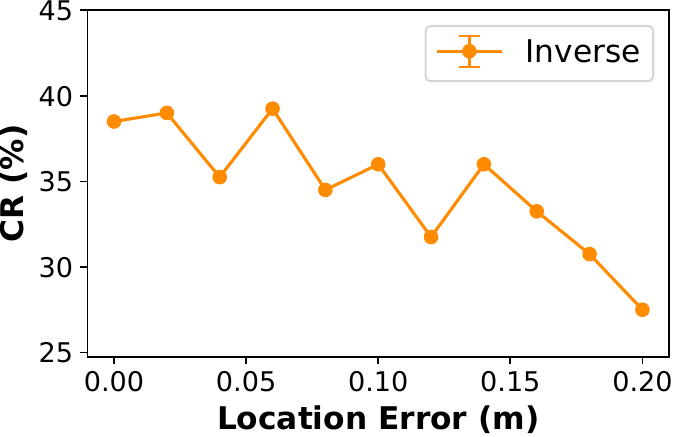}
        \vspace{-1.5em}
        \caption{Attack robustness against object displacement.}
        \label{fig:loc_error} 
    \end{minipage}
    \hfill
    \begin{minipage}{0.48\columnwidth}
        \includegraphics[width=\linewidth]{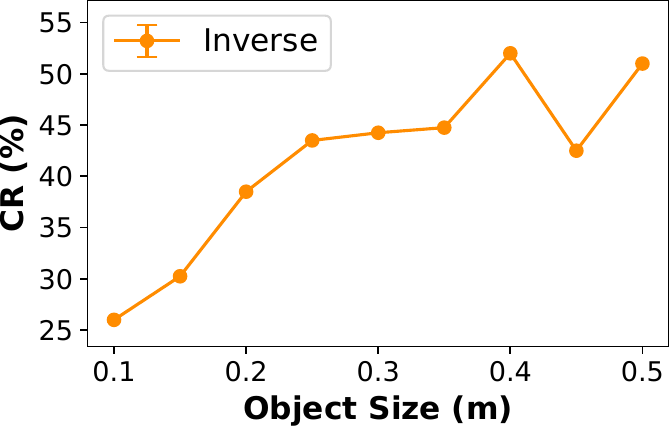}
        \vspace{-1.5em}
        \caption{Attack robustness against object size.}
        \label{fig:object_size_cr}
    \end{minipage}
    \hfill
    \vspace{-1em}
\end{figure}

\vspace{-0.5em}
In this section, we evaluate our attack using a public real-world autonomous driving dataset. This data-driven evaluation quantifies the effectiveness and robustness of our attack. 

\vspace{-1em}
\subsection{Experiment Setting}\label{sec:digital_exp_setting}

\textbf{Dataset.} 
We select 100 driving scenes from the nuScenes dataset~\cite{caesar2020nuscenes}, a large-scale autonomous driving dataset consisting of extensive data and labels covering both perception and prediction. The scene selection criteria are as follows. We select driving scenes aligned with our specified attack scenario, in which the ego vehicle approaches a static vehicle parked on the roadside. The parked vehicle is viewed as the adversarial vehicle; the ego vehicle serves as the victim AV. 
In each scene, we ensure that the victim AV starts with a minimum separation of 20 meters from the adversarial vehicle and reaches an attack point at a distance of less than 10 meters.
During the attack planning, we leverage the annotated key frames captured at $2\,\text{Hz}$ in the nuScenes dataset to identify the adversarial location set. 
We then evaluate these locations on frames that simulate various velocities of the victim AV, selected via a kinematic model from the unannotated nuScenes dataset sampled at $20\,\text{Hz}$.
We simulate variations in the victim AV's velocities at 1.5x, 1.25x, 0.75x, and 0.5x of the original recorded speed.

\textbf{AD Models.} 
We employ representative models for the victim AV's AD system. We use PIXOR~\cite{yang2018pixor} for LiDAR-based detection, CenterPoint Tracker~\cite{yin2021center} for tracking, 
Trajectron++~\cite{salzmann2020trajectron++} for trajectory prediction, and a Model Predictive Control (MPC)-based planner~\cite{chen2022interactive} for motion planning.
Note that Trajectron++ is a widely used and open-source prediction model on the nuScenes dataset. 
We train both PIXOR and Trajectron++ using their default configurations.

\textbf{Adversarial Object.}
Following~\cite{zhu2021can}, we simulate adversarial objects at three adversarial locations. Each object is represented by a random point cluster with a radius of $0.2\,\text{m}$. The number of points in each cluster is set to 4.
During the location sampling step to derive $C^{\text{box}}$, the search space is a $4 \times 4 \times 1\,\text{m}^3$ cube above the adversarial vehicle. In the location refinement, we reduce the search space to a sphere with a radius of $0.1\,\text{m}$, allowing for more precise adjustments.

\textbf{Evaluation Metrics.} 
We employ the following three metrics to measure the attack performance:

\textit{Average Trajectory Distance (ATD)}: This is the average distance between the predicted trajectory of the adversarial vehicle and the victim AV's original planned trajectory. It inversely quantifies the efficacy for the interference of the adversarial vehicle's trajectory to the victim AV. A smaller ATD suggests a larger interference.

\textit{Planning-Response Error (PRE)}: The average displacement error between the victim AV's planned trajectories in the presence and absence of attack, respectively. A greater PRE signifies an increased tendency for the victim AV to deviate from its original action.

\textit{Collision Rate (CR)}: The proportion of the adversarial vehicle's predicted trajectories that intersect with the victim AV's planned trajectories in any future frame. Such intersections imply hazardous driving actions by the victim AV.


\textbf{Baseline: Brute-Force Sampling.} 
This straightforward baseline samples numerous locations around the adversarial vehicle and identifies the location set that results in the smallest ATD as the adversarial location set. We employ this customized baseline because there is no existing attack that can achieve the goal of this work.

\vspace{-1em}
\subsection{Attack Effectiveness}\label{sec:digital_exp_attack_effectiveness}
\vspace{-0.5em}

Fig.~\ref{fig:eval_errorbar} shows the error bars representing the mean and standard deviation of ATD, PRE, and CR values achieved by three repeated experiments. ``The number of queries'' refers to the maximum number of queries the attacker can make to the victim's AD system. For instance, if the number of queries is set to 100, the attacker is limited to sample at most 100 locations around the adversarial vehicle and restricted to a total of 100 queries across all interactions with the victim AV's detection, tracking, prediction, and planning models. 
From the results, we can see that our method with 2,000 queries achieves the best attack performance, with the lowest ATD ($4.7\,\text{m}$), the highest PRE ($2.2\,\text{m}$), and the highest CR (63\%). We can also observe that the attack performance is better for both brute-force sampling and our inverse attacks when the number of queries is larger. Notably, under the same number of queries, our method consistently outperforms the brute-force sampling. In particular, our method with 400 queries results in a higher CR than the brute-force sampling with 2,000 queries, indicating a reduction of the attack overhead by a factor of five.

To illustrate the attack impact, we categorize the victim AV's driving behaviors under attacks into four categories, i.e., {\em unchanged}, {\em sudden brake}, {\em sudden acceleration}, and {\em lane change}. 
Note that the victim AV's original driving behavior is moving forward within its current lane. Thus, the latter three behaviors are considered hazardous because they deviate from the victim AV's original driving intent.
The categorization is based on the metrics of Maximum Lateral Deviation (MLaD) and Longitudinal Deviation (MLoD). MLaD measures the lateral deviation between the victim AV's planned trajectories in the presence and absence of attack, while MLoD measures the longitudinal deviation.
A trajectory with an MLoD greater or smaller than $\pm 2\,\text{m}$ is categorized as a sudden acceleration or brake. An MLaD greater or smaller than $\pm 1\,\text{m}$ indicates a left or right lane change. 
Figs.~\ref{fig:vis_brake}, ~\ref{fig:vis_acc}, and ~\ref{fig:vis_lane_change} visualize three representative scenes of sudden brake, sudden acceleration, and lane change, respectively. 
Fig.~\ref{fig:attack_impact_categorisation} displays the categorization results for 400 victim AV's planned trajectories under attacks. These trajectories are generated on 100 driving scenes at four different velocities. The number of queries is set to 100. The results show that our method induces hazardous driving behaviors of the victim AV for more than 50\% of the trajectories, while the brute-force method results in 12\% fewer hazardous trajectories. 
This suggests that our attack not only induces errors in trajectory prediction but also forces the victim AV's planning module to adopt unsafe maneuvers as corrective measures to avert potential collisions with the adversarial vehicle's incorrectly predicted future trajectory.

\vspace{-0.5em}
\subsection{Attack Robustness}
\textbf{Robustness against Object Displacement.}
In real-world deployment, precisely placing objects at pre-computed locations may be challenging. To evaluate the robustness of our attack against object displacement errors, we randomly shift the identified adversarial locations and evaluate our attack's effectiveness. In this experiment, we use the adversarial location set generated by our attack method with a limit of 100 queries. Fig.~\ref{fig:loc_error} illustrates CR as the location shift changes from 0 to $0.2\,\text{m}$ in intervals of $0.02\,\text{m}$. The corresponding ATD and PRE results are presented in Appendix~\ref{sec:experiment_dataset_appendix}.
The results show that ATD gradually increases, while PRE and CR decrease, as the location shift increases.
This is because when the shift exceeds the grid cell/voxel size of the object detector, the additional point clusters caused by the objects might be assigned to a different grid cell/voxel than the originally intended one, leading to a detection bounding box that deviates from the desired state perturbation. However, within the grid cell/voxel size limit of $0.1\,\text{m}$, the location sets generated by our attack demonstrate relative robustness, making them deployable in real-world scenarios.

\textbf{Robustness against Object Size.} In this work, the attacker launches the attack by placing objects, such as cardboards or boxes, at designated locations. These objects vary in size and shape. To evaluate robustness of our attack against object size variations, we generate random point clusters with varying radius, ranging from $0.1\,\text{m}$ to $0.5\,\text{m}$, and position them at same adversarial locations.  
Following the setting in~\cite{zhu2021can}, we set the number of points in each cluster proportional to the square of the object size. 
Fig.~\ref{fig:object_size_cr} illustrates the CR under different object sizes. The corresponding ATD and PRE results are presented in Appendix~\ref{sec:experiment_dataset_appendix}. We can observe that a smaller object size results in weaker attack performance, indicated by a higher ATD and lower PRE and CR. This is because smaller objects induce fewer points, leading to limited perturbing effect in detection results. Object sizes in the range of $0.2\,\text{m}$ to $0.4\,\text{m}$ demonstrate stable, optimal performance across all metrics. However, the performance declines and becomes much more random beyond this range. This is likely due to larger objects occupying numerous grid cells/voxels, resulting in a significant deviation from the original detection results.

\vspace{-1em}
\subsection{Attack Transferability}\label{sec:attack_transferability}
\vspace{-0.5em}
We also evaluate the transferability of our attack under the black-box scenario. In this experiment, we assume the victim AV uses AgentFormer as its trajectory prediction model. We assume a black-box condition where the attacker generates locations of adversarial objects based on Trajectron++ and uses these locations to compromise the target model, i.e., AgentFormer. The results (see Appendix~\ref{sec:experiment_dataset_appendix}) indicate that our attack effectively compromises a different predictor, achieving a maximum CR of 31\% with 500 queries. Similar to the white-box results, our attack consistently outperforms the brute-force approach across different numbers of queries, demonstrating its superior transferability and efficiency.







\begin{figure}[!t]
    \centering
    \includegraphics[width=0.85\columnwidth]{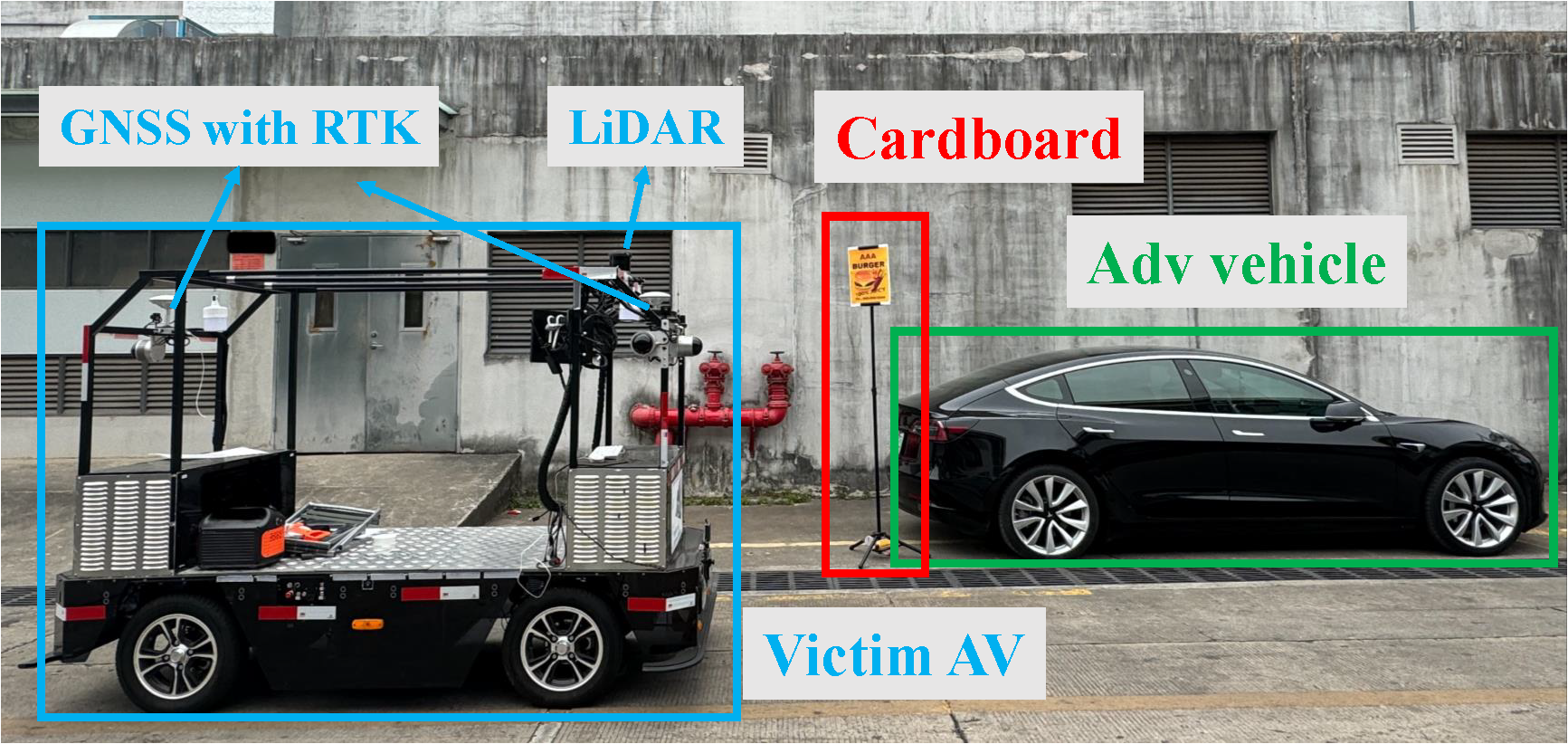}
    \vspace{-0.5em}
    \caption{Experimental setup for physical world attack.}
    \label{fig:phy_exp_setup}
    \vspace{-0.5em}
\end{figure}

\vspace{-0.5em}
\section{Experiments in Physical World}\label{sec:experiment_physical}
\begin{figure*}[!t]
    \centering
    \subfigure[Left-side scenario with adversarial objects deployed (victim AV's view).]{
        \label{fig:left_attack_scene}
        \includegraphics[width=0.23\textwidth]{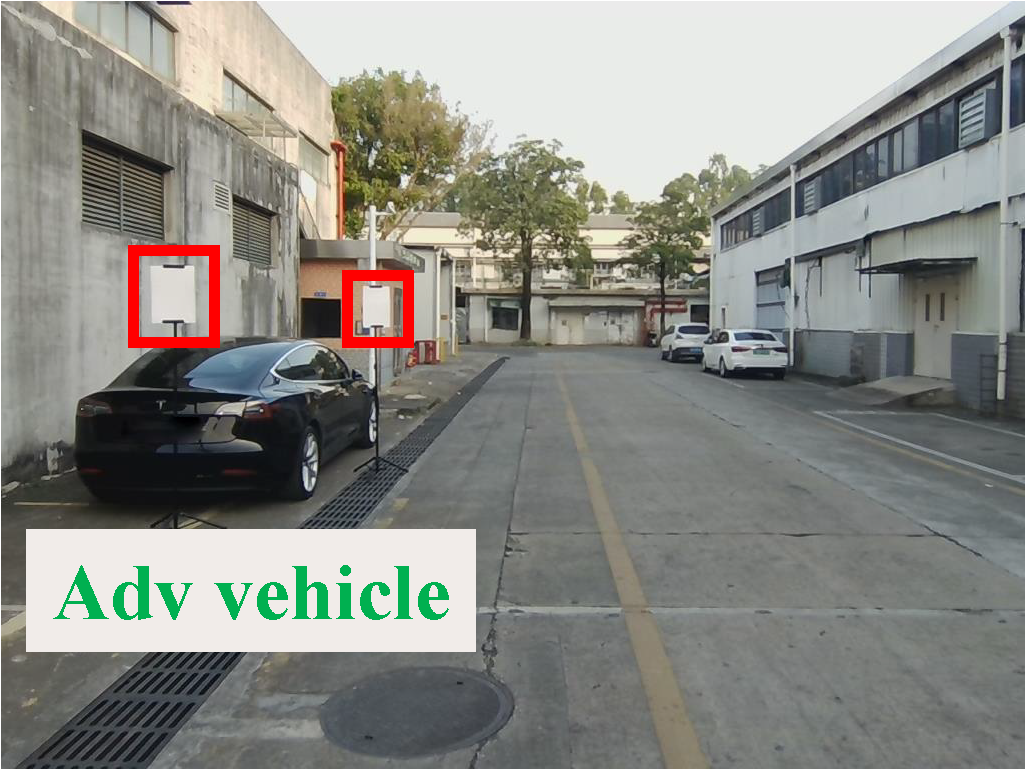}
    }\hspace{3pt}
    \subfigure[Result of left-side scenario when no adversarial objects are deployed.]{
        \label{fig:left_clean_res}
        \includegraphics[width=0.23\textwidth]{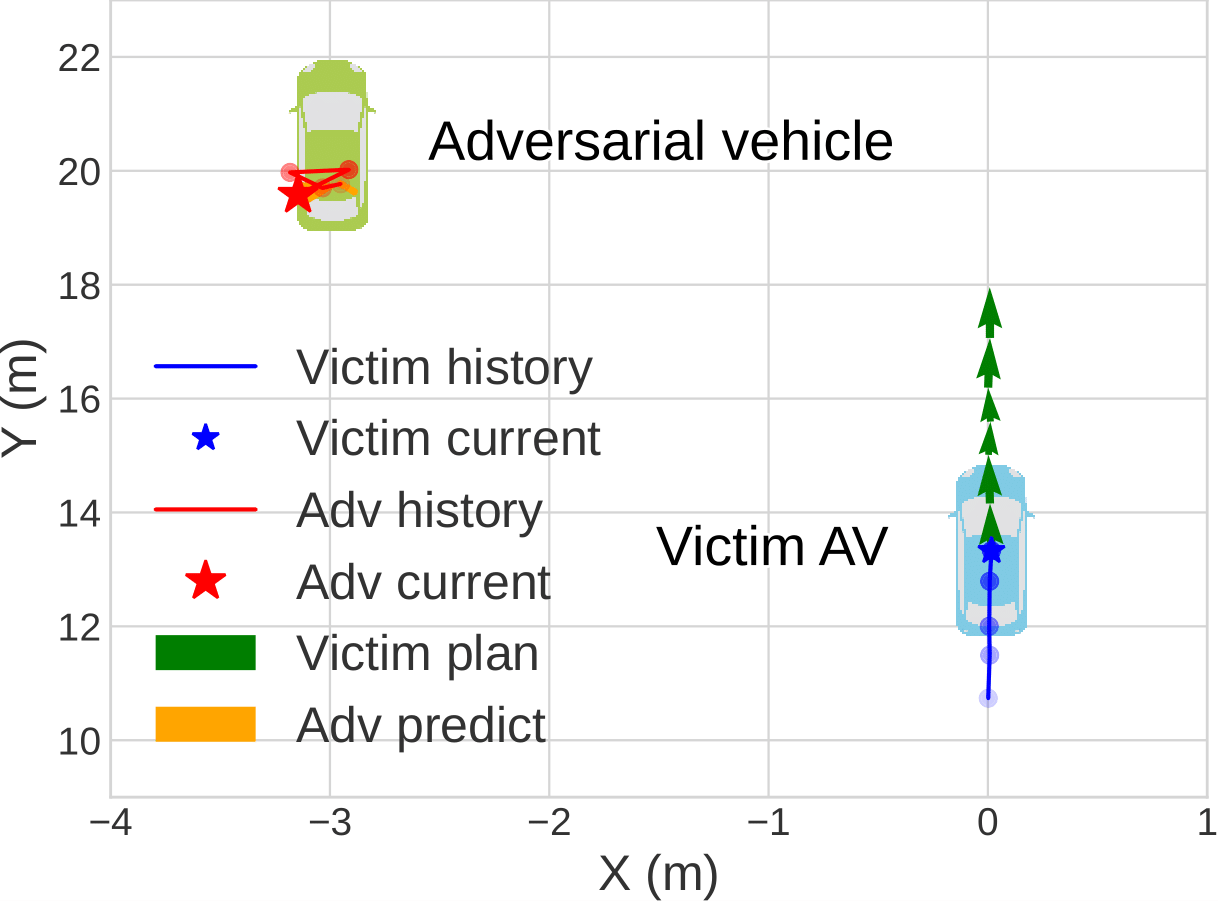}
    }\hspace{3pt}
    \subfigure[Result of left-side attack(ours) at a velocity of $5\,\text{km/h}$. Success ratio: 5/5.]{
        \label{fig:left_attack_res_mod}
        \includegraphics[width=0.23\textwidth]{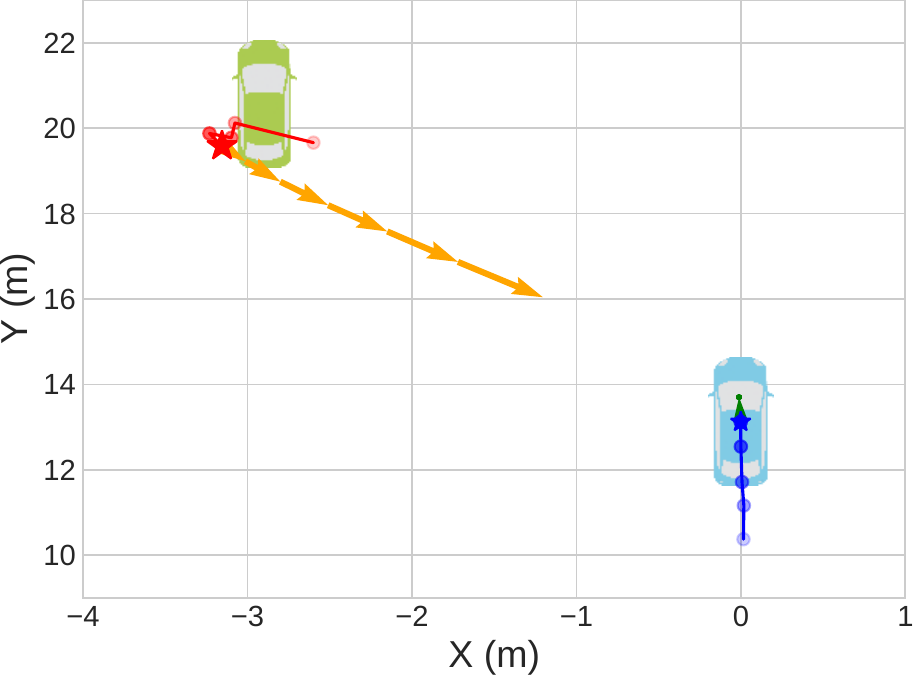}
    }\hspace{3pt}
    \subfigure[Result of left-side attack(ours) at a velocity of $10\,\text{km/h}$. Success ratio: 4/5.]{
        \label{fig:left_attack_res_high}
        \includegraphics[width=0.23\textwidth]{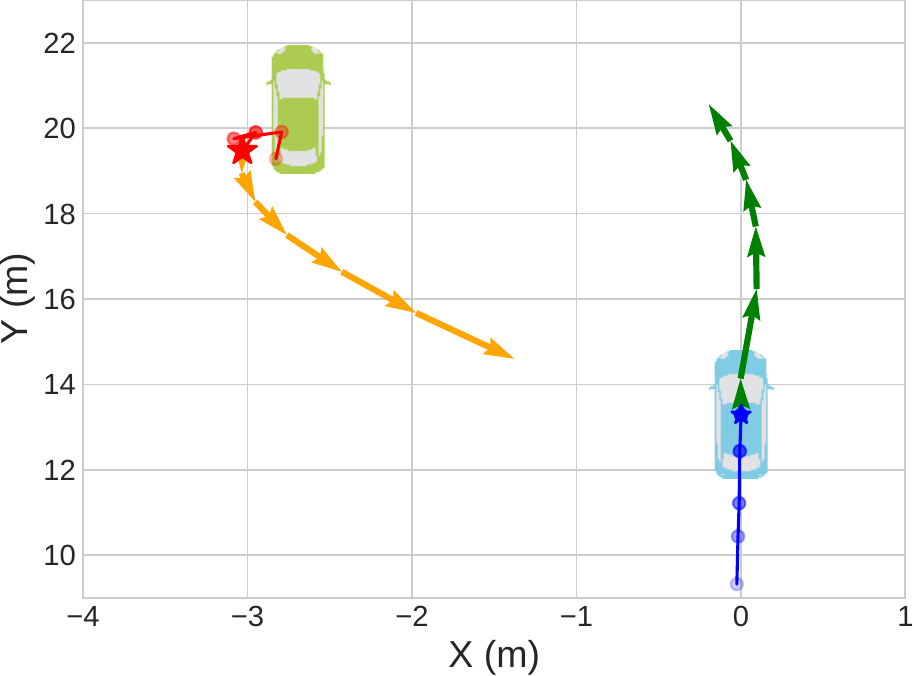}
    }
    
    \subfigure[Right-side scenario with adv objects deployed (victim AV's view).]{
        \label{fig:right_attack_scene}
        \includegraphics[width=0.23\textwidth]{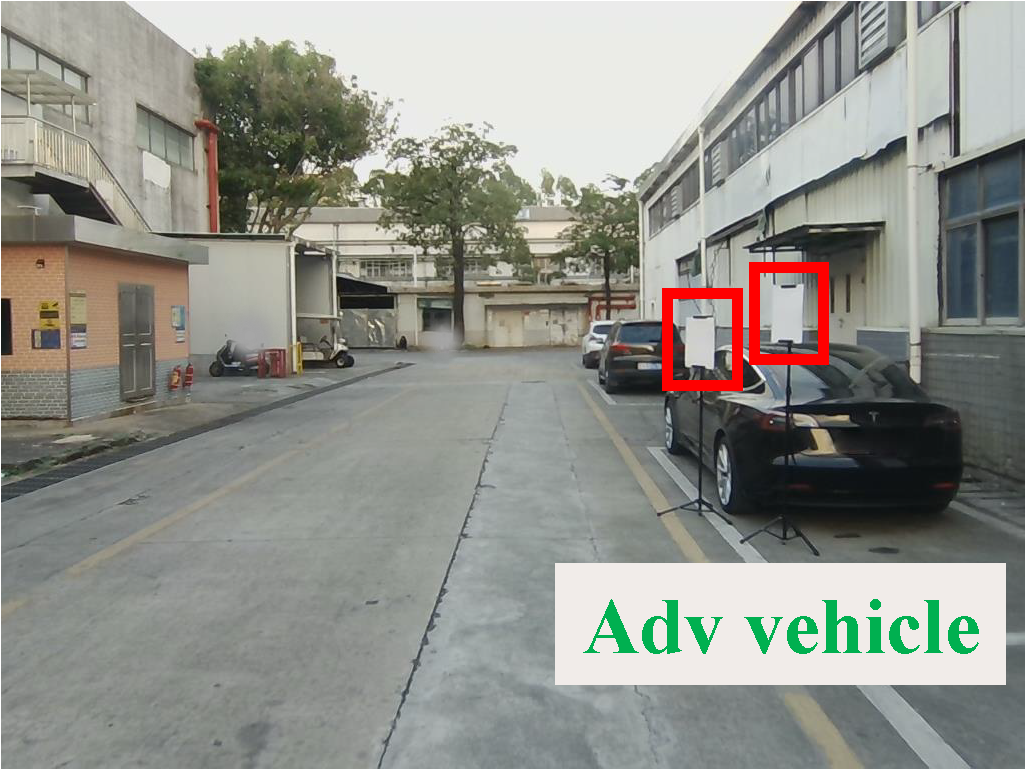}
    }\hspace{3pt}
    \subfigure[Result of right-side scenario when no adversarial objects are deployed.]{
        \label{fig:right_clean_res}
        \includegraphics[width=0.23\textwidth]{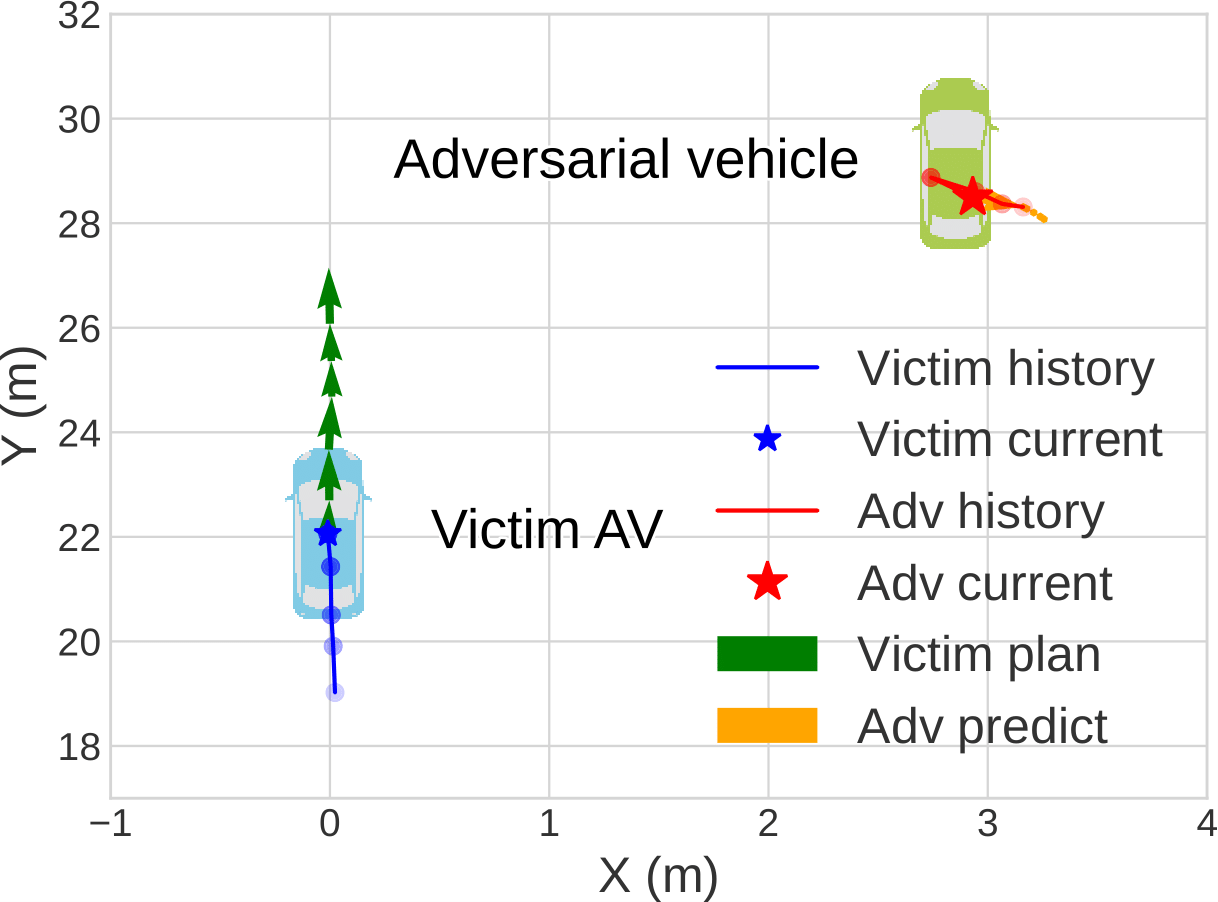}
    }\hspace{3pt}
    \subfigure[Result of right-side attack(ours) at a velocity of $5\,\text{km/h}$. Success ratio: 4/5.]{
        \label{fig:right_attack_res_mod}
        \includegraphics[width=0.23\textwidth]{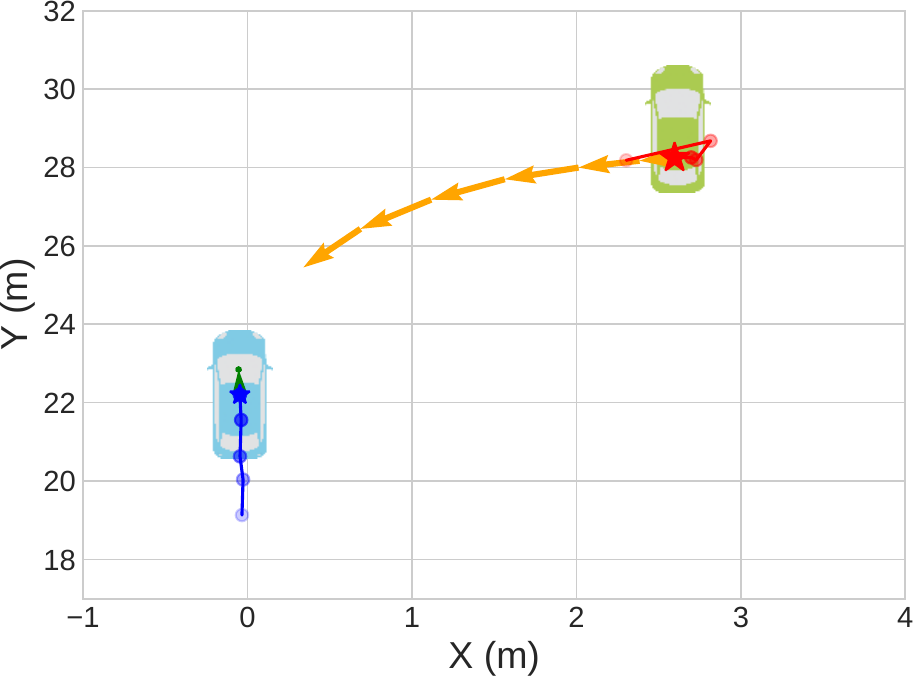}
    }\hspace{3pt}
    \subfigure[Result of right-side attack(ours) at a velocity of $10\text{km/h}$.Success ratio:5/5.]{
        \label{fig:right_attack_res_high}
        \includegraphics[width=0.23\textwidth]{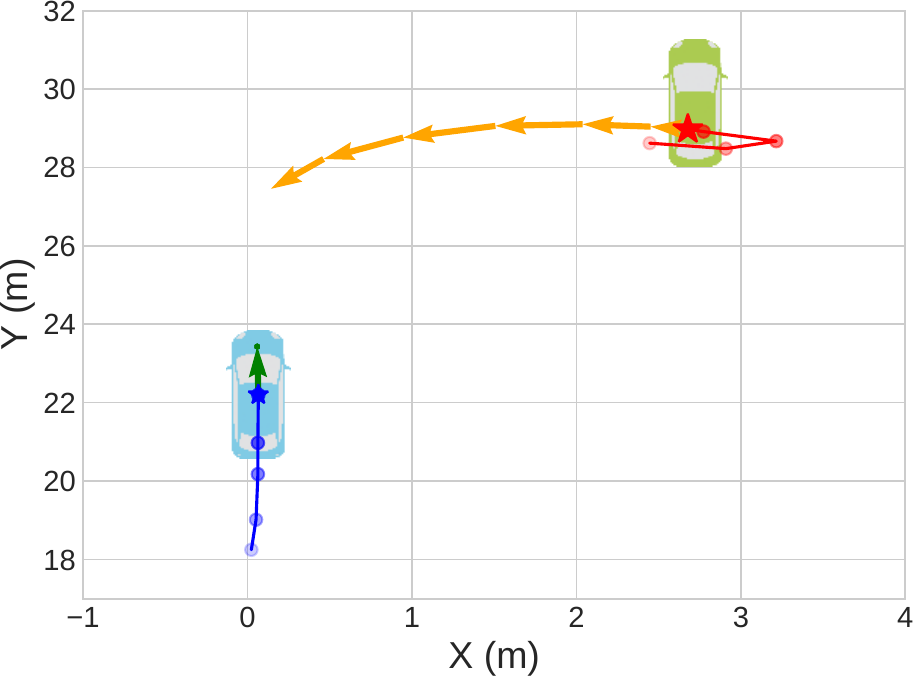}
    }
    \vspace{-0.7em}
    \caption{Two real-world scenarios with a parked black Tesla car as the adversarial vehicle. The adversarial objects (cardboards) placed at the locations computed by our inverse attack are highlighted with red boxes. The attack results are illustrated in grid maps, where the blue car represents the victim AV and the green car represents the adversarial vehicle.}
    \label{fig:physical_experiments}
    \vspace{-0.7em}
    
\end{figure*}

\subsection{Experimental Settings}\label{sec:physical_exp_setting}

\textbf{Testbed Car.}
As illustrated in Fig.~\ref{fig:phy_exp_setup}, our testbed car includes a LiDAR and a Global Navigation Satellite System receiver (GNSS) with Real-Time Kinematic (RTK). 
The testbed car serves as both the attacker's data collection car during the attack planning stage and the victim AV during the evaluation. 
We integrate the commonly used Robosense Helios-32 LiDAR for perception, featuring 32 lines and a $10\,\text{Hz}$ frame rate. The LiDAR is mounted on the top front center of the testbed car at the height of $1.6\,\text{m}$ from the ground. We use the BYNAV X1 high Precision GNSS/INS Receiver with RTK to construct accurate coordinates for prediction and planning, achieving centimeter-level precision in vehicle localization by correcting GPS errors with differential signals. Our testbed car uses the same models as in the dataset-based experiments for perception, prediction, and planning.

\textbf{Adversarial Vehicle and Objects.} Our proposed attack does not have any requirements on the size or color of the adversarial vehicle, which can be any car parked on the roadside.
In our experiments, we use a black Tesla Model 3 as the adversarial vehicle. 
For the adversarial objects, we utilize two cardboards measuring $0.3\,\text{m}$ in width and $0.42\,\text{m}$ in height, equivalent to the standard A3 paper size. During the attack planning phase, to simulate realistic point clusters of cardboards akin to those captured in the real world, we use the ray-casting method~\cite{mit_graphics} to sample points. In the attack deployment phase, each cardboard is mounted on a tripod for flexible placement at various locations around the adversarial vehicle. The cardboards are tilted at a 45\degree{} angle to face the side of the victim AV.
While our experiments use white cardboards for better visibility, they can be substituted with ubiquitous roadside objects, such as billboards, for stealthiness.

\textbf{Attack Planning and Deployment.}
During the attack planning phase, we establish a start point as the origin of the global coordinate system. 
The testbed car, acting as the attacker's data collection car, drives from the start point and passes by the attack point at a velocity of $5\,\text{km/h}$, simultaneously collecting the LiDAR point cloud and the corresponding GNSS data. 
The input states to the prediction model are computed in this global coordinate system, originating at the start point. Specifically, we transform the GNSS data in geodetic coordinates (latitude, longitude, altitude) into ENU (East, North, Up) coordinates based on this origin. This conversion enables us to obtain the states of victim AV and combine these coordinates with detection results to obtain the adversarial vehicle's states.
Based on the collected data in the absence of attack, we derive two adversarial locations aiming to mislead the victim AV's trajectory prediction. Considering the practicality of placing the cardboards, we limit the location search space to the vicinity of the adversarial vehicle. Specifically, we define the side search region to be $5.2 \times 0.4 \times 0.8\,\text{m}^3$, matching the length of the adversarial vehicle, and the rear/front search regions to be $0.4 \times 2.2 \times 0.8 \,\text{m}^3$, aligning with the vehicle's width.
During the attack deployment, the cardboards are positioned at the decided adversarial locations. The testbed car, now acting as the victim AV, drives from the start point established during the attack planning phase towards the attack point at various velocities to collect LiDAR point cloud and GNSS data. The collected data under attack are then fed into the victim AV's AD system for evaluation.

\vspace{-1em}
\subsection{Attack Effectiveness}
\vspace{-0.5em}
In this section, we evaluate our attack's effectiveness in two real-world scenarios, where the victim AV drives on the road with an adversarial vehicle parked either on its left or right side, referred to as the ``left-side scenario'' and the ``right-side scenario'', respectively. We compare our attack with the random location attack and the brute-force sampling attack. 
An attack is viewed as successful when the PRE exceeds a threshold of $1\,\text{m}$, which is a significant change in the victim AV's planning decision due to the erroneously predicted trajectory.

\textbf{Inverse Attack (Ours).}
In these experiments, we place objects at the adversarial locations identified by our attack during the attack planning phase, using data collected at a velocity of $5\,\text{km/h}$. We then evaluate the effectiveness of these locations when the victim AV drives at velocities of $5\,\text{km/h}$ or $10\,\text{km/h}$, with slight variations for each trial, capturing the uncertainties encountered in real driving environments.
Figs.~\ref{fig:left_attack_scene} and \ref{fig:right_attack_scene} present the scenarios when our attack is deployed. Corresponding LiDAR point clouds perceived by the victim AV under attack can be found in Appendix~\ref{sec:experiment_physical_appendix}.
In non-attack scenarios, the adversarial vehicle is predicted static and thus poses no threat to the victim AV's trajectory planning, as depicted in Figs.~\ref{fig:left_clean_res} and \ref{fig:right_clean_res}.
In the left-side scenario, our attack succeeds in five out of five trials at an evaluation velocity of $5\,\text{km/h}$. Fig.~\ref{fig:left_attack_res_mod} shows a successful attack where our attack misleads the victim AV at a velocity of $5\,\text{km/h}$ to predict a future trajectory heading towards itself, forcing the victim AV to an emergency brake. 
Our attack, utilizing adversarial locations planned using data at a velocity of $5\,\text{km/h}$, succeeds in four out of five trials at an evaluation velocity of $10\,\text{km/h}$. Fig.~\ref{fig:left_attack_res_high} presents a successful example, where our attack induces a similar prediction error as in the scene at a velocity of $5\,\text{km/h}$ in Fig.~\ref{fig:left_attack_res_mod}, demonstrating our attack's velocity insensitivity.
In the right-side scenario, the planned adversarial locations achieve four successes out of five trials at an evaluation velocity of $5\,\text{km/h}$ and five out of five at $10\,\text{km/h}$.
Example successful attack scenes at velocities of $5\,\text{km/h}$ and $10\,\text{km/h}$ are shown in Figs.~\ref{fig:right_attack_res_mod} and \ref{fig:right_attack_res_high}, with similar trajectory prediction errors 
and the same sudden brake decision taken by the victim AV in both scenes.
In both left- and right-side scenarios, successful attack examples at different evaluation velocities demonstrate similar perception errors at the current time step, as marked by red stars in Fig.~\ref{fig:physical_experiments}.
This indicates the effectiveness of single-point attack in inducing prediction error at the current frame. 
Detailed quantitative results, including comparisons of ATD and PRE, are presented in Appendix~\ref{sec:experiment_physical_appendix}.

\begin{figure}[!t]
    \centering
    
    \subfigure[Result of left-side random location attack. Success ratio: 0/5.]{
        \label{fig:random_attack_res_left}
        \includegraphics[width=0.45\columnwidth]{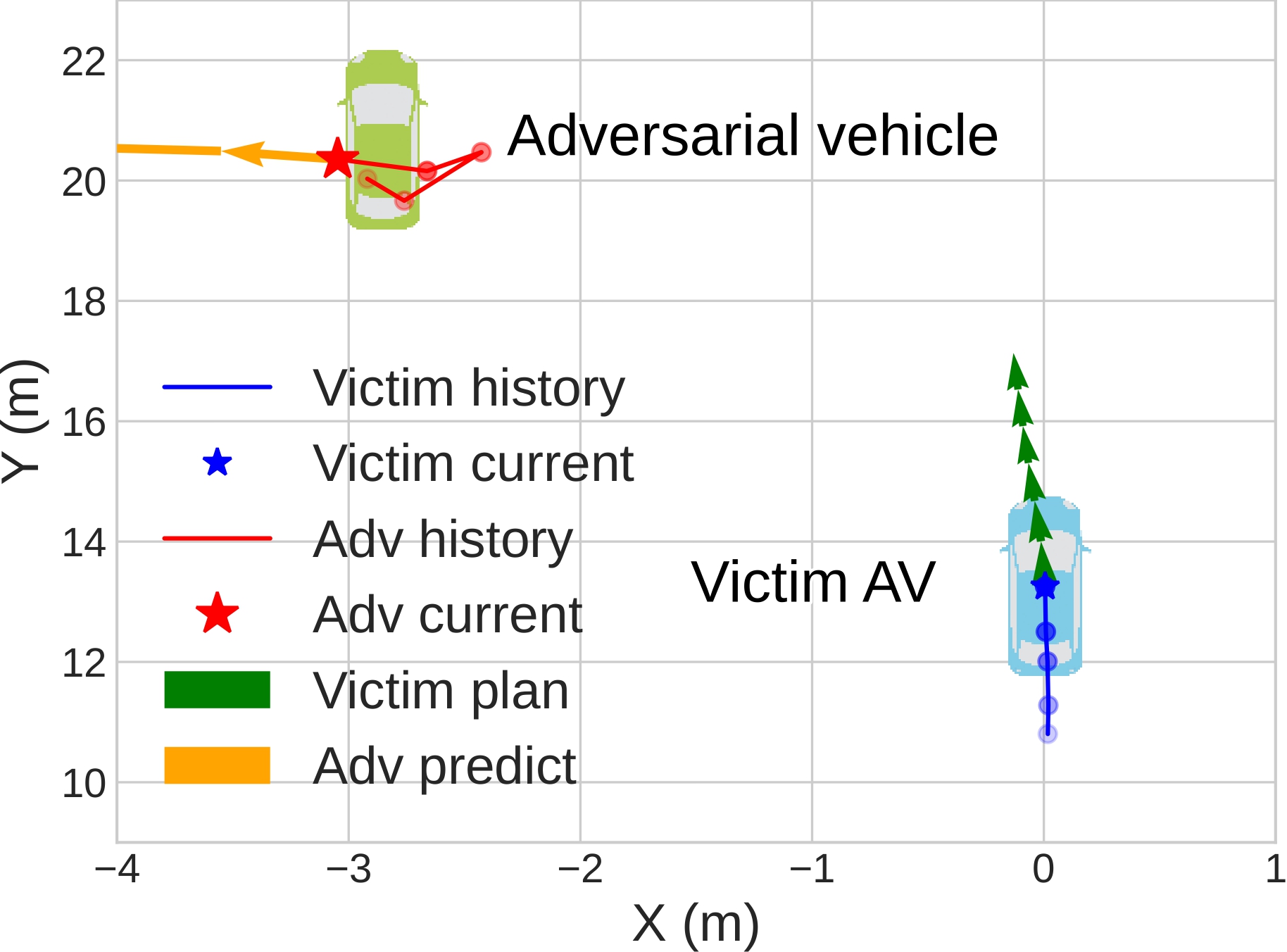}
    }
    \subfigure[Result of right-side random location attack. Success ratio: 0/5.]{
        \label{fig:random_attack_res_right}
        \includegraphics[width=0.45\columnwidth]{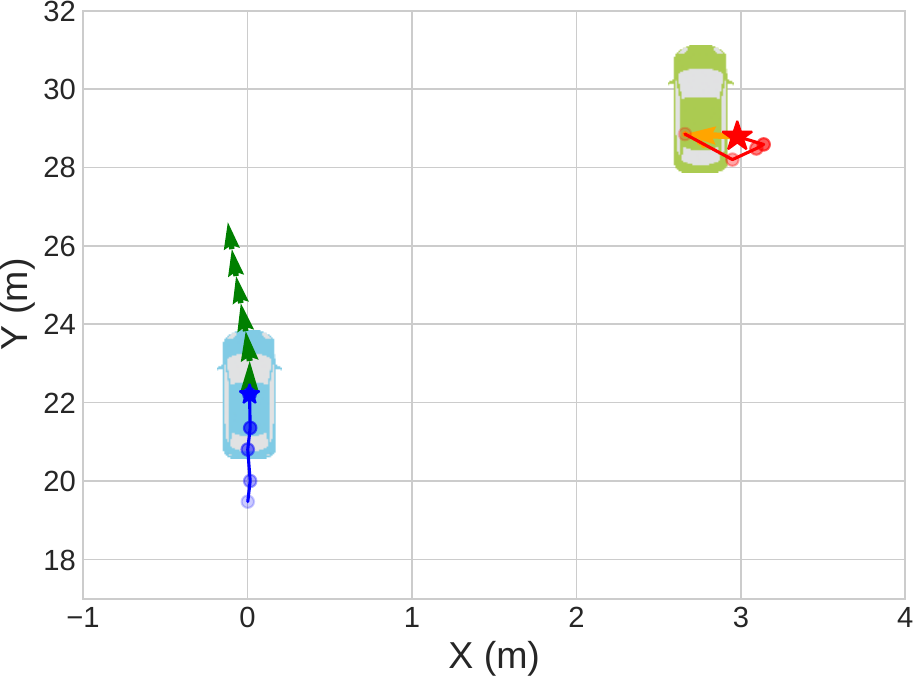}
    }
    \vspace{-1em}
    \caption{Random location attack results.}
    \label{fig:random_attack}
    \vspace{-0.5em}
\end{figure}

\textbf{Baseline 1: Random Location Attack.} 
In this set of experiments, adversarial cardboards are randomly placed around the adversarial vehicle. 
Some example scenes are shown in Appendix~\ref{sec:experiment_physical_appendix}. 
In both left-side and right-side random location attack scenarios, our evaluation yields no successful attacks out of five trials.
Fig.~\ref{fig:random_attack_res_left} illustrates a left-side attack example where cardboards, randomly placed on the right side of the adversarial vehicle, cause a predicted trajectory of the adversarial vehicle moving to the left of the victim AV, posing a limited threat.
Fig.~\ref{fig:random_attack_res_right} shows a right-side attack example with cardboards randomly placed at the rear, leading to a static predicted trajectory of the adversarial vehicle with no threat to the victim AV.
The results demonstrate that, objects placed at random locations may affect the predicted trajectories of the adversarial vehicle, while failing to achieve the attack goal defined in Eq.~\ref{eq:problem_definition}. This underscores the importance of strategic object placement at adversarial locations, which is facilitated by our inverse attack.

\textbf{Baseline 2: Brute-force Sampling Attack.} 
We evaluate the attack effectiveness of the brute-force sampling method using locations derived from data at a velocity of $5\,\text{km/h}$ in the planning phase. The evaluation is conducted in the left-side scenario at velocities of $5\,\text{km/h}$ and $10\,\text{km/h}$.
This method achieves a success ratio of three out of five trials
at both velocities.
The reason for the reduced attack success ratio compared with our inverse attack, even under the less challenging condition at a velocity of $5\,\text{km/h}$, is likely due to small fluctuations in the victim AV's velocity between trials.
These variations can render velocity-sensitive adversarial locations ineffective, leading to attack failures due to minor discrepancies from the original scene in the attack planning phase.

\begin{figure}[!t]
    \centering
    \subfigure[Result of coordinate shift (0.1m). Success ratio: 3/5.]{
        \label{fig:loc_error_0.1_res}
        \includegraphics[width=0.45\columnwidth]{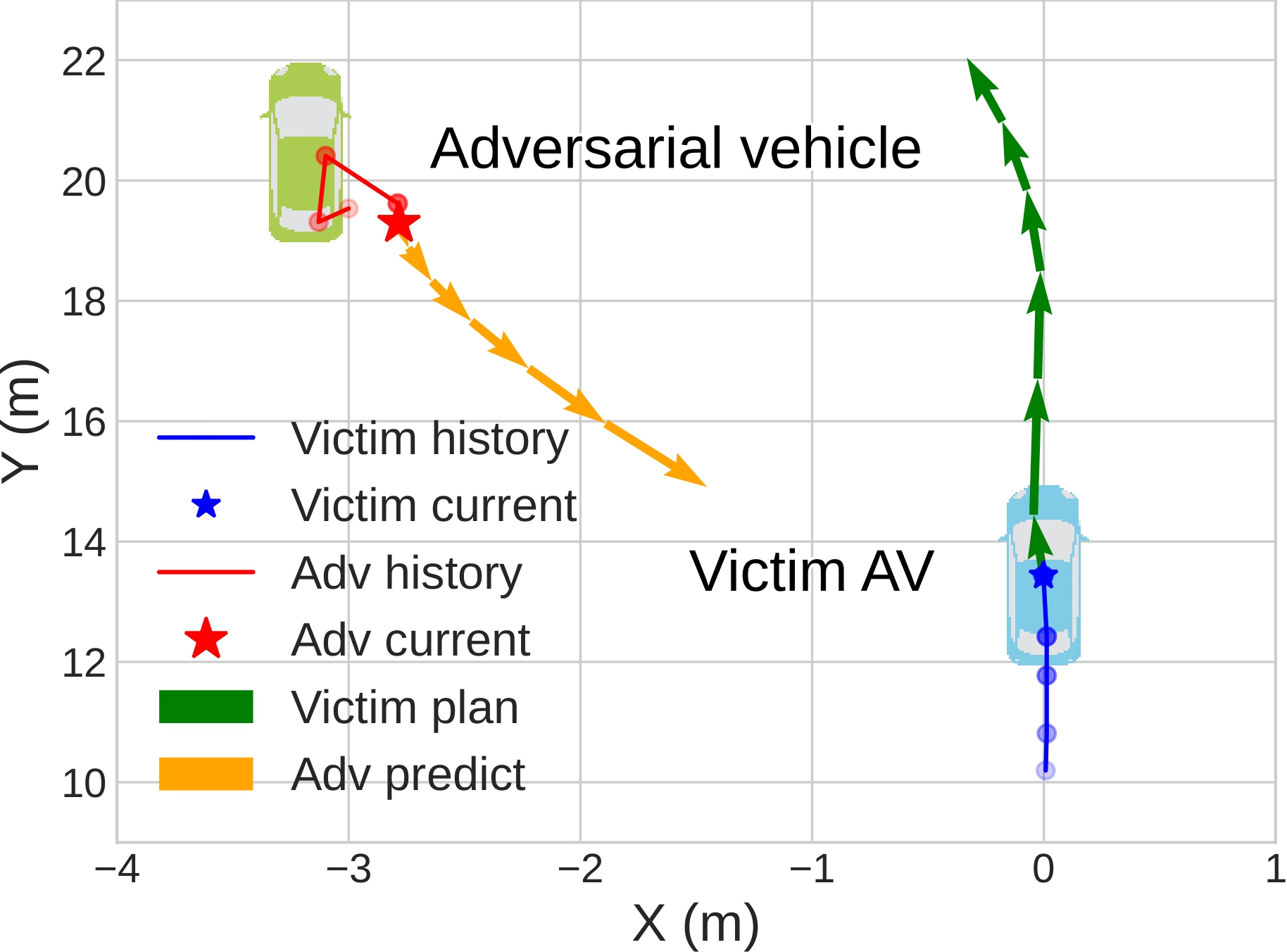}
    }\hspace{0pt}
    \subfigure[Result of orientation shift. Success ratio: 5/5.]{
        \label{fig:loc_error_orientation_res}
        \includegraphics[width=0.45\columnwidth]{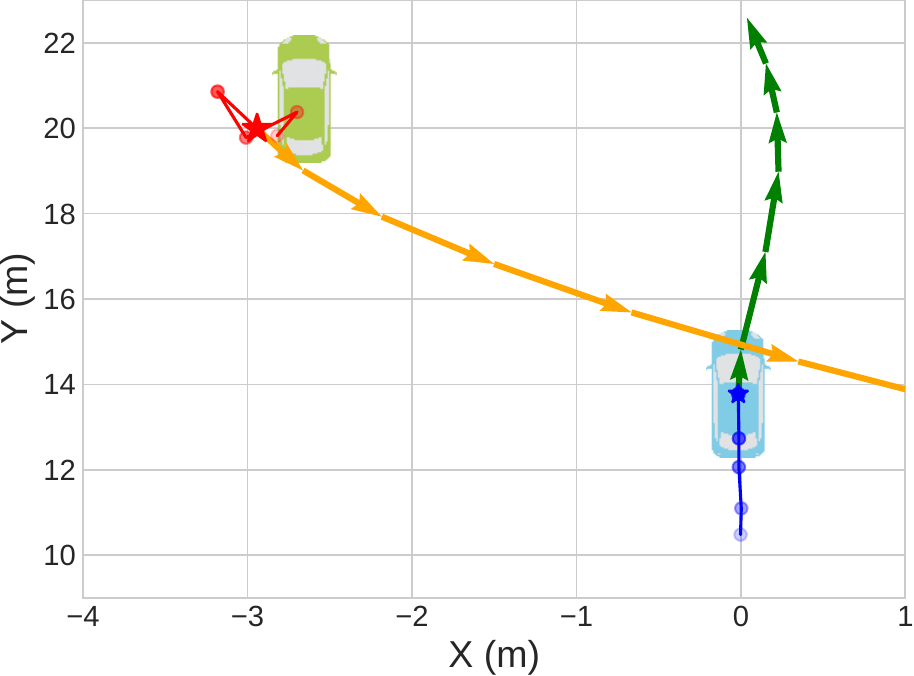}
    }\hspace{0pt}
    \vspace{-1em}
    \caption{Robustness against object displacement errors induced by shifts in object coordinates or orientations.}
    \label{fig:loc_error_realworld}
    \vspace{-1em}
\end{figure}

\vspace{-0.5em}
\subsection{Attack Robustness}
\textbf{Robustness against Object Displacement.}
When deploying cardboards at the adversarial locations, there may be deviations in their 3D coordinates and orientations. To evaluate our attack's robustness against these deviations, we conduct experiments to randomly shift each cardboard's 3D coordinates and orientation, respectively. 
The experiments are performed in the left-side attack scenario, with the victim AV driving at a velocity of $5\,\text{km/h}$. Some example scenes are illustrated in Appendix~\ref{sec:experiment_physical_appendix}. 
To simulate the shifts, we move each cardboard $0.1\,\text{m}$ from its planned location in random directions. Results show that shifting cardboards' coordinates by $0.1\,\text{m}$ leads to three successful attacks out of five trials. Fig.~\ref{fig:loc_error_0.1_res} illustrates an example successful attack, achieving a similar attack result as in the original attack scene in Fig.~\ref{fig:left_attack_res_high}.
This robustness is due to the majority of points, even after shifting, still falling within the same grid cell or voxel after LiDAR detection processing. However, when the shift increases to $0.3\,\text{m}$, which exceeds the grid cell/voxel size, the attack success ratio drops to one out of five trials.
To simulate orientation shifts, the cardboards are rotated randomly within $[-30\degree, 30\degree]$. In all five trials, our attack is successful. Fig.~\ref{fig:loc_error_orientation_res} shows a successful attack.
This robustness may stem from the observation that, despite substantial orientation shifts, the points on a cardboard mostly remain within the original grid or shift to an adjacent grid, resulting in similar perception and prediction outcomes.
To summarize, the adversarial locations identified by our inverse attack remain effective within a reasonable range of object displacement errors induced by coordinate and orientation shifts.

\begin{figure}[!t]
    \centering
    \subfigure[Result of victim AV in left driving direction. Success ratio: 2/5.]{
        \label{fig:direction_left_res}
        \includegraphics[width=0.45\columnwidth]{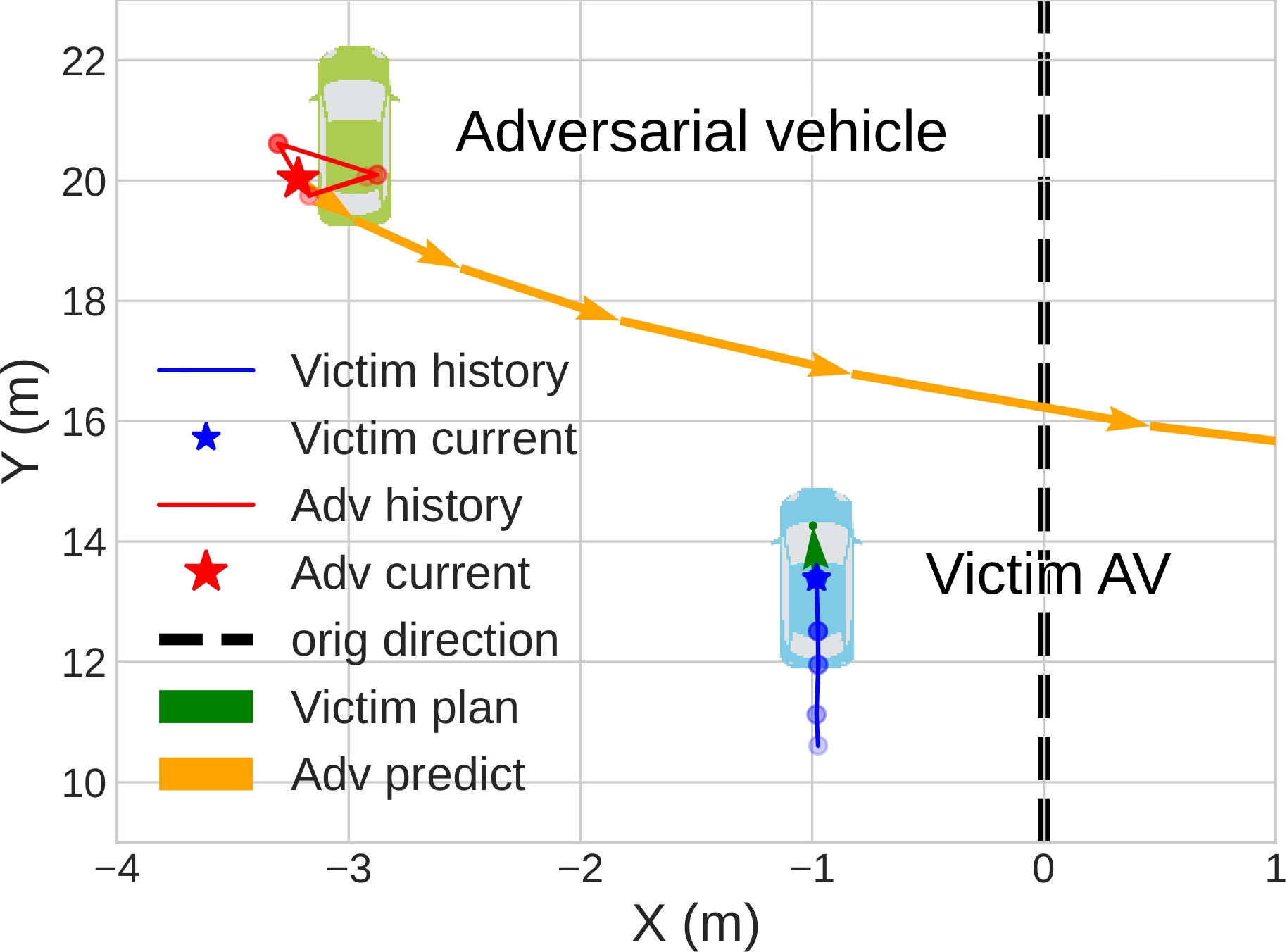}
    }\hspace{0pt}
    \subfigure[Result of victim AV in right driving direction. Success ratio: 3/5.]{
        \label{fig:direction_right_res}
        \includegraphics[width=0.45\columnwidth]{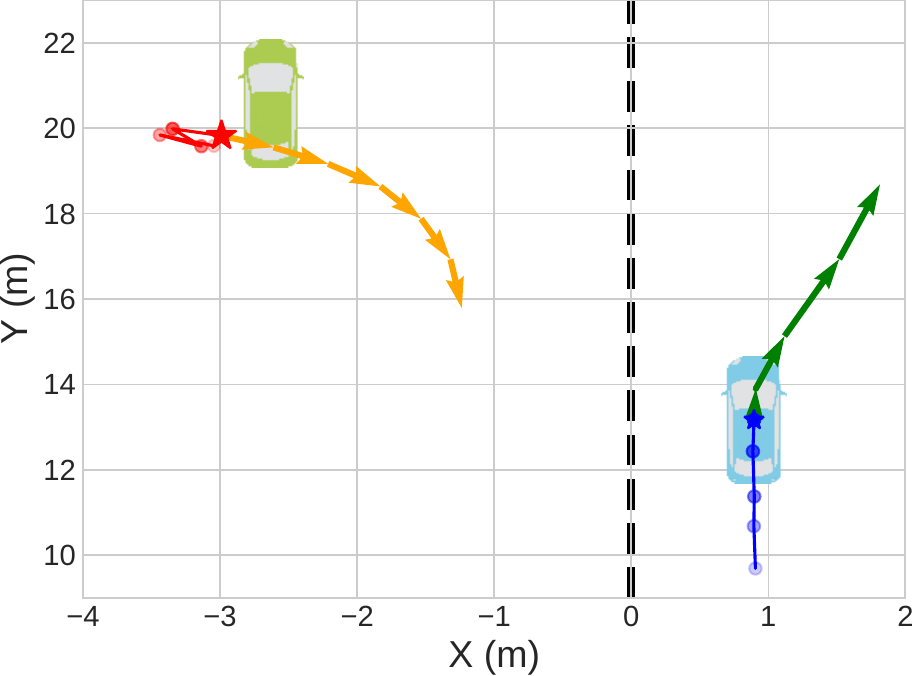}
    }\hspace{0pt}
    \vspace{-1em}
    \caption{Robustness against various driving directions of the victim AV.}
    \label{fig:diff_direction}
    \vspace{-1em}
\end{figure}

\textbf{Robustness against Varied Driving Direction.}
During the evaluation
, the victim AV remains in the same lane as in the planning phase. However, it may not maintain a perfectly straight path and occasionally steer slightly to the left or right.
To evaluate the robustness of our inverse attack against the variations in victim AV's driving direction, we conduct an experiment in the left-side scenario at a velocity of $5\,\text{km/h}$, with example scenes shown in Appendix~\ref{sec:experiment_physical_appendix}. 
With the planned adversarial locations unchanged, we shift the victim AV's starting and attack point by $1\,\text{m}$ to the left or right of their original positions, simulating the victim AV going slightly rightward or leftward. When shifting the victim AV's driving direction to the left, two out of five attack trials are successful. When shifting to the right, the attack success ratio is three out of five. Figs.~\ref{fig:direction_left_res} and~\ref{fig:direction_right_res} show example successful attack results. When shifting to the left direction, the victim AV is forced to brake urgently due to the reduced distance from the predicted trajectory of the adversarial vehicle, while shifting to the right direction leads to a lane change to avoid intersection with the adversarial vehicle's predicted trajectory.

\vspace{-0.5em}
\section{Discussions}
\subsection{A Potential Defense}
This work shows that a lack of awareness for adversarial vehicle states can lead to hazardous results, which our proposed attack exploits. As depicted in Fig.~\ref{fig:perturb_distrib}, in a sequence of adversarial states under attack, the coordinates are typically perturbed within a reasonable range ($\pm 1\text{m}$), but the heading shows significant volatility. For instance, an adversarial vehicle's heading might abruptly reverse from one frame to the next, which violates the vehicle physical dynamics.
Inspired by this, we propose a \textit{heading-centric adversarial state detector} based on the kinematic bicycle model as the defense for our proposed attack. It scrutinizes the discrepancy between observed and physically plausible heading changes across consecutive frames. 
Given a wheelbase $L_{w}$ and a maximum steering angle $\theta_{\text{max}}$, following the kinematic bicycle model, the expected maximum heading change is denoted as $\Delta h_{\text{max}} = \frac{v \Delta t \tan(\theta_{\text{max}})}{L_{w}}$, 
where $v$ represents the vehicle's velocity and $\Delta t$ is the time interval between two consecutive frames. 
A scene is labeled adversarial if the observed heading changes in at least two frames exceed the calculated $\Delta h_{\text{max}}$. This serves as a criterion for detecting abnormal behaviors, such as those of a vehicle under object-based attacks.
As the defense, the detected adversarial states are replaced with the states predicted by the kinematic bicycle model using preceding states.
We evaluate the detector on 400 attack scenes under the same experiment setting as in Section~\ref{sec:experiment_dataset}.
We set $L_{w}$ at $2.5\,\text{m}$ and $\theta_{\text{max}}$ at $1\,\text{rad}$, following standard vehicle configurations. The results show that, $44.5\%$ (178/400) attacked scenes are detected. With defense, the ATD changes from $5.45\,\text{m}$ to $6.03\,\text{m}$, PRE from $1.51\,\text{m}$ to $1.15\,\text{m}$, and CR from $40\%$ to $25\%$.

\vspace{-0.5em}
\subsection{Other Possible Defenses}\label{sec:discussion_defense}
\vspace{-0.5em}
Based on the nature of our attack mechanisms, three types of existing general defense approaches might be applied. We discuss and evaluate the effectiveness of such defenses against our attack strategy.

\textit{Spatial-temporal consistency checks}: Recent studies~\cite{percepguard, Muller_VOGUES, cho2023adopt} have explored spatial-temporal consistency checks to detect perception attacks. PercepGuard~\cite{percepguard} is a representative method that leverages the spatial-temporal properties of bounding boxes across different classes to detect anomalies. In our experiments, PercepGuard detects anomalies in 4 out of 400 attack scenes, with a limit of 100 queries, under the same settings as in the dataset-based experiments. Unlike typical attacks that aim to induce object disappearance, creation, or misclassification, our attack subtly perturbs bounding boxes, presenting a challenge for spatial-temporal checks.

\textit{Adversarial training}: Following adversarial training~\cite{goodfellow2014explaining}, we fine-tune the Trajectron++ predictor with adversarial states generated by our attacks for an additional 10 epochs. After training, the CR decreases from 40\% to 35\%, while the mean Final Displacement Error increases from 2.18m to 2.22m in clean scenes. This illustrates a trade-off between robustness and accuracy in the trajectory prediction model. Moreover, effective adversarial training requires the defender to know the attack methods, which is not always possible.

\textit{MC-dropout}: We also explore MC-dropout~\cite{gal2016dropout}, a technique that introduces randomness into the perception model during inference, thus disrupting the matching between bounding box perturbations and adversarial states. Implementing MC-dropout with 5 runs led to a reduction in the collision rate from 40\% to 18\%. However, this defense significantly increases the computational overhead during inference, impacting the real-time performance capabilities of AD systems.

In conclusion, existing methods can partially defend against our attack but also introduce some additional overhead. Our proposed lightweight defense method checks key adversarial states against vehicle dynamics, achieving a good balance of efficacy and cost.




\vspace{-0.5em}
\subsection{Limitations and Future Work}\label{sec:discussion_limit_future_work}
\textbf{Single-Point Attack Impact.} A limitation of our attack is its diminishing impact once the victim AV passes the attack point. For example, a victim AV brakes suddenly at the attack point may resume normal behavior in subsequent frames. This occurs because the current frame at the attack point becomes one of the historical frames as the victim moves, being less effective as shown in finding \textbf{F1} in Section~\ref{sec:findings}. To address this limitation, developing a temporally effective attack to maintain consistent impact over time is interesting.

\textbf{Different Attack Scenarios.}
Another limitation is that our attack focuses on the driving scenario where a victim AV passes a roadside-parked adversarial vehicle.
Future work can explore other common scenarios, such as an adversarial vehicle stopped at an intersection, potentially impacting more vehicles in dense traffic scenarios. Moreover, it is interesting to extend our attack to scenarios where both the adversarial vehicle and the victim AV are moving. In these cases, drones can act as movable adversarial objects, utilizing advanced drone localization and tracking technologies.
This approach opens up an opportunity for a continuous and dynamic attack, adapting to the changing positions of the adversarial vehicle.

\textbf{Attack Multi-sensor Fusion.}\label{sec:discussion_multi_sensor_fusion}
Our attack can be extended to multi-sensor fusion systems by optimizing existing perception attacks for different sensing modalities. For example, we can use adversarial objects with specific shapes and textures to compromise both LiDAR and camera perceptions, akin to techniques discussed in~\cite{zhu2024malicious, cao2021invisible}. Specifically, the shape and texture attributes of the adversarial objects can be randomly sampled to create the bounding box perturbation set. Then, our prediction-side attack methodology can be applied to generate a set of adversarial states. Lastly, a matching and refining phase is applied, where the shape and texture parameters are adjusted to optimize the attack’s effectiveness across sensors.

\vspace{-0.5em}
\section{Conclusion}
\vspace{-0.5em}
In this paper, we investigate the possibility of compromising trajectory prediction in autonomous driving systems via physically realizable object-based LiDAR attacks on perception modules.
We find that prediction models are vulnerable to single-point attacks. Building on this insight, we introduce a novel two-stage attack framework that efficiently identifies an effective, velocity-insensitive, and feasible adversarial location set. By strategically placing objects at these locations, the attacker can manipulate the victim vehicle to predict a wrong trajectory of a nearby agent, potentially inducing hazardous victim AV responses.
We evaluate the proposed attack framework on both a public autonomous driving dataset and a custom-built real testbed car. The results demonstrate that our proposed attack consistently achieves the highest threat across various driving conditions and shows robustness against different types of noise.

\vspace{-0.5em}
\section{Acknowledgements}
\vspace{-0.5em}
We thank our shepherd and reviewers for their valuable feedback on the paper. We also thank Qian Xu and Bingyuan Huang from City University of Hong Kong for their generous assistance with the physical experiment setup. This work is supported by Hong Kong Research Grant Council under GRF project 11216323, by the US National Science Foundation under grant CNS-2120369, and by the National Research Foundation Singapore and DSO National Laboratories under the AI Singapore Programme (AISG Award No: AISG2-GC-2023-006). 


\bibliographystyle{plain}
\bibliography{usenix}

\begin{thebibliography}{10}

\bibitem{autoware_ai}
Autoware.ai.
\newblock \url{https://www.autoware.ai}.

\bibitem{baidu_apollo}
Baidu apollo.
\newblock \url{http://apollo.auto}.

\bibitem{mit_graphics}
Intro to rendering, ray casting.
\newblock \url{https://ocw.mit.edu/courses/electrical-engineering-and-computer-science/6-837-computer-graphics-fall-2012/lecture-notes/MIT6837F12_Lec11.pdf}.

\bibitem{athalye2018synthesizing}
Anish Athalye, Logan Engstrom, Andrew Ilyas, and Kevin Kwok.
\newblock Synthesizing robust adversarial examples.
\newblock In {\em International conference on machine learning}, pages 284--293. PMLR, 2018.

\bibitem{autoware_universe}
{Autoware Foundation}.
\newblock Autoware.universe.
\newblock \url{https://github.com/autowarefoundation/autoware.universe}.

\bibitem{caesar2020nuscenes}
Holger Caesar, Varun Bankiti, Alex~H Lang, Sourabh Vora, Venice~Erin Liong, Qiang Xu, Anush Krishnan, Yu~Pan, Giancarlo Baldan, and Oscar Beijbom.
\newblock nuscenes: A multimodal dataset for autonomous driving.
\newblock In {\em Proceedings of the IEEE/CVF conference on computer vision and pattern recognition}, pages 11621--11631, 2020.

\bibitem{cao2023you}
Yulong Cao, S~Hrushikesh Bhupathiraju, Pirouz Naghavi, Takeshi Sugawara, Z~Morley Mao, and Sara Rampazzi.
\newblock You can't see me: Physical removal attacks on $\{$LiDAR-based$\}$ autonomous vehicles driving frameworks.
\newblock In {\em 32nd USENIX Security Symposium (USENIX Security 23)}, pages 2993--3010, 2023.

\bibitem{cao2021invisible}
Yulong Cao, Ningfei Wang, Chaowei Xiao, Dawei Yang, Jin Fang, Ruigang Yang, Qi~Alfred Chen, Mingyan Liu, and Bo~Li.
\newblock Invisible for both camera and lidar: Security of multi-sensor fusion based perception in autonomous driving under physical-world attacks.
\newblock In {\em 2021 IEEE Symposium on Security and Privacy (SP)}, pages 176--194. IEEE, 2021.

\bibitem{cao2022advdo}
Yulong Cao, Chaowei Xiao, Anima Anandkumar, Danfei Xu, and Marco Pavone.
\newblock Advdo: Realistic adversarial attacks for trajectory prediction.
\newblock In {\em European Conference on Computer Vision}, pages 36--52. Springer, 2022.

\bibitem{cao2019adversarial}
Yulong Cao, Chaowei Xiao, Benjamin Cyr, Yimeng Zhou, Won Park, Sara Rampazzi, Qi~Alfred Chen, Kevin Fu, and Z~Morley Mao.
\newblock Adversarial sensor attack on lidar-based perception in autonomous driving.
\newblock In {\em Proceedings of the 2019 ACM SIGSAC conference on computer and communications security}, pages 2267--2281, 2019.

\bibitem{cao2023robust}
Yulong Cao, Danfei Xu, Xinshuo Weng, Zhuoqing Mao, Anima Anandkumar, Chaowei Xiao, and Marco Pavone.
\newblock Robust trajectory prediction against adversarial attacks.
\newblock In {\em Conference on Robot Learning}, pages 128--137. PMLR, 2023.

\bibitem{carlini2018audio}
Nicholas Carlini and David Wagner.
\newblock Audio adversarial examples: Targeted attacks on speech-to-text.
\newblock In {\em 2018 IEEE security and privacy workshops (SPW)}, pages 1--7. IEEE, 2018.

\bibitem{chen2021unified}
Xuesong Chen, Canmiao Fu, Feng Zheng, Yong Zhao, Hongsheng Li, Ping Luo, and Guo-Jun Qi.
\newblock A unified multi-scenario attacking network for visual object tracking.
\newblock In {\em Proceedings of the AAAI Conference on Artificial Intelligence}, volume~35, pages 1097--1104, 2021.

\bibitem{chen2022scept}
Yuxiao Chen, Boris Ivanovic, and Marco Pavone.
\newblock Scept: Scene-consistent, policy-based trajectory predictions for planning.
\newblock In {\em Proceedings of the IEEE/CVF Conference on Computer Vision and Pattern Recognition}, pages 17103--17112, 2022.

\bibitem{chen2022interactive}
Yuxiao Chen, Ugo Rosolia, Wyatt Ubellacker, Noel Csomay-Shanklin, and Aaron~D Ames.
\newblock Interactive multi-modal motion planning with branch model predictive control.
\newblock {\em IEEE Robotics and Automation Letters}, 7(2):5365--5372, 2022.

\bibitem{cho2023adopt}
Minkyoung Cho, Yulong Cao, Zixiang Zhou, and Z~Morley Mao.
\newblock Adopt: Lidar spoofing attack detection based on point-level temporal consistency.
\newblock {\em arXiv preprint arXiv:2310.14504}, 2023.

\bibitem{dolgov2010path}
Dmitri Dolgov, Sebastian Thrun, Michael Montemerlo, and James Diebel.
\newblock Path planning for autonomous vehicles in unknown semi-structured environments.
\newblock {\em The international journal of robotics research}, 29(5):485--501, 2010.

\bibitem{gal2016dropout}
Yarin Gal and Zoubin Ghahramani.
\newblock Dropout as a bayesian approximation: Representing model uncertainty in deep learning.
\newblock In {\em international conference on machine learning}, pages 1050--1059. PMLR, 2016.

\bibitem{goodfellow2014explaining}
Ian~J Goodfellow, Jonathon Shlens, and Christian Szegedy.
\newblock Explaining and harnessing adversarial examples.
\newblock {\em arXiv preprint arXiv:1412.6572}, 2014.

\bibitem{hau2021object}
Zhongyuan Hau, Kenneth~T Co, Soteris Demetriou, and Emil~C Lupu.
\newblock Object removal attacks on lidar-based 3d object detectors.
\newblock {\em arXiv preprint arXiv:2102.03722}, 2021.

\bibitem{hau2022using}
Zhongyuan Hau, Soteris Demetriou, and Emil~C Lupu.
\newblock Using 3d shadows to detect object hiding attacks on autonomous vehicle perception.
\newblock In {\em 2022 IEEE Security and Privacy Workshops (SPW)}, pages 229--235. IEEE, 2022.

\bibitem{jha2020ml}
Saurabh Jha, Shengkun Cui, Subho Banerjee, James Cyriac, Timothy Tsai, Zbigniew Kalbarczyk, and Ravishankar~K Iyer.
\newblock Ml-driven malware that targets av safety.
\newblock In {\em 2020 50th annual IEEE/IFIP international conference on dependable systems and networks (DSN)}, pages 113--124. IEEE, 2020.

\bibitem{jia2020robust}
Shuai Jia, Chao Ma, Yibing Song, and Xiaokang Yang.
\newblock Robust tracking against adversarial attacks.
\newblock In {\em Computer Vision--ECCV 2020: 16th European Conference, Glasgow, UK, August 23--28, 2020, Proceedings, Part XIX 16}, pages 69--84. Springer, 2020.

\bibitem{jia2020fooling}
Yunhan~Jia Jia, Yantao Lu, Junjie Shen, Qi~Alfred Chen, Hao Chen, Zhenyu Zhong, and Tao~Wei Wei.
\newblock Fooling detection alone is not enough: Adversarial attack against multiple object tracking.
\newblock In {\em International Conference on Learning Representations (ICLR'20)}, 2020.

\bibitem{jin2023pla}
Zizhi Jin, Xiaoyu Ji, Yushi Cheng, Bo~Yang, Chen Yan, and Wenyuan Xu.
\newblock Pla-lidar: Physical laser attacks against lidar-based 3d object detection in autonomous vehicle.
\newblock In {\em 2023 IEEE Symposium on Security and Privacy (SP)}, pages 1822--1839. IEEE, 2023.

\bibitem{kuwata2009real}
Yoshiaki Kuwata, Justin Teo, Gaston Fiore, Sertac Karaman, Emilio Frazzoli, and Jonathan~P How.
\newblock Real-time motion planning with applications to autonomous urban driving.
\newblock {\em IEEE Transactions on control systems technology}, 17(5):1105--1118, 2009.

\bibitem{lee2019physical}
Mark Lee and Zico Kolter.
\newblock On physical adversarial patches for object detection.
\newblock {\em arXiv preprint arXiv:1906.11897}, 2019.

\bibitem{lee2017desire}
Namhoon Lee, Wongun Choi, Paul Vernaza, Christopher~B Choy, Philip~HS Torr, and Manmohan Chandraker.
\newblock Desire: Distant future prediction in dynamic scenes with interacting agents.
\newblock In {\em Proceedings of the IEEE conference on computer vision and pattern recognition}, pages 336--345, 2017.

\bibitem{liang2020learning}
Ming Liang, Bin Yang, Rui Hu, Yun Chen, Renjie Liao, Song Feng, and Raquel Urtasun.
\newblock Learning lane graph representations for motion forecasting.
\newblock In {\em Computer Vision--ECCV 2020: 16th European Conference, Glasgow, UK, August 23--28, 2020, Proceedings, Part II 16}, pages 541--556. Springer, 2020.

\bibitem{liu2021multimodal}
Yicheng Liu, Jinghuai Zhang, Liangji Fang, Qinhong Jiang, and Bolei Zhou.
\newblock Multimodal motion prediction with stacked transformers.
\newblock In {\em Proceedings of the IEEE/CVF Conference on Computer Vision and Pattern Recognition}, pages 7577--7586, 2021.

\bibitem{luo2020probabilistic}
Chenxu Luo, Lin Sun, Dariush Dabiri, and Alan Yuille.
\newblock Probabilistic multi-modal trajectory prediction with lane attention for autonomous vehicles.
\newblock In {\em 2020 IEEE/RSJ International Conference on Intelligent Robots and Systems (IROS)}, pages 2370--2376. IEEE, 2020.

\bibitem{ma2023slowtrack}
Chen Ma, Ningfei Wang, Qi~Alfred Chen, and Chao Shen.
\newblock Slowtrack: Increasing the latency of camera-based perception in autonomous driving using adversarial examples.
\newblock {\em arXiv preprint arXiv:2312.09520}, 2023.

\bibitem{percepguard}
Yanmao Man, Raymond Muller, Ming Li, Z.~Berkay Celik, and Ryan Gerdes.
\newblock That person moves like a car: Misclassification attack detection for autonomous systems using spatiotemporal consistency.
\newblock In {\em 32nd USENIX Security Symposium (USENIX Security 23)}, pages 6929--6946, Anaheim, CA, August 2023. USENIX Association.

\bibitem{muller2022physical}
Raymond Muller, Yanmao Man, Z~Berkay Celik, Ming Li, and Ryan Gerdes.
\newblock Physical hijacking attacks against object trackers.
\newblock In {\em Proceedings of the 2022 ACM SIGSAC Conference on Computer and Communications Security}, pages 2309--2322, 2022.

\bibitem{Muller_VOGUES}
Raymond Muller, Yanmao Man, Ryan Gerdes, Ming Li, Jonathan Petit, and Z.~Berkay Celik.
\newblock Vogues: Validation of object guise using estimated components.
\newblock In {\em 33rd USENIX Security Symposium (USENIX Security '24 Fall)}, 2024.

\bibitem{npr2023traffic}
{NPR}.
\newblock San francisco's driverless cars are causing traffic jams, city officials say.
\newblock \url{https://www.npr.org/2023/08/26/1195695051/driverless-cars-san-francisco-waymo-cruise}, August 2023.

\bibitem{papernot2017practical}
Nicolas Papernot, Patrick McDaniel, Ian Goodfellow, Somesh Jha, Z~Berkay Celik, and Ananthram Swami.
\newblock Practical black-box attacks against machine learning.
\newblock In {\em Proceedings of the 2017 ACM on Asia conference on computer and communications security}, pages 506--519, 2017.

\bibitem{petit2015remote}
Jonathan Petit, Bas Stottelaar, Michael Feiri, and Frank Kargl.
\newblock Remote attacks on automated vehicles sensors: Experiments on camera and lidar.
\newblock {\em Black Hat Europe}, 11(2015):995, 2015.

\bibitem{salzmann2020trajectron++}
Tim Salzmann, Boris Ivanovic, Punarjay Chakravarty, and Marco Pavone.
\newblock Trajectron++: Dynamically-feasible trajectory forecasting with heterogeneous data.
\newblock In {\em Computer Vision--ECCV 2020: 16th European Conference, Glasgow, UK, August 23--28, 2020, Proceedings, Part XVIII 16}, pages 683--700. Springer, 2020.

\bibitem{sato2023wip}
Takami Sato, Yuki Hayakawa, Ryo Suzuki, Yohsuke Shiiki, Kentaro Yoshioka, and Qi~Alfred Chen.
\newblock Wip: Practical removal attacks on lidar-based object detection in autonomous driving.
\newblock In {\em ISOC Symposium on Vehicle Security and Privacy (VehicleSec)}, 2023.

\bibitem{scheel2022urban}
Oliver Scheel, Luca Bergamini, Maciej Wolczyk, B{\l}a{\.z}ej Osi{\'n}ski, and Peter Ondruska.
\newblock Urban driver: Learning to drive from real-world demonstrations using policy gradients.
\newblock In {\em Conference on Robot Learning}, pages 718--728. PMLR, 2022.

\bibitem{sharif2016accessorize}
Mahmood Sharif, Sruti Bhagavatula, Lujo Bauer, and Michael~K Reiter.
\newblock Accessorize to a crime: Real and stealthy attacks on state-of-the-art face recognition.
\newblock In {\em Proceedings of the 2016 acm sigsac conference on computer and communications security}, pages 1528--1540, 2016.

\bibitem{shi2019pointrcnn}
Shaoshuai Shi, Xiaogang Wang, and Hongsheng Li.
\newblock Pointrcnn: 3d object proposal generation and detection from point cloud.
\newblock In {\em Proceedings of the IEEE/CVF conference on computer vision and pattern recognition}, pages 770--779, 2019.

\bibitem{shin2017illusion}
Hocheol Shin, Dohyun Kim, Yujin Kwon, and Yongdae Kim.
\newblock Illusion and dazzle: Adversarial optical channel exploits against lidars for automotive applications.
\newblock In {\em Cryptographic Hardware and Embedded Systems--CHES 2017: 19th International Conference, Taipei, Taiwan, September 25-28, 2017, Proceedings}, pages 445--467. Springer, 2017.

\bibitem{sun2020towards}
Jiachen Sun, Yulong Cao, Qi~Alfred Chen, and Z~Morley Mao.
\newblock Towards robust $\{$LiDAR-based$\}$ perception in autonomous driving: General black-box adversarial sensor attack and countermeasures.
\newblock In {\em 29th USENIX Security Symposium (USENIX Security 20)}, pages 877--894, 2020.

\bibitem{suya2020hybrid}
Fnu Suya, Jianfeng Chi, David Evans, and Yuan Tian.
\newblock Hybrid batch attacks: Finding black-box adversarial examples with limited queries.
\newblock In {\em 29th USENIX Security Symposium (USENIX Security 20)}, pages 1327--1344, 2020.

\bibitem{tan2023targeted}
Kaiyuan Tan, Jun Wang, and Yiannis Kantaros.
\newblock Targeted adversarial attacks against neural network trajectory predictors.
\newblock In {\em Learning for Dynamics and Control Conference}, pages 431--444. PMLR, 2023.

\bibitem{tsai2020robust}
Tzungyu Tsai, Kaichen Yang, Tsung-Yi Ho, and Yier Jin.
\newblock Robust adversarial objects against deep learning models.
\newblock In {\em Proceedings of the AAAI Conference on Artificial Intelligence}, volume~34, pages 954--962, 2020.

\bibitem{tu2020physically}
James Tu, Mengye Ren, Sivabalan Manivasagam, Ming Liang, Bin Yang, Richard Du, Frank Cheng, and Raquel Urtasun.
\newblock Physically realizable adversarial examples for lidar object detection.
\newblock In {\em Proceedings of the IEEE/CVF Conference on Computer Vision and Pattern Recognition}, pages 13716--13725, 2020.

\bibitem{velodyne_idriverplus_2019}
{Velodyne Lidar}.
\newblock Idriverplus building smart autonomous vehicles with velodyne lidar technology.
\newblock 7 2019.

\bibitem{vueron_vueone}
{Vueron}.
\newblock Vueone -- {VUERON} | {LiDAR} solution provider.
\newblock \url{https://vueron.com/vueone/}, 2023.

\bibitem{wan2022too}
Ziwen Wan, Junjie Shen, Jalen Chuang, Xin Xia, Joshua Garcia, Jiaqi Ma, and Qi~Alfred Chen.
\newblock Too afraid to drive: Systematic discovery of semantic dos vulnerability in autonomous driving planning under physical-world attacks.
\newblock In {\em Network and Distributed System Security (NDSS) Symposium}, April 2022.

\bibitem{xie2017adversarial}
Cihang Xie, Jianyu Wang, Zhishuai Zhang, Yuyin Zhou, Lingxi Xie, and Alan Yuille.
\newblock Adversarial examples for semantic segmentation and object detection.
\newblock In {\em Proceedings of the IEEE international conference on computer vision}, pages 1369--1378, 2017.

\bibitem{yan2020hijacking}
Xiyu Yan, Xuesong Chen, Yong Jiang, Shu-Tao Xia, Yong Zhao, and Feng Zheng.
\newblock Hijacking tracker: A powerful adversarial attack on visual tracking.
\newblock In {\em ICASSP 2020-2020 IEEE International Conference on Acoustics, Speech and Signal Processing (ICASSP)}, pages 2897--2901. IEEE, 2020.

\bibitem{yan2018second}
Yan Yan, Yuxing Mao, and Bo~Li.
\newblock Second: Sparsely embedded convolutional detection.
\newblock {\em Sensors}, 18(10):3337, 2018.

\bibitem{yang2018pixor}
Bin Yang, Wenjie Luo, and Raquel Urtasun.
\newblock Pixor: Real-time 3d object detection from point clouds.
\newblock In {\em Proceedings of the IEEE conference on Computer Vision and Pattern Recognition}, pages 7652--7660, 2018.

\bibitem{yin2021center}
Tianwei Yin, Xingyi Zhou, and Philipp Krahenbuhl.
\newblock Center-based 3d object detection and tracking.
\newblock In {\em Proceedings of the IEEE/CVF conference on computer vision and pattern recognition}, pages 11784--11793, 2021.

\bibitem{yuan2021agentformer}
Ye~Yuan, Xinshuo Weng, Yanglan Ou, and Kris~M Kitani.
\newblock Agentformer: Agent-aware transformers for socio-temporal multi-agent forecasting.
\newblock In {\em Proceedings of the IEEE/CVF International Conference on Computer Vision}, pages 9813--9823, 2021.

\bibitem{zeng2021lanercnn}
Wenyuan Zeng, Ming Liang, Renjie Liao, and Raquel Urtasun.
\newblock Lanercnn: Distributed representations for graph-centric motion forecasting.
\newblock In {\em 2021 IEEE/RSJ International Conference on Intelligent Robots and Systems (IROS)}, pages 532--539. IEEE, 2021.

\bibitem{zeng2019end}
Wenyuan Zeng, Wenjie Luo, Simon Suo, Abbas Sadat, Bin Yang, Sergio Casas, and Raquel Urtasun.
\newblock End-to-end interpretable neural motion planner.
\newblock In {\em Proceedings of the IEEE/CVF Conference on Computer Vision and Pattern Recognition}, pages 8660--8669, 2019.

\bibitem{zhang2022adversarial}
Qingzhao Zhang, Shengtuo Hu, Jiachen Sun, Qi~Alfred Chen, and Z~Morley Mao.
\newblock On adversarial robustness of trajectory prediction for autonomous vehicles.
\newblock In {\em Proceedings of the IEEE/CVF Conference on Computer Vision and Pattern Recognition}, pages 15159--15168, 2022.

\bibitem{zhang2021argot}
Zihan Zhang, Mingxuan Liu, Chao Zhang, Yiming Zhang, Zhou Li, Qi~Li, Haixin Duan, and Donghong Sun.
\newblock Argot: Generating adversarial readable chinese texts.
\newblock In {\em Proceedings of the Twenty-Ninth International Conference on International Joint Conferences on Artificial Intelligence}, pages 2533--2539, 2021.

\bibitem{zhou2022hivt}
Zikang Zhou, Luyao Ye, Jianping Wang, Kui Wu, and Kejie Lu.
\newblock Hivt: Hierarchical vector transformer for multi-agent motion prediction.
\newblock In {\em Proceedings of the IEEE/CVF Conference on Computer Vision and Pattern Recognition}, pages 8823--8833, 2022.

\bibitem{zhu2021adversarial}
Yi~Zhu, Chenglin Miao, Foad Hajiaghajani, Mengdi Huai, Lu~Su, and Chunming Qiao.
\newblock Adversarial attacks against lidar semantic segmentation in autonomous driving.
\newblock In {\em Proceedings of the 19th ACM conference on embedded networked sensor systems}, pages 329--342, 2021.

\bibitem{zhu2024malicious}
Yi~Zhu, Chenglin Miao, Hongfei Xue, Yunnan Yu, Lu~Su, and Chunming Qiao.
\newblock Malicious attacks against multi-sensor fusion in autonomous driving.
\newblock In {\em Proceedings of the 30th Annual International Conference on Mobile Computing and Networking}, pages 436--451, 2024.

\bibitem{zhu2021can}
Yi~Zhu, Chenglin Miao, Tianhang Zheng, Foad Hajiaghajani, Lu~Su, and Chunming Qiao.
\newblock Can we use arbitrary objects to attack lidar perception in autonomous driving?
\newblock In {\em Proceedings of the 2021 ACM SIGSAC Conference on Computer and Communications Security}, pages 1945--1960, 2021.

\end{thebibliography}

\appendix
\section*{Appendix}
\renewcommand{\thesubsection}{\Alph{subsection}}

\subsection{Generalizability Across Models}\label{sec:appendix_generalizability}
\begin{figure}[!ht]
    \centering
    \includegraphics[width=0.45\columnwidth]{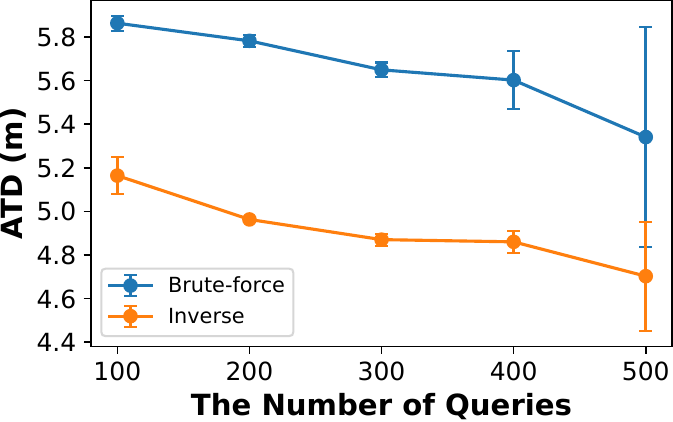}\\
    \vspace{1em}
    \hfill
    \includegraphics[width=0.45\columnwidth]{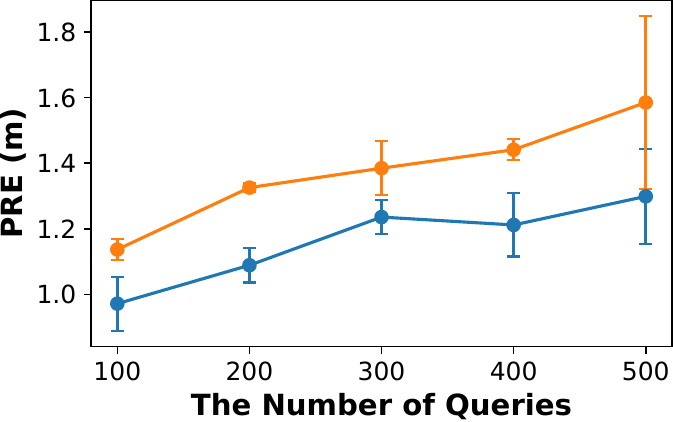}
    \hfill
    \includegraphics[width=0.45\columnwidth]{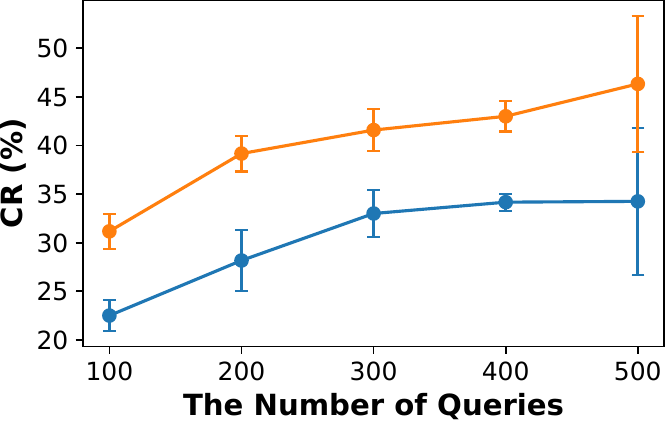}
    \hfill
    \vspace{-0.5em}
    \caption{ATD, PRE and CR under brute-force sampling attack and our inverse attack on victim AV using AgentFormer trajectory predictor.}
    \label{fig:eval_errorbar_appendix}
    
\end{figure}

To extend our analysis beyond Trajectron++, the primary trajectory prediction model utilized in this study, we apply our attack to AgentFormer~\cite{yuan2021agentformer}, another advanced trajectory predictor, to assess its generalizability. We follow the same experimental settings as those used in the dataset-based experiments, and the AgentFormer model is also trained on the nuScenes dataset using its default configuration. Fig.~\ref{fig:eval_errorbar_appendix} compares the Average Trajectory Distance (ATD), Planning-Response Error (PRE), and Collision Rate (CR) metrics between our inverse attack method and the brute-force sampling method when applied to the victim AV system utilizing AgentFormer. The number of queries ranges from 100 to 500. The results show that our inverse attack consistently outperforms the brute-force sampling method on three metrics across all query limitations. These outcomes, similar to those presented in Fig.~\ref{fig:eval_errorbar}, further demonstrating the generalizability of our attack. 

\subsection{Supplementary Experiments}
\subsubsection{Experiments on Dataset}\label{sec:experiment_dataset_appendix}
\begin{figure}[!t]
\centering
\includegraphics[width=0.45\columnwidth]{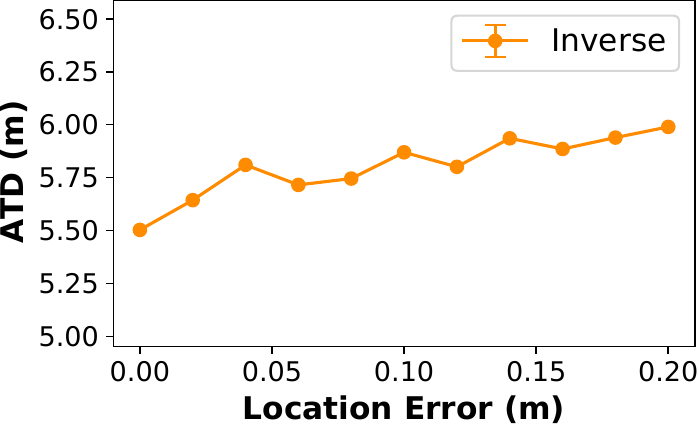} 
\hfill
\includegraphics[width=0.45\columnwidth]{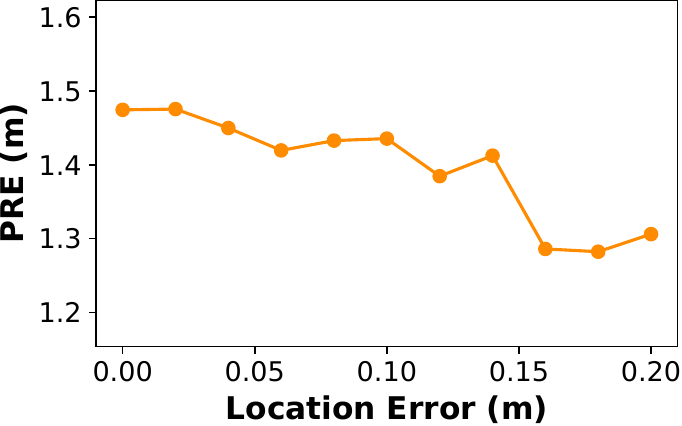} 
\vspace{-0.5em}
\caption{Attack robustness against object displacement.}
\label{fig:loc_error_appendix}
\end{figure}

\begin{figure}[!t]
\centering
\includegraphics[width=0.45\columnwidth]{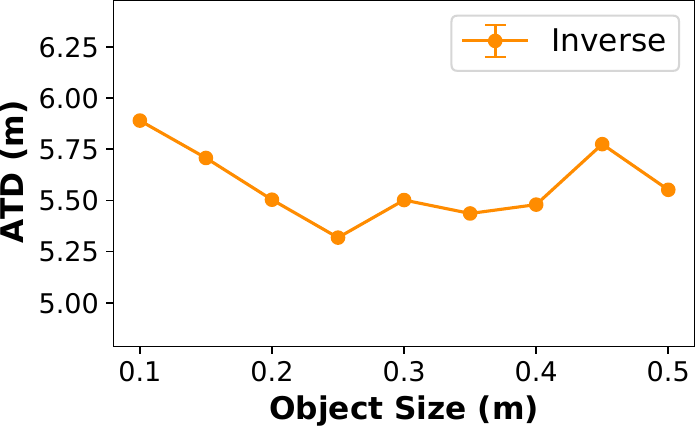} 
\hfill
\includegraphics[width=0.45\columnwidth]{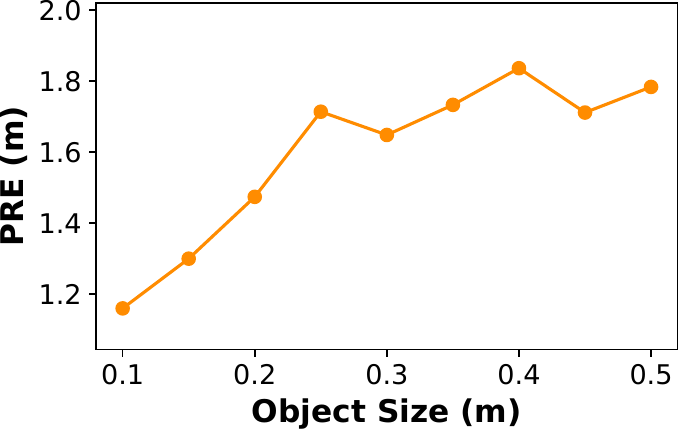}
\vspace{-0.5em}
\caption{Attack robustness against object size.}
\label{fig:object_size_appendix}
\end{figure}

\begin{figure}[!t]
\centering
\includegraphics[width=0.45\columnwidth]{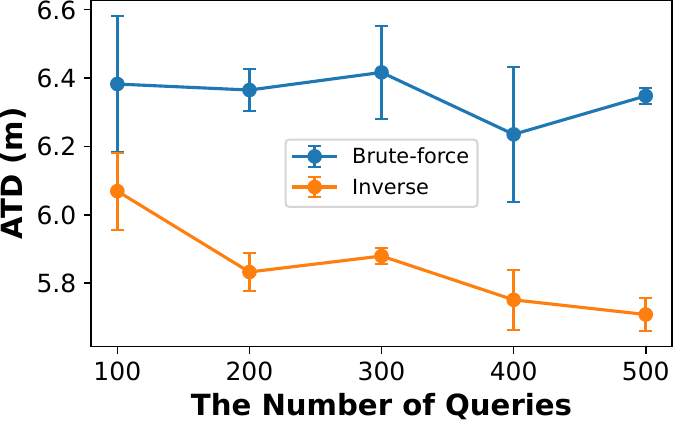}\\
\vspace{1em}
\hfill
\includegraphics[width=0.45\columnwidth]{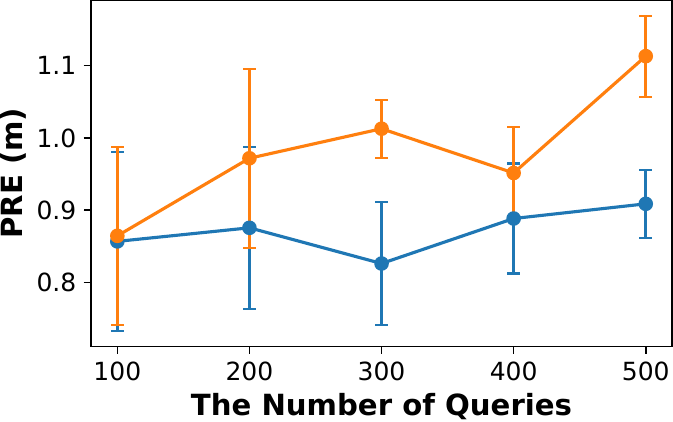}
\hfill
\includegraphics[width=0.45\columnwidth]{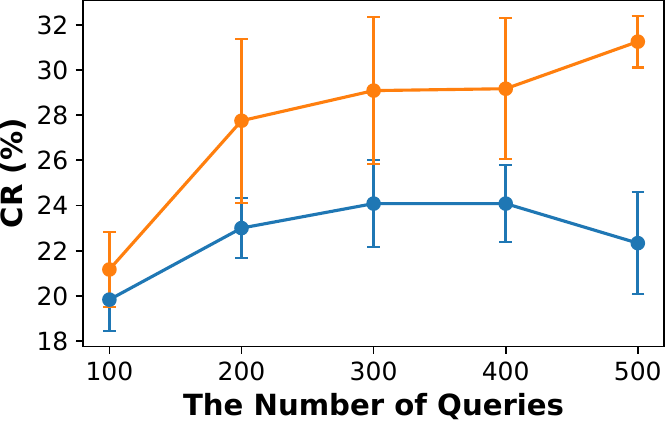}
\hfill
\vspace{-0.5em}
\caption{ATD, PRE, and CR under brute-force sampling attack and our inverse attack in the black-box scenario. The attack is planned using Trajectron++ predictor and evaluated on the AgentFormer model.}
\label{fig:appendix_eval_errorbar_blackbox}
\end{figure}

Fig.~\ref{fig:loc_error_appendix} shows the Average Trajectory Distance (ATD) and Planning-Response Error (PRE) of our inverse attack method under varied object displacements.
Fig.~\ref{fig:object_size_appendix} demonstrates how different object sizes affect the ATD and PRE metrics of our inverse attack method.
Fig.~\ref{fig:appendix_eval_errorbar_blackbox} illustrates the ATD, PRE, and CR metrics in the attack transferability experiments, where the attack is planned using the Trajectron++ predictor and evaluated on the AgentFormer model.


\subsubsection{Experiments in Physical World}\label{sec:experiment_physical_appendix}
\begin{figure}[!t]
    \centering
    \hfill
    \subfigure[Testbed car equipped with LiDAR and GNSS.]{
        \includegraphics[width=0.4\columnwidth]{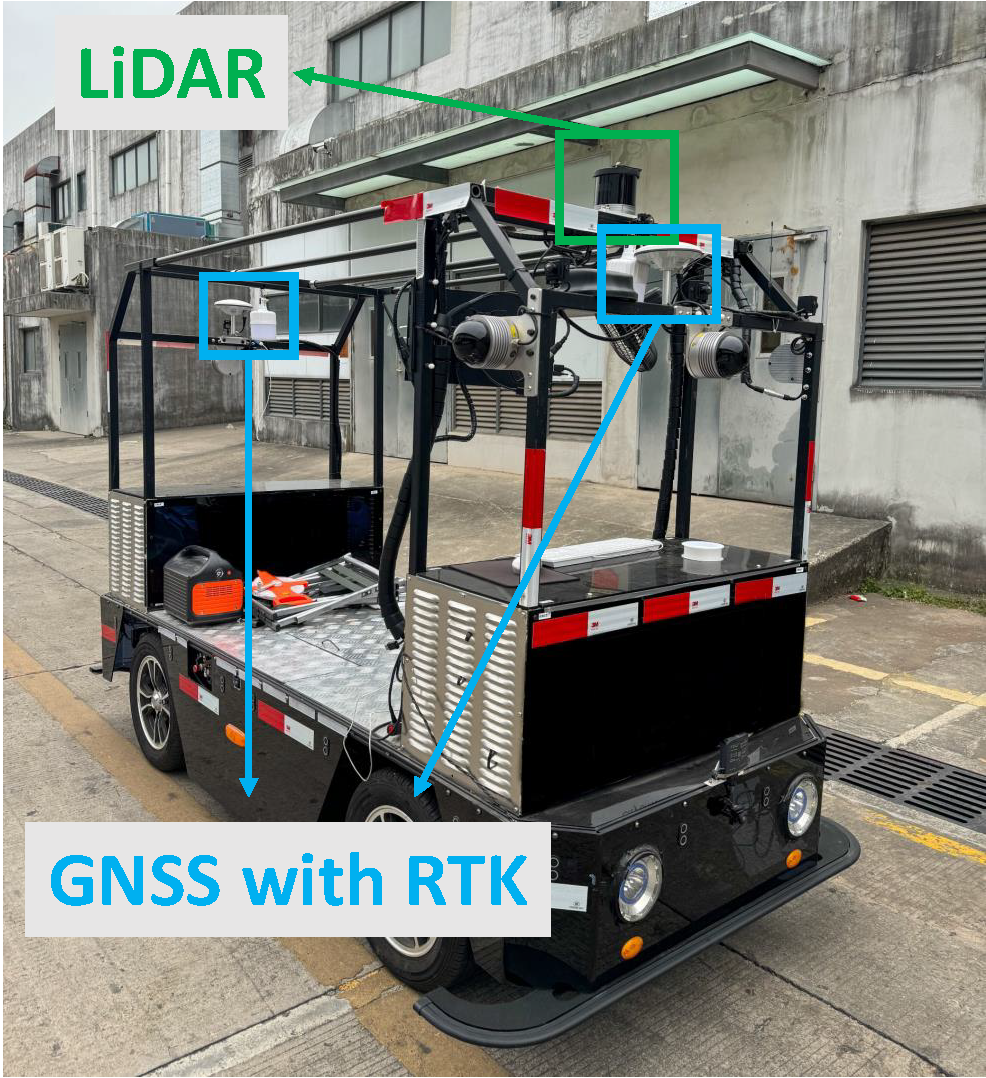}
    }
    \hfill
    \subfigure[Cardboards affixed to tripods.]{
        \includegraphics[width=0.31\columnwidth]{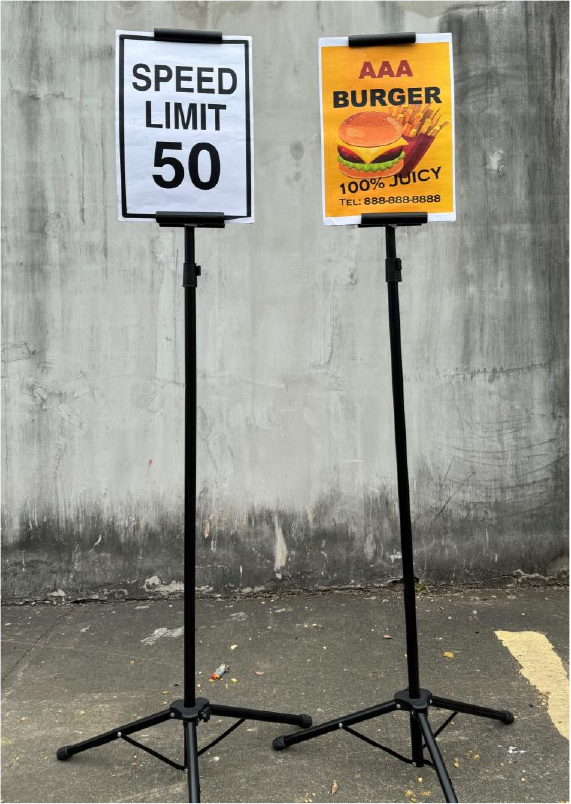}
    }
    \hfill
    \vspace{-0.5em}
    \caption{Physical-world experiment setup.}
    \label{fig:setup_appendix}
\end{figure}

\textbf{Experimental Settings}
Fig.~\ref{fig:setup_appendix} provides a detailed view of the testbed car and cardboards used in the physical-world experiment. The top LiDAR, marked in the green box, is responsible for collecting point cloud data. The GNSS Receiver with RTK, highlighted in blue, enables precise localization of the testbed car, facilitating the construction of an accurate coordinate system essential for prediction and planning. Regarding the cardboards, we design them to mimic traffic signs and billboards, which are common in driving scenarios, thereby enhancing the stealthiness of our attack.

\begin{figure}[!t]
    \centering
    \hfill
    \subfigure[Point cloud view of the left-side attack scene.]{
        \label{fig:left_attack_pc}
        \includegraphics[width=0.45\columnwidth]{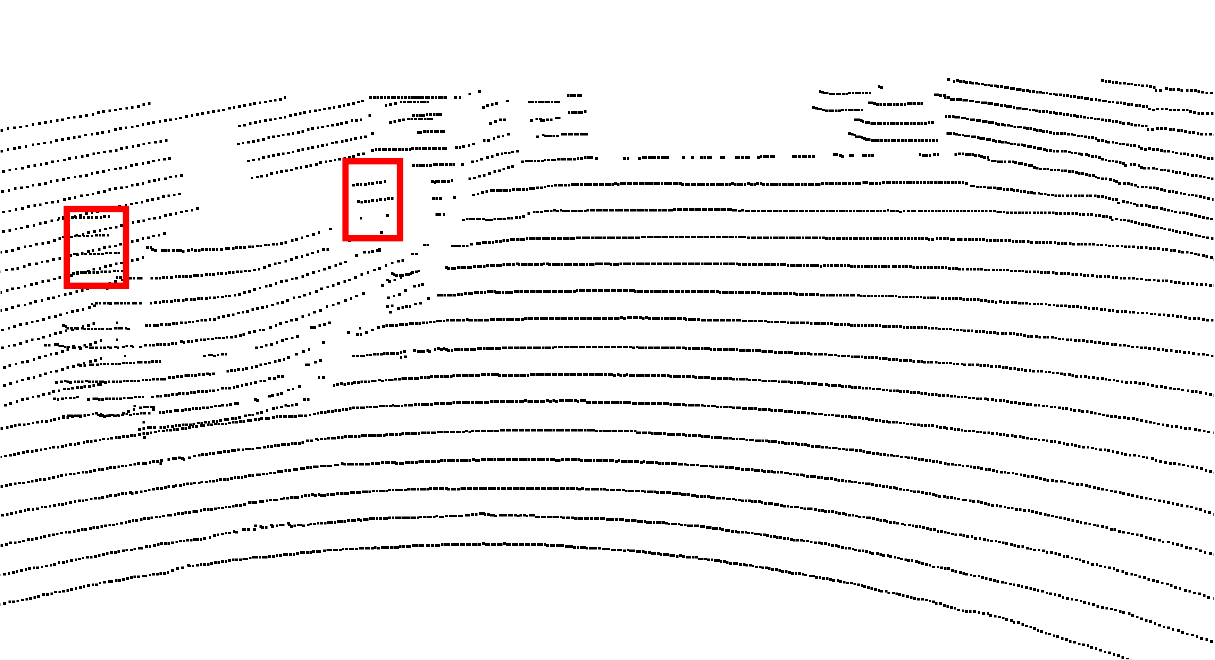}
    }
    \subfigure[Point cloud view of the right-side attack scene.]{
        \label{fig:right_attack_pc}
        \includegraphics[width=0.45\columnwidth]{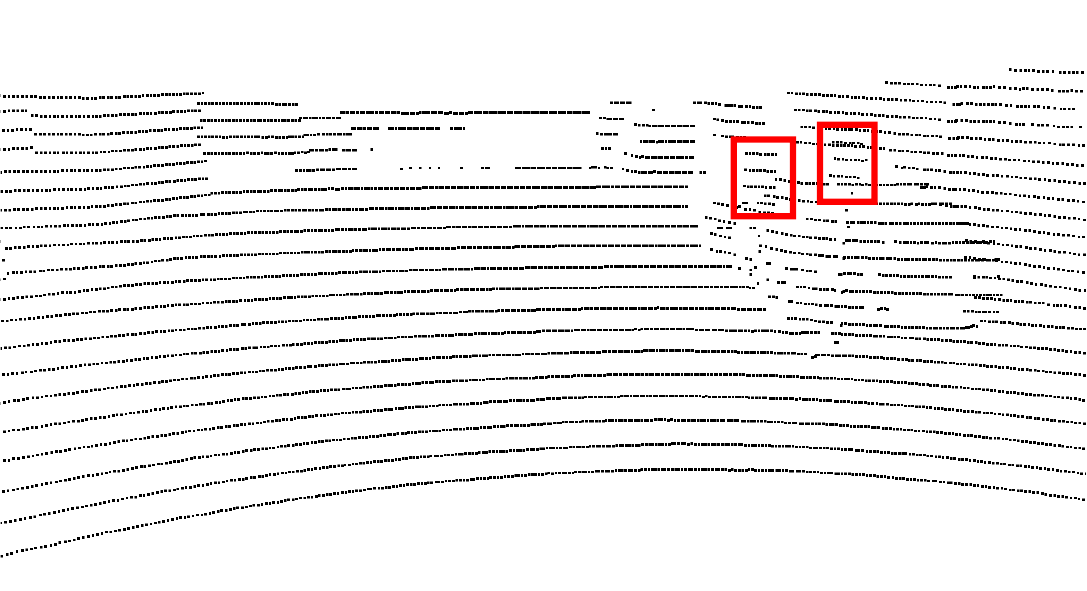}
    }
    \hfill
    \vspace{-0.5em}
    \caption{LiDAR point cloud view of two real-world attack scenarios.}
    \label{fig:pointcloud_view_appendix}
    
\end{figure}

\begin{table}[!t]
    \centering
    \caption{ATD and PRE under clean scenario and our inverse attack at velocities of $5\,\text{km/h}$ and $10\,\text{km/h}$.}
    \label{tab:phy_eval_metrics}
    \vspace{0.5em}
    \resizebox{0.9\columnwidth}{!}{
    \begin{tabular}{@{}c|ccc|ccc@{}}
    \toprule
    \multirow{2}{*}{Scenario} & \multicolumn{3}{c|}{Left} & \multicolumn{3}{c}{Right} \\
                           & Clean  & $5\,\text{km/h}$  & $10\,\text{km/h}$  & Clean  & $5\,\text{km/h}$  & $10\,\text{km/h}$  \\ \midrule
    ATD                 & 4.78      & 3.50  & 4.12  & 4.71      & 3.14  & 3.62  \\
    PRE                 & 0.00      & 1.36  & 2.01  & 0.00      & 1.17  & 1.83  \\ \bottomrule
    \end{tabular}
    }
\end{table}

\textbf{Attack Effectiveness} 
Figs.~\ref{fig:left_attack_pc} and~\ref{fig:right_attack_pc} depict the LiDAR point cloud views captured by the victim AV (our testbed car) in the left- and right-side attack scenarios, respectively. Notably, cardboards, outlined in red boxes, contribute additional point clusters to the collected LiDAR point cloud data. These point clusters serve as the basis of our attack, inducing perception and prediction errors.
Table~\ref{tab:phy_eval_metrics} presents the ATD and PRE metrics in the absence and presence of our inverse attack, across evaluation velocities of $5\,\text{km/h}$ and $10\,\text{km/h}$, within both the left- and right-side scenarios. The results indicate that, compared to the clean scenario (without attack), our inverse attack significantly reduces ATD and increases PRE, quantitatively demonstrating its effectiveness.

\begin{figure}[!t]
    \centering
    \hfill
    \subfigure[Left-side scenario under random location attack.]{
        \includegraphics[width=0.4\columnwidth]{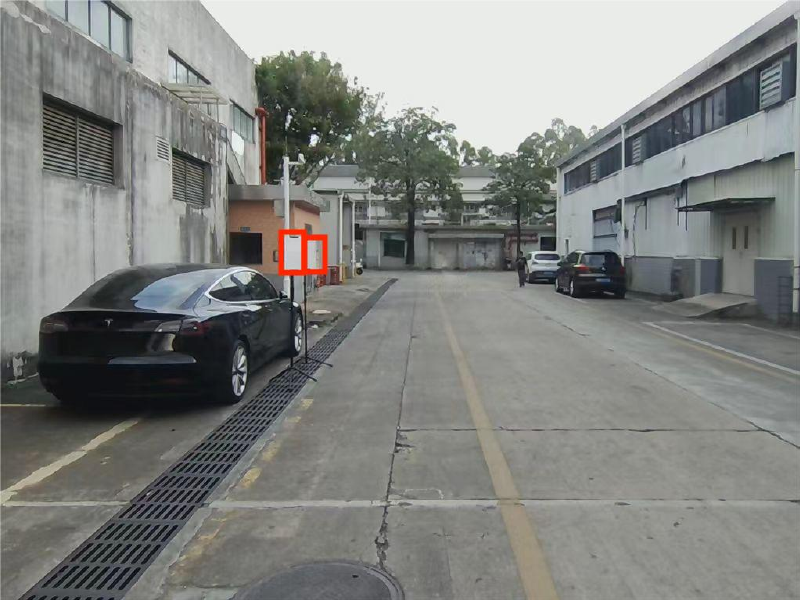}
    }
    \hfill
    \subfigure[Right-side scenario under random location attack.]{
        \includegraphics[width=0.4\columnwidth]{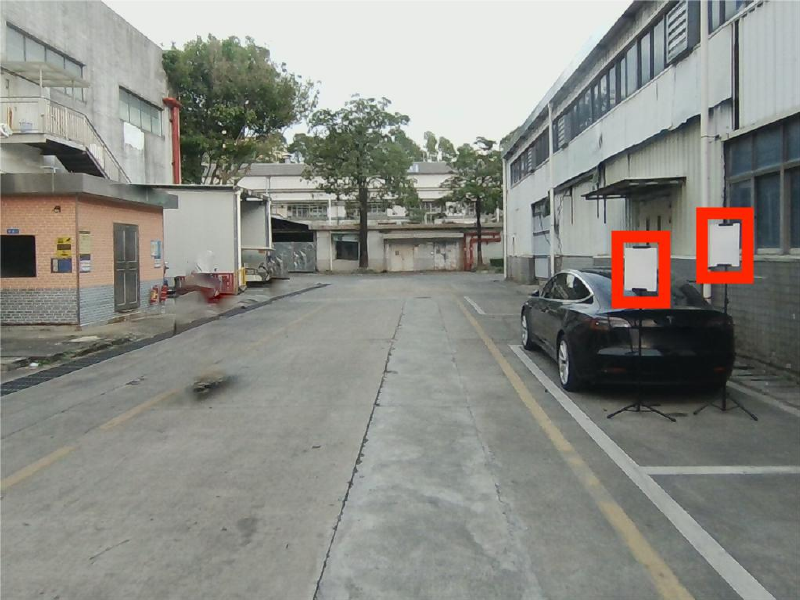}
    }
    \hfill
    \vspace{-0.5em}
    \caption{Random location attack scenes.}
    \label{fig:random_loc_scene_appendix}
    
\end{figure}

\begin{figure}[!t]
    \centering
    \hfill
    \subfigure[Orientation shift scene.]{
        \label{fig:loc_error_orientation_scene}
        \includegraphics[width=0.4\columnwidth]{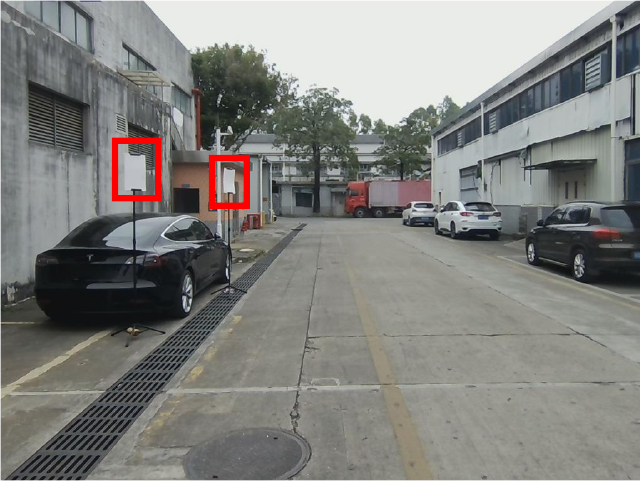}
    }
    \hfill
    \subfigure[Coordinate shift (0.3m) scene.]{
        \label{fig:loc_error_03_scene}
        \includegraphics[width=0.4\columnwidth]{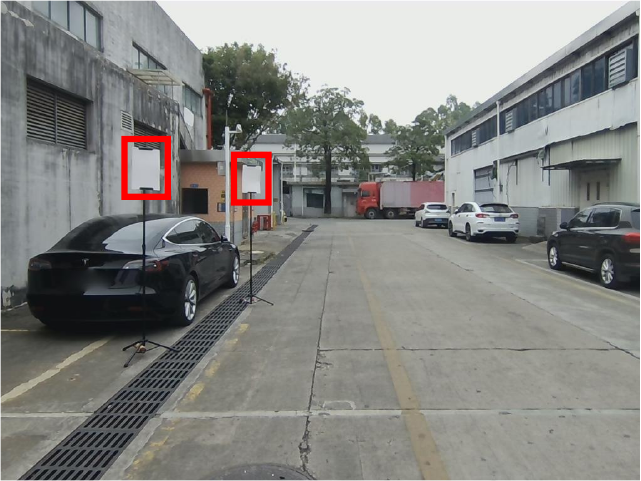}
    }
    \hfill
    \vspace{-0.5em}
    \caption{Scenes with object displacement errors induced by shifts in object coordinates or orientations.}
    \label{fig:loc_error_realworld_appendix}
    
\end{figure}

\begin{figure}[!t]
    \centering
    \hfill
    \subfigure[Driving direction to left.]{
        \label{fig:direction_left_scene}
        \includegraphics[width=0.4\columnwidth]{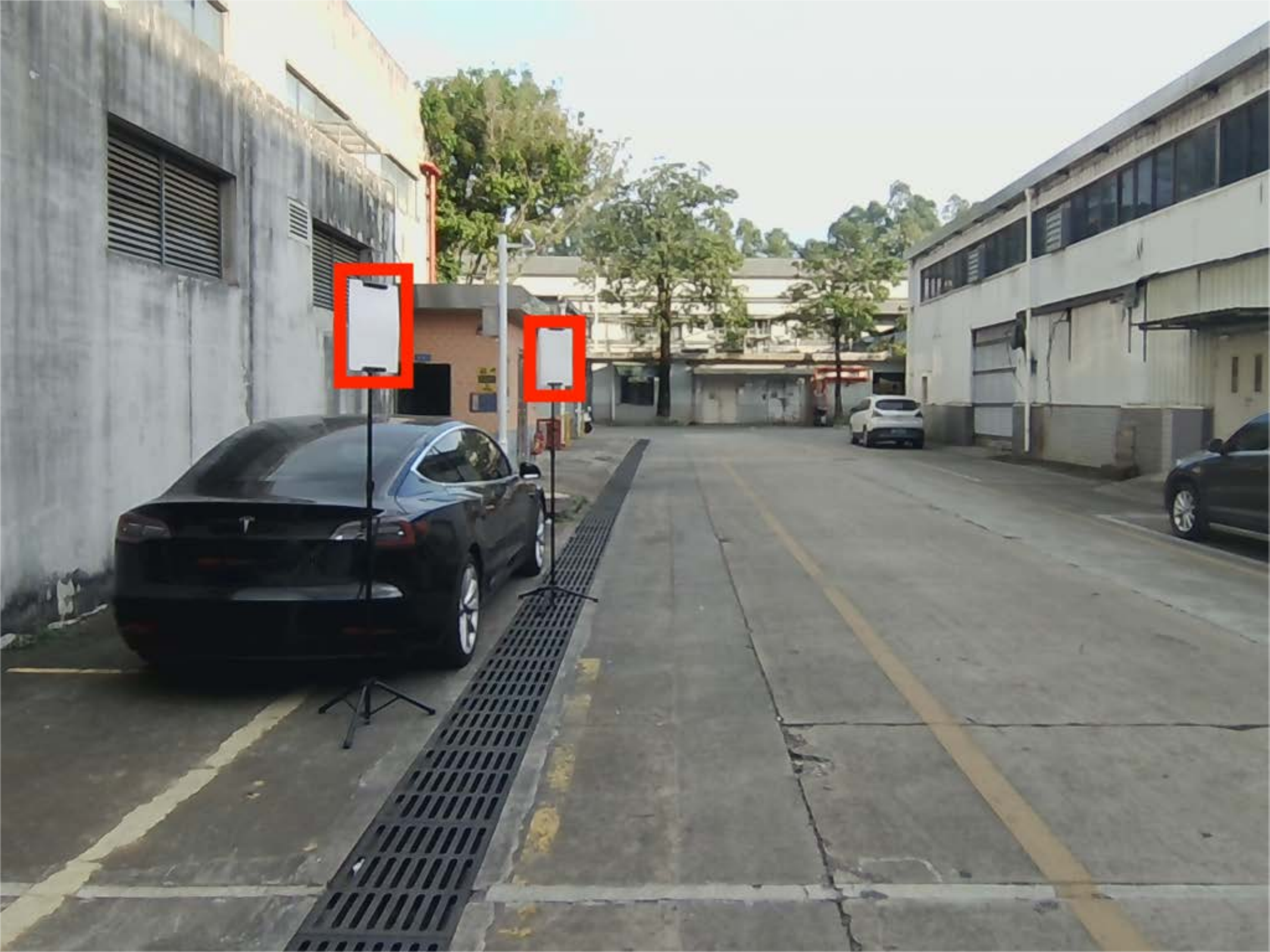}
    }
    \hfill
    \subfigure[Driving direction to right.]{
        \label{fig:direction_right_scene}
        \includegraphics[width=0.4\columnwidth]{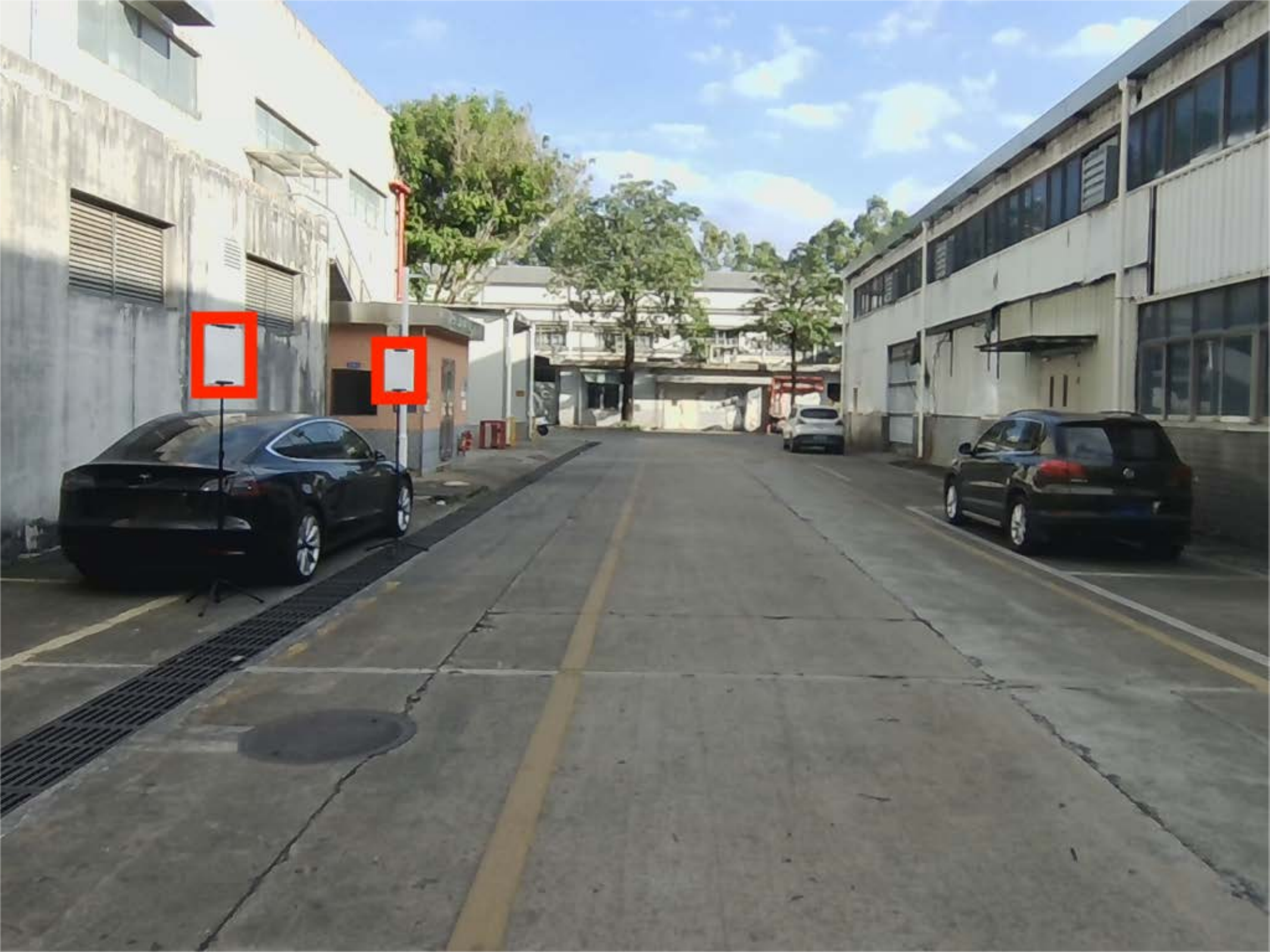}
    }
    \hfill
    \vspace{-1em}
    \caption{Scenes under various driving directions of the victim AV.}
    \label{fig:diff_direction_appendix}
    \vspace{-1em}
\end{figure}

\textbf{Attack Robustness}
Fig.~\ref{fig:random_loc_scene_appendix} shows example scenes of the random location attack, with their corresponding results presented in Fig.~\ref{fig:random_attack}. In each trial, we sample a varied set of adversarial locations beyond those depicted, including positions beside and behind the adversarial vehicle.

Fig.~\ref{fig:loc_error_realworld_appendix} illustrates example scenes with object displacements. Specifically, in Fig.~\ref{fig:loc_error_orientation_scene}, the cardboards are adjusted to face forward, deviating from their initial orientations toward the victim vehicle.
In Fig.~\ref{fig:loc_error_03_scene}, the coordinates of cardboards are shifted by $0.3\,\text{m}$ from the predetermined adversarial locations, notably moving them toward the road.
In real-world deployment, attackers equipped with distance-measuring devices, such as a tape measure, can precisely place objects at specific adversarial locations.

The example scenes depicting various driving directions of the victim AV are presented in Fig.~\ref{fig:diff_direction_appendix}, with the associated attack results shown in Fig.~\ref{fig:diff_direction}.
\end{document}